\documentclass[aps,pra,reprint,amsmath,amssymb,superscriptaddress,onecolumn,longbibliography,nofootinbib,notitlepage]{revtex4-2}
\usepackage{mathtools}
\usepackage{amsmath}
\usepackage{siunitx}
\usepackage[hidelinks]{hyperref}
\usepackage{svg}
\usepackage{float}
\usepackage[braket, qm]{qcircuit}
\usepackage{bbm}
\usepackage{bm}
\usepackage{lipsum} 
\usepackage{tocloft}
\usepackage{titletoc}
\usepackage{graphicx}
\usepackage{multirow}
\graphicspath{{./Figures/}}
\usepackage{lipsum} 
\usepackage[most]{tcolorbox} 
\usepackage{bbding}
\newtcolorbox{redbox}[1]{colback=red!5!white,colframe=red!75!black,fonttitle=\bfseries,title=#1}
\usepackage{rotating}
\usepackage[shortlabels]{enumitem}

\newcommand{\nocontentsline}[3]{}
\let\origcontentsline\addcontentsline
\newcommand\stoptoc{\let\addcontentsline\nocontentsline}
\newcommand\resumetoc{\let\addcontentsline\origcontentsline}

\newcommand{\avg}[1]{\langle #1 \rangle}

\newcommand{\rhou}{\hat{\rho}}
\newcommand{\Usense}{\mathcal{U}_{\rm sense}}
\newcommand{\Uqs}{\mathcal{U}_{\rm QS}}
\newcommand{\Uqcs}{\mathcal{U}_{\rm QCS}}

\newcommand{\Fqs}{\mathcal{F}_{\rm QS}}
\newcommand{\Fqcs}{\mathcal{F}_{\rm QCS}}
\newcommand{\xqs}{\bm{x}_{\rm QS}}
\newcommand{\xqcs}{\bm{x}_{\rm QCS}}
\newcommand{\Xqs}{\bm{X}_{\rm QS}}
\newcommand{\Xqcs}{\bm{X}_{\rm QCS}}
\newcommand{\Xbqs}{\bm{\bar{X}}_{\rm QS}}

\newcommand{\Udec}{\mathcal{U}_{\rm meas}}
\newcommand{\Uenc}{\mathcal{U}_{\rm probe}}
\newcommand{\Ucoh}[1]{\mathcal{U}^{(#1)}_{{\rm coh}}}

\newcommand{\XOR}{\texttt{XOR}}
\newcommand{\logspirals}{\texttt{Logspirals}}
\newcommand{\spirals}{\texttt{Spirals}}
\newcommand{\circles}{\texttt{Circles}}

\newcommand{\F}{\mathcal{F}}
\newcommand{\Ft}{\mathcal{F}^{\star}}
\newcommand{\MSE}{{\rm MSE}}
\newcommand{\EMSE}{\mathbb{E}[{\rm MSE}]}

\newcommand{\QCSA}{QCSA}

\begin{document}
\title{Quantum computational sensing using quantum signal processing,\\quantum neural networks, and Hamiltonian engineering}

\author{Saeed~A.~Khan}
\thanks{Equal contribution.}
\affiliation{School of Applied and Engineering Physics, Cornell University, Ithaca, NY 14853, USA}

\author{Sridhar~Prabhu}
\thanks{Equal contribution.}
\affiliation{School of Applied and Engineering Physics, Cornell University, Ithaca, NY 14853, USA}
\affiliation{Department of Physics, Cornell University, Ithaca, NY 14853, USA}

\author{Logan~G.~Wright}
\thanks{Present address: Department of Applied Physics, Yale University, New Haven, CT 06520, USA}
\affiliation{School of Applied and Engineering Physics, Cornell University, Ithaca, NY 14853, USA}

\author{Peter~L.~McMahon}
\email{To whom correspondence should be addressed: sk3239@cornell.edu, svp36@cornell.edu, pmcmahon@cornell.edu}
\affiliation{School of Applied and Engineering Physics, Cornell University, Ithaca, NY 14853, USA}
\affiliation{Kavli Institute at Cornell for Nanoscale Science, Cornell University, Ithaca, NY 14853, USA}

\begin{abstract}
Combining quantum sensing with quantum computing can lead to quantum computational sensors that are able to more efficiently extract task-specific information from physical signals than is possible otherwise. Early examples of quantum computational sensing (QCS) have largely focused on protocols where only a single sensing operation appears before measurement---with an exception being the recent application of Grover's algorithm to signal detection. In this paper we present, in theory and numerical simulations, the application of two quantum algorithms---quantum signal processing and quantum neural networks---to various binary and multiclass machine-learning classification tasks in sensing. Here sensing operations are interleaved with computing operations, giving rise to nonlinear functions of the sensed signals. We have evaluated tasks based on static and time-varying signals, including a classification task that requires distinguishing magnetic-field signals sensed by up to 7 spatially separated qubits, where the task dataset was obtained from experimentally recorded spatiotemporal magnetoencephalography signals. Our approach to optimizing the circuit parameters in a QCS protocol takes into account quantum sampling noise and allows us to engineer protocols that can yield accurate results with as few as just a single measurement shot. In all cases, we have been able to show a regime of operation where a quantum computational sensor can achieve higher accuracy than a conventional quantum sensor for a given budget of sensing time, with a simulated accuracy advantage of $>$20 percentage points for some tasks. We also present protocols for performing nonlinear tasks using Hamiltonian-engineered bosonic systems and quantum signal processing with hybrid qubit-bosonic systems, and empirically show an advantage when the received signal has a limited mean photon number. Overall, we have shown that substantial quantum computational-sensing advantages can be obtained even if the quantum system is small, including few-qubit systems, systems comprising a single qubit and a single bosonic mode, and even just a single qubit alone---raising the prospects for experimental proof-of-principle and practical realizations. Altogether, our methods and results advance our understanding of how we can achieve quantum computational-sensing advantages for nonlinear tasks and provide further motivation for finding ways to fruitfully adapt quantum algorithms to coherently process sensed signals prior to measurement.
\end{abstract}

\maketitle

\section{Introduction}
\label{sec:intro}

Quantum computational sensing (QCS)~\cite{perspective}---the use of quantum systems to sense and simultaneously compute functions of physical signals---provides a promising avenue to obtain an advantage over conventional quantum sensing for specific tasks. A variety of quantum computational sensing protocols have been proposed to realize this quantum computational-sensing advantage (QCSA), for tasks ranging from linear function approximation~\cite{eldredge2018optimal, zhuang_physical-layer_2019} to classification~\cite{banchi_quantum-enhanced_2020, sinanan-singh_single-shot_2024}, to signal detection treated as a search problem~\cite{allen_quantum_2025}, across various sensing platforms from qubit-based~\cite{qian_heisenberg-scaling_2019} to bosonic sensor networks~\cite{zhuang_physical-layer_2019, bringewatt_optimal_2024}, to hybrid qubit--bosonic platforms~\cite{liao_quantum-enhanced_2024, sinanan-singh_single-shot_2024}. 

With the exception of the search protocol of Ref.~\cite{allen_quantum_2025}, these QCS schemes contain just a single sensing operation within the quantum circuit prior to measurement. As has been explored in quantum computing, disconnected from quantum sensing---for example, in variational approaches to quantum machine learning \cite{cerezo2021variational} and in quantum signal processing \cite{martyn2021grand}---it is often possible to realize more sophisticated functions if the data inputs are encoded multiple times, interleaved with computing operations. We will exploit this insight in designing protocols for QCS by allowing multiple sensing operations prior to measurement.

Additionally, in the existing QCS schemes where the task is cast as a supervised learning problem and the QCS circuit's parameters are trained \cite{zhuang_physical-layer_2019,banchi_quantum-enhanced_2020,liao_quantum-enhanced_2024}, the training procedures use knowledge of the exact probability distribution of measurement outcomes or exact expectation values. We will also address this.

\begin{figure}[t]
    \centering
    \includegraphics{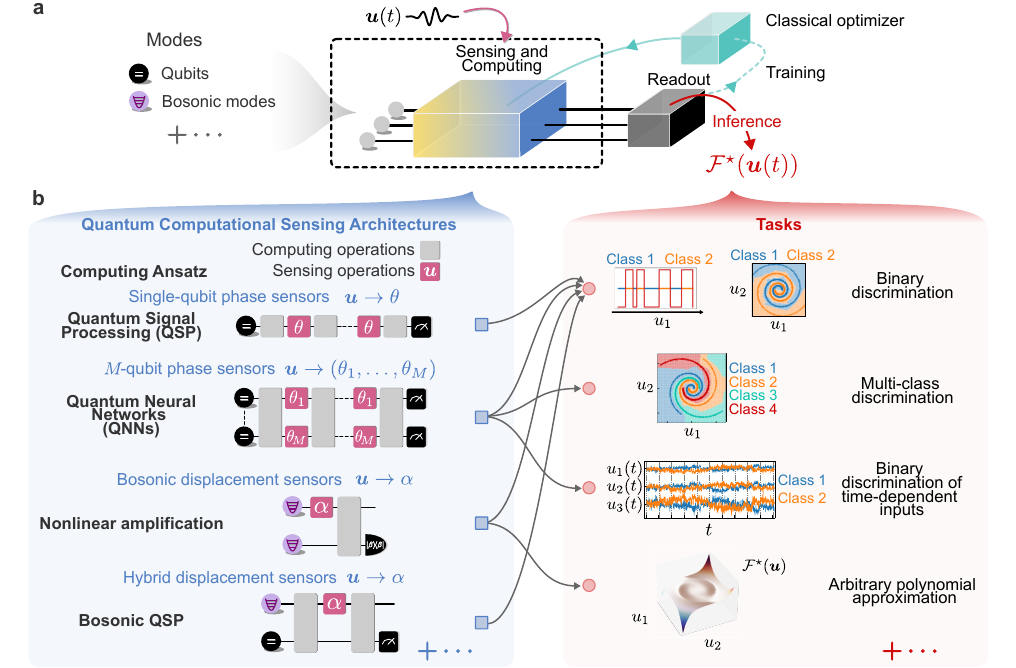}
    \caption{\textbf{Designing quantum computational sensing protocols for general tasks.} \textbf{a}, Quantum computational sensing uses a quantum system to sense and simultaneously extract desired features of a received physical signal; formally this can be described as the computing of a function $\Ft$ of the received signal $\bm{u}(t)$. The quantum system can comprise any quantum `mode', for example qubits, bosonic modes, and so on. We define a general architecture for this quantum system to be deployed as a quantum computational sensor, and provide both analytic and supervised learning (training-based) approaches for its efficient optimization. \textbf{b}, QCS protocols across distinct sensing platforms that we optimize for a variety of tasks, as depicted. }
    \label{fig:one}
\end{figure}

In this paper, we demonstrate an approach to optimizing QCS protocols, tested across a variety of tasks to illustrate its generality~(see Fig.~\ref{fig:one}). We consider both protocols with a single sensing operation performed prior to measurement and a generalization that allows multiple sensing operations to be interleaved with computing operations before a single measurement. We have explored QCS protocols with multiple sensing operations prior to measurement that have a circuit ansatz structure inspired by quantum signal processing~\cite{martyn2021grand, sinanan-singh_single-shot_2024} and quantum neural networks~\cite{cerezo2021variational}. Within the category of protocols having a single sensing operation prior to measurement, we also adapt bosonic quantum signal processing~\cite{sinanan-singh_single-shot_2024} and Hamiltonian engineering to engineer new QCS protocols.

All the protocols we study have free parameters that tune the function of the sensed parameters that the QCS circuit and measurement implements. For a given task, these parameters need to be chosen so that the quantum computational sensor performs the appropriate function. We choose the parameters either using analytic methods when the exact computation required is known, or---in most cases in this paper---using supervised learning when the task is defined by a training dataset rather than by a prespecified function. Our supervised-learning approach performs both training and inference using finitely-sampled measurement results---even as few as a single shot---without assuming access to exact measurement outcome probabilities or expectation values. We demonstrate the generality of our approach in two ways: by applying it to various sensing platforms, and by applying it to various tasks. We consider qubit-based, bosonic-mode-based, as well as hybrid qubit--bosonic hardware platforms. The tasks we address range from binary classification of static signals~\cite{zhuang_physical-layer_2019, liao_quantum-enhanced_2024, sinanan-singh_single-shot_2024} to more sophisticated tasks such as multiclass discrimination, tasks requiring higher-order nonlinearities, and tasks based on physical signals that are multidimensional and functions of time. For each task, we compare the performance of the QCS protocol for it against an appropriate conventional QS protocol, quantifying the QCSA.


We adopt a unifying description of quantum computational sensing in the language of function approximation: while the measurements of a conventional quantum sensor are used to directly estimate a sensed physical signal $\bm{u}(t)$\footnote{We use boldface notation to indicate multidimensional quantities.}, the measurements of a quantum computational sensor are designed to allow us to directly estimate some (generally nonlinear) target function of the sensed signal, $\Ft(\bm{u}(t))$ (see Fig.~\ref{fig:one}). This motivates the use of quantum operations that enable the computation of functions of the parameters that a quantum sensor senses. The computing operations are optimized to obtain QCS protocols that are specific to a given task. If the desired $\Ft$ for a specific task is known, analytic methods can be used for this optimization step (we provide an example of one such protocol). However, for most tasks, we do not have an explicit description of $\Ft$ is; we instead show that a supervised-learning approach, which requires no prior knowledge of the function $\Ft$, can be used to instead optimize a QCS protocol based on a training dataset. Crucially, we show that this approach is possible even using noisy measurements from the quantum computational sensor.

For single-qubit phase sensors, we show that we can adapt the quantum-signal-processing algorithm~\cite{martyn2021grand} as an ansatz for a QCS protocol that can be trained by supervised learning to perform classification of sensed phases. By analyzing low-dimensional binary classification tasks that are nonlinear but still interpretable, we show how increasing computing operations allows a more accurate construction of a target function $\Ft$, enabling more powerful QCS protocols in our architecture. Our use of supervised learning for QCS protocol optimization demonstrates that these target functions need not be known at the outset, and can be successfully inferred from training data. For multi-qubit sensors, we show how circuits inspired by quantum neural networks achieve the same objective. These tasks also allow us to analyze fundamental aspects of QCS, such as the role of quantum sampling noise and task complexity. We also apply a quantum-neural-network-based QCS protocol to complex tasks over spatiotemporally-varying signals. If we consider a signal sampled at many discrete times as a high-dimensional vector, then the function $\Ft$ is computed on a very high-dimensional input---and yet we found that it was possible to successfully train a QCS protocol for classification of spatiotemporal signals.

We also explore protocols based on a single sensing operation prior to measurement. For displacement sensing using bosonic sensors, we devise a protocol that directly uses the strong connection between QCS and function approximation. By harnessing a nonlinear quantum non-demolition Hamiltonian, we show that a bosonic quantum computational sensor can be engineered such that its measurement directly provides an estimate of an arbitrarily-complex polynomial of the sensed displacement. The Hamiltonian parameters required for a given polynomial can be determined analytically. The QCS protocol outperforms a conventional approach that estimates both quadratures of the sensed displacement and computes the desired polynomial using classical postprocessing. For hybrid qubit-bosonic sensors, we consider a generalization of the bosonic quantum signal processing algorithm~\cite{sinanan-singh_single-shot_2024} for displacement sensing, using it to devise QCS protocols for nonlinear binary classification tasks.

The rest of this paper is organized as follows. In Sec.~\ref{sec:theory}, we provide a brief introduction to QCS in terms of function approximation, and introduce the general architecture of QCS protocols that we analyze in the remainder of the paper. Sec.~\ref{sec:qubits} presents results for quantum computational phase-sensing using qubit-based sensors, from tasks that use a single qubit sensor, to multi-qubit sensing of multi-dimensional signals, and ultimately single-shot processing of spatiotemporal signals. Sec.~\ref{sec:bosonic} discusses QCS of signals using bosonic modes for arbitrary polynomial approximation. Sec.~\ref{sec:hybrid} describes QCS protocols for multivariate signals using hybrid sensors comprising a bosonic mode coupled to a single qubit. We conclude with several directions for future work, including prospects for experimental implementations.

\section{Quantum Computational Sensing}
\label{sec:theory}

In this section we provide a brief introduction to the paradigm of quantum computational sensing, and present our platform-independent architecture for quantum computational sensing protocols that will be used in the remainder of this work. In doing so, it will prove useful to recap how a quantum system is typically used for \textit{conventional} quantum sensing (QS) of an unknown physical signal $\bm{u}$. 

A quantum system can be used to sense a signal $\bm{u}$ if there is a physical interaction between the system and the signal. The interaction leads to a $\bm{u}$-dependent unitary operation $\Usense(\bm{u})$, whose action on the quantum system makes its state dependent on the signal $\bm{u}$. In practice, a quantum sensor in an initial state $\ket{\psi_0}$ is typically prepared prior to the sensing operation via the action of an `encoding' unitary $\Uenc$; such an operation can be used to prepare a probe state that makes the quantum system maximally sensitive to the signal. Following the sensing step, a general `decoding' operation $\Udec$ can be performed; the final state of the quantum sensor following the sensing protocol is thus given by $\ket{\psi(\bm{u})} = \Udec~\Usense(\bm{u})~\Uenc \ket{\psi_0} \equiv \Uqs \ket{\psi_0}$, where we have introduced $\Uqs$ as a shorthand for the entire quantum sensing protocol. $\Udec$ can be interpreted as setting the basis for measurements that follow the execution of the unitary operations. Finally, any information about $\bm{u}$ must be extracted via measurements performed on the quantum sensor state $\ket{\psi(\bm{u})}$; we denote the expected value (ensemble average) of measurement results as a vector $\xqs(\bm{u})$, whose dimension depends on the specific measurement scheme.  

The aim of conventional QS is to estimate $\bm{u}$~\cite{degen2017quantum}, to which end a quantum sensor is engineered such that $\xqs(\bm{u}) \propto \bm{u}$. Note, however, that being a quantum system, measurements $\Xqs$ of such a quantum sensor have outcomes that are fundamentally stochastic, and the mean over repeated measurements yields the aforementioned expected values, $\mathbb{E}[\Xqs] = \xqs$. Averaging multiple measurement results reduces noise, which we define using the bar notation, $\Xbqs = \frac{1}{S}\sum \Xqs$. Nevertheless, for any finite number of measurements $S$, quantum sampling noise limits the information extractable about $\bm{u}$ from a quantum sensor. 

However, constructing an estimate of a physical signal $\bm{u}$ itself is not the final objective in some applications. Instead, one may be interested in extracting specific features of the received signal, an objective that can be described as estimating a function $\Ft$ of $\bm{u}$ (whose output can generally be multidimensional). For conventional QS, we simply have $\Ft(\bm{u}) = \bm{u}$. However, we may only wish to determine whether the received signal was above or below a predetermined threshold $u_{\rm th}$; then, $\Ft(\bm{u}) = {\rm sign}(u_i - u_{\rm th})$. For classification tasks, inputs $\bm{u}$ belong to one of $j=1,\ldots,C$ discrete classes with a certain probability. Classification amounts to identifying the likeliest class label of a sensed signal $\bm{u}$, formally $\Ft = {\rm argmax}_{j \in 1,\ldots C}~{\rm Prob}({\rm Class}=j|\bm{u})$. In other situations, one may be interested in computing the autocorrelation of a signal over time (where $\Ft(\bm{u}) = \bm{u}_i(t)\bm{u}_i(t')$), or spatiotemporal correlations of signals received by different sensors (where $\Ft(\bm{u}) = \bm{u}_i(t)\bm{u}_j(t')$). Of course, a conventional quantum sensor can be used for such tasks, by first obtaining a noisy estimate of $\bm{u}$, and subsequently performing classical postprocessing of this estimate to construct an approximation of the desired function $\Ft$. However, the fidelity of such an approach will be limited by the precision of the initial estimate, and can also depend on the complexity of $\Ft$ (this is also true of the processing of stochastic variables generated by completely classical processes; see Appendix~\ref{app:gauss}). 

Quantum computational sensing (QCS) instead proposes the use of a quantum system to sense \textit{and} simultaneously process a physical signal $\bm{u}$, so that measurements performed on the system directly extract the features desired for a specific task (see Fig.~\ref{fig:one}a). Explicitly, we define a quantum computational sensor as performing a protocol $\Uqcs$ where $\ket{\psi(\bm{u})} = \Uqcs \ket{\psi_0}$ such that the corresponding measurement results $\xqcs$ compute some function of $\bm{u}$. Ideally, this function is the desired target function itself, namely $\xqcs \propto \Ft$ (although this is not always necessary, as we will see). Such an approach has been shown to provide an advantage over conventional quantum sensing followed by classical postprocessing for the same resources, across a variety of tasks and sensing platforms~\cite{perspective}, and is referred to as a quantum computational-sensing advantage (QCSA).

The precise nature of the functions $\mathcal{F}$ that can be computed by a quantum computational sensor depends on the structure of the quantum circuit $\Uqcs$.\footnote{In this work we discuss QCS protocols in the circuit model, i.e., assuming discrete-time operations, but QCS protocols could also be constructed by engineering continuous-time unitary evolution.} However, the complexity of the functions a protocol can compute is just one aspect of quantum computational sensing. Being a quantum system, measurements $\Xqcs$ from a quantum computational sensor also exhibit quantum sampling noise. Engineering $\Uqcs$ to realize a desired target $\Ft$ invariably modifies the sampling-noise properties. This dependence leads to a complex interplay between the functions that a quantum computational sensor can be used to compute, and the precision with which they can be computed, a relationship we will analyze with several examples in this paper. 

We consider a general structure for $\Uqcs$ given by:
\begin{align}
    \Uqcs = \Udec\left[ \prod_{l=1}^L \Usense(\bm{u}^{(l)})~\Ucoh{l} \right],
    \label{eq:cohprocessing}
\end{align}
which allows for $L\geq1$ sensing operations interleaved with $L$ coherent processing operations $\Ucoh{l}$, before a single measurement is performed. Most generally, the signal sensed during different sensing operations can also be different, indexed $\bm{u}^{(l)}$; this can describe, for example, the quantum computational sensing of time-varying signals $\bm{u}(t)$, where $\bm{u}^{(l)} = \bm{u}(t_l)$. For $L=1$, Eq.~(\ref{eq:cohprocessing}) describes a QCS structure where a single sensing operation is preceded by a probe state preparation step $\Ucoh{1} \to \Uenc$, and is followed by preparation of a measurement basis $\Udec$---this is a structure that has been used in a variety of QCS protocols to date~\cite{perspective}. For $L>1$, on the other hand, this QCS structure allows for multiple sensing operations to be performed and the sensed signals to be processed coherently prior to a measurement being executed.

\subsection{Summary of main results}

In this paper, we develop several QCS protocols with the general structure given by Eq.~(\ref{eq:cohprocessing}), for a range of complex tasks. Our work has two main thrusts. First, we focus on nonlinear tasks, which formally require computing a nonlinear target function $\Ft(\bm{u})$. Using a conventional quantum sensor---which estimates $\bm{u}$ directly---such tasks require nonlinear postprocessing that can lead to a reduction in SNR, and---for discrimination tasks---an increase in classification error. We show that it is possible to achieve an advantage by performing some or all of the nonlinear processing in the quantum system before measurement. Secondly, we focus (primarily, but not exclusively) on tasks where $\Ft$ is \textit{a priori} unknown. In this setting, we demonstrate a general technique for training $\Uqcs$ to devise QCS protocols in the setting of supervised learning. Perhaps the most important aspect of our approach is that we conduct training using finitely-sampled measurement results, without assuming knowledge of the density matrix of the quantum system, expectation values of observables, or probabilities of measurement outcomes. We also show that our training procedure is compatible with the inclusion of classical postprocessing (e.g., using classical neural networks) of measurement results where both the quantum system and the classical postprocessor are trained together (end-to-end training). We demonstrate our approach to develop the QCS protocols listed below, and to identify the advantage they can deliver over conventional QS protocols:

\begin{enumerate}[(1),noitemsep,leftmargin=3\parindent] 
\item using a single-qubit quantum computational sensor for single-variable binary classification tasks, based on quantum signal processing (QSP) and optimized using supervised learning~(Sec.~\ref{subsec:1q}). We also provide analytic results for single-qubit QS performance as a function of task complexity, and comparison to QCS protocols~(Sec.~\ref{subsec:taskcomp}).
\item using multi-qubit quantum computational sensors for binary and multi-class discrimination tasks, based on quantum neural networks (QNNs) and optimized using supervised learning~(Sec.~\ref{subsec:2q}). 
\item using multi-qubit quantum computational sensors for the binary classification of time-varying signals obtained from a real dataset, based on QNNs and optimized using supervised learning~(Sec.~\ref{sec:meg}).
\item using bosonic quantum computational sensors for arbitrary polynomial approximation, based on Hamiltonian engineering and using an analytic method to derive protocol parameters~(Sec.~\ref{sec:bosonic}).
\item using hybrid qubit-bosonic mode quantum computational sensors for binary classification tasks, based on bosonic QSP and optimized using supervised learning~(Sec.~\ref{sec:hybrid}).
\item using a single-qubit quantum computational sensor for multi-variate binary classification, based on multivariable-QSP and optimized using supervised learning~(Appendix~\ref{app:mvbc}).
\end{enumerate}

\section{Quantum computational sensing using qubit-based sensors}
\label{sec:qubits}

We begin by considering a simple, illustrative task for which a function of a scalar unknown phase $u \to \theta$ is ultimately the object of interest, as opposed to the value of $\theta$ itself: binary classification. As we will show, quantum computational sensing with just a single qubit can provide an advantage for binary classification of the sensed phase $\theta$. We then demonstrate advantage for more complex tasks such as multi-class discrimination, and over multi-parameter and even time-varying signals $\bm{\theta}(t)$ using quantum computational sensors comprising several qubits.

\subsection{QCS using a single qubit for binary classification tasks}
\label{subsec:1q}


\begin{figure}[h!]
    \centering
    \includegraphics{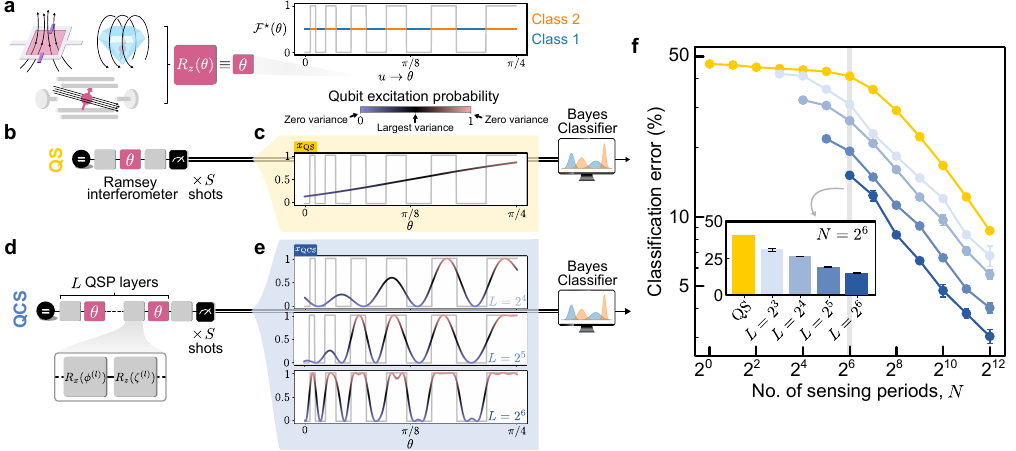}
    \caption{\textbf{Quantum computational sensing using a single qubit.} \textbf{a}, Single-variable binary classification task in the context of sensing. The signal---here, a phase---is sensed by a quantum sensor via a physical interaction that can be realized in a variety of quantum sensing platforms; some examples depicted in the left panel include magnetic field sensing using SQUIDs or NV centers in diamond, or electric field sensing using trapped ions. The phase is indexed by one of two class labels, as shown in the right panel. \textbf{b}, Conventional quantum sensing of the phase $\theta$ using a single qubit as a Ramsey interferometer. \textbf{c}, Expected measurement outcomes $x_{\rm QS}$ from the conventional quantum sensor. Finitely sampled measurement results $\bar{X}_{\rm QS}$ are passed through a Bayes classifier to predict the class label. \textbf{d}, Quantum computational sensing using a single qubit, which coherently performs $L$ sensing operations interleaved with $L$ trainable, coherent processing operations, as depicted. \textbf{e}, Expected measurement outcomes $x_{\rm QCS}$ from a trained quantum computational sensor for varying $L$. Finitely sampled measurement results $\bar{X}_{\rm QCS}$ are again passed through a Bayes classifier to predict the class label. \textbf{f}, Classification error as a function of $N$, the number of sensing periods; $N=S$ for the QS scheme, while $N=L\times S$ for the QCS scheme. Error bars indicate performance differences due to variation in training of QCS protocols; where not visible, they are smaller than the height of plot markers. Inset: classification error for fixed $N=2^6$. The error \% difference between the QS protocol and the single-shot ($L=2^6$) QCS protocol is $41.1\%-15.0\%=26.1$ percentage points.}
    \label{fig:1d}
\end{figure}


We begin by introducing binary classification tasks in the context of phase sensing. A signal to be sensed can interact with a quantum system via a physical interaction $\Usense$, imparting a phase $\theta$ onto its quantum state; without loss of generality, we take this interaction to impart a $z$-rotation $R_z(\theta)$. For a qubit-based sensor, this physical interaction can take a variety of forms; some of the most popular examples depicted in the left panel of Fig.~\ref{fig:1d}a include magnetic field sensing using SQUID loops in superconducting circuits or using NV centers in diamond, or electric field sensing using trapped ions. Unlike standard parameter estimation tasks, however, the unknown phase $\theta$ being sensed is now indexed by one of two possible class labels. The objective of the classification task is then to determine this \textit{a priori} unknown label. For concreteness, we consider the one-dimensional binary classification shown in the right panel of Fig.~\ref{fig:1d}a, where $\theta$ can take a value $\theta \in [0,\frac{\pi}{4}]$, with each value in this range belonging to either Class 1 (blue) or Class 2 (orange). In the language of function approximation, it is possible to describe this task as one of approximating a function $\Ft(\theta)$ that outputs the class label (hence either 0 or 1). This target function defines the boundary between the classes, and is depicted in gray in Fig.~\ref{fig:1d}a. Note that $\Ft(\theta)$ is highly nonlinear in $\theta$. 

As mentioned earlier, such a task can be performed by first estimating $\theta$ using a conventional quantum sensor, and subsequently using classical postprocessing of this estimate to predict the class label (computing the target function $\Ft(\theta)$ in the process). For this single-variable task, the required computation can even be visualized, as follows. If the estimate $\theta$ falls into any of the blue regions in Fig.~\ref{fig:1d}a, the class label is predicted to be Class 1, while if it falls into any of the orange regions, the class label is predicted to be Class 2. 

We begin by discussing this conventional approach for the present task, depicted in Fig.~\ref{fig:1d}b. Here $\theta$ is sensed using a single-qubit Ramsey interferometer. The measurement result--the scalar $x_{\rm QS}(\theta)$, which is simply the excitation probability of the measured qubit---is shown in Fig.~\ref{fig:1d}c. Note that $x_{\rm QS}(\theta)$ is approximately linear in $\theta$, and more importantly describes a one-to-one map; as a result, knowledge of $x_{\rm QS}(\theta)$ can be used to uniquely estimate the unknown $\theta$, as required for parameter estimation. The estimated $\theta$ can be further processed to predict the class label, as mentioned before.

The difficulty with this approach, however, lies in the realization that for any finite number of shots $S$ we only have access to the stochastic, shot-averaged measurement result $\bar{X}_{\rm QS}$, which provides a noisy estimate of $\theta$ with a precision limited by $S$; this stochastic variable is what must be used to determine the underlying class label. This stochasticity---due to intrinsic quantum sampling noise---will lead to a finite classification error. In this single-variable case using a single qubit system, the minimum possible classification error can be precisely quantified. This is done by determining the optimal classifier to be used for the stochastic outputs $\bar{X}_{\rm QS}$, namely a Bayes classifier~\cite{devroye_bayes_1996}, which is constructed using knowledge of the exact probability distribution of $\bar{X}_{\rm QS}(\theta)$. In Fig.~\ref{fig:1d}f we plot the classification error obtained using this Bayes classifier as a function of the number of shots $S$ used to compute $\bar{X}_{\rm QS}$; for the QS scheme this also equals the total number of sensing periods $N=S$ (see Appendix~\ref{app:bayes} for full details of the implementation of the Bayes classifier). We see that for a single shot, the conventional QS scheme is able to recover very limited information about the underlying class label, achieving a classification error barely below the random guessing level of 50\%. Note that our use of an optimal classifier ensures that this is not a limitation of the classical postprocessing step, but a fundamental restriction enforced by quantum sampling noise and the stochasticity of quantum measurements. With increasing $S$, the variance in $\bar{X}_{\rm QS}$ decreases and the classification error can be reduced, as expected.

We now consider a QCS approach to this task: can engineering a more targeted relation between the measurement result $x_{\rm QCS}$ of a quantum computational sensor and the sensed phase $\theta$ yield a lower classification error than the conventional QS approach? To this end, we first identify a scheme that can equip a quantum-computational sensor to compute more general, including nonlinear functions of a sensed input. Here we take inspiration from the field of quantum algorithms, where a known protocol can be used to compute arbitrary functions using a single qubit: the Quantum Signal Processing (QSP) algorithm~\cite{martyn2021grand}. The QSP algorithm for a single qubit interleaves repeated applications of an input unitary parameterized by a fixed parameter with non-commuting single-qubit rotations, to compute polynomial functions of the input parameter; the complexity of computable functions increases with the number of input calls. Here, we adopt a QSP-inspired protocol for single-variable QCS: within the general structure of Eq.~(\ref{eq:cohprocessing}), $L>1$ sensing operations $\Usense(\theta)$ sensing the static ($l$-independent) phase $\theta$ are interleaved with $L$ coherent processing operations $\Ucoh{l}$. The processing operations are made up of single-qubit rotations $R_x(\phi^{(l)})$ that are orthogonal to the sensing ($z$) axis---essential for QSP---together with a $z$-rotation $R_z(\zeta^{(l)})$; $\phi^{(l)}$, $\zeta^{(l)}$ are trainable rotation angles for the $l$th layer (see Fig.~\ref{fig:1d}d). Increasing $L$ increases the depth of the QCS protocol and thus the complexity of computable functions.

Importantly, $\Uqcs$ must be engineered or optimized for the task at hand; this requires determining the angles parameterizing $\Ucoh{l}$. In applications of QSP, this is done by optimizing the angles with the aim of producing a known target function~\cite{martyn2021grand}. However for completely general tasks (examples of which we will consider in this paper) the target function may not always be known; in such cases, supervised learning can be used to optimize the QCS protocol using a labeled training dataset without requiring explicit knowledge of $\Ft$. More importantly, the optimization of parameters in the QSP algorithm is performed assuming access to the actual qubit state and hence the expected measurement result $x_{\rm QCS}$. However, in a practical setting one only has access to measurement results obtained using a finite number of shots $S$. Consequently, in this work the training of $\Uqcs$ is performed under the constraints of finite sampling: we minimize the cross entropy loss over a training set comprising stochastic measurement results $\bar{X}_{\rm QCS}$ constructed using $S$ shots~(full details can be found in Appendix~\ref{app:svbc}). 

The qualitative difference between QCS and QS can already be seen via the dependence of expected measurement result $x_{\rm QCS}$ on $\theta$ for trained $\Uqcs$, shown in Fig.~\ref{fig:1d}e for select layer numbers $L$. We immediately see that $x_{\rm QCS}$ is no longer a one-to-one function of $\theta$. Consequently, it is not even possible to use such a sensor to estimate $\theta$ uniquely. However, the relaxation of this constraint allows $x_{\rm QCS}$ to be much more general, and importantly to be nonlinear in $\theta$, just like the target function $\Ft(\theta)$. In particular, we see that the trained QCS protocols produce $x_{\rm QCS}(\theta)$ that provide an increasingly more faithful approximation to the nonlinear decision boundary defined by $\Ft$ as $L$ increases. This is also consistent with the QSP-inspired nature of our protocol, where the complexity of functions that can be approximated increases with depth, here analogous to the number of sensing operations $L$. However, this nonlinear processing would not be useful if, in the process of engineering $\Uqcs$ to provide the requisite $x_{\rm QCS}$, the resulting sampling noise $\zeta_{\rm QCS}$ was also increased. We have depicted the variance of finitely-sampled measurement outcomes due to this sampling noise via the color of the $x_{\rm QCS}$ curves in Fig.~\ref{fig:1d}d. Due to the binomial nature of sampling of a single qubit state, this variance is minimized when $x_{\rm QCS} \to 0, 1$. For classification tasks, $\Ft$ naturally takes the form of binary functions for which the variance would naturally be minimized. This emphasizes how the nature of quantum sampling noise for a given QCS platform can strongly determine the nature and complexity of functions $\Ft$ it may be best suited to estimating. 

We now return to our original question: can this QCS protocol provide an advantage over the conventional QS strategy? For a fair comparison, we note that we cannot simply allow the quantum computational sensor access to an arbitrary number of sensing operations $L$. In the sensing context, access to repeated sensing operations is a resource, which we characterize as the total number of sensing periods $N$; for this QCS protocol with $L$ sensing operations per a single measurement, now $N=L\times S$ for $S$ measurement shots. For fixed $N$, $L$ can therefore at most equal $N$, in which case the QCS approach must perform classification using a single shot. The final result is shown via the classification error obtained using a Bayes classifier for both QS and QCS approaches, in Fig.~\ref{fig:1d}f: the QCS approach is able to obtain a consistently lower classification error than the conventional QS approach given access to the same number of sensing periods $N$. The classification error also reduces with increasing $L$, due to the aforementioned ability to compute an increasingly more accurate approximation of $\Ft$. The inset of Fig.~\ref{fig:1d}f compares the QS and QCS performance at a fixed $N = 2^6$. The QCS error again decreases as $L$ increases, even though for fixed $N = L \times S$ this means the number of measurement shots $S$ available to the quantum computational sensor also decreases. In fact, the lowest QCS error of $15.1\%$ is achieved for $L=2^6$, which corresponds to the single-shot regime, and provides a substantial improvement over the QS error of $41.0\%$. Finally, we note that even intermediate values of $L$ do provide a (smaller) QCSA: this is a feature we observe quite generally, and is very promising for near-term experimental demonstrations with limited quantum computing resources.

\subsection{Dependence on task complexity}
\label{subsec:taskcomp}

By identifying an advantage for tasks beyond the confines of parameter estimation, QCS opens up a multitude of directions for exploring quantum advantages: in principle, any task that requires the processing of a sensed signal $\bm{u}$ to compute a function $\Ft(\bm{u})$---which can be arbitrary---defines a new task for which a QCSA could be pursued. Then, a natural question emerges: does the extent of achievable QCSA depend on the nature of the task? The very fact that tasks that require the estimation of an arbitrary $\Ft$ can be considered makes this question difficult to answer in complete generality. Nevertheless, for a carefully-designed selection of tasks, we are able to systematically explore the dependence of QCSA on task complexity; consequently, we identify that the extent of QCSA can increase as the difficulty of the task increases. 

We still consider single-variable binary classification tasks of the form of Fig.~\ref{fig:1d}a, for sensed phases over a fixed domain $\theta \in [0,\frac{\pi}{4}]$, and where $R$ noncontiguous regions define the $\theta$ values that belong to each class ($R=6$ in Fig.~\ref{fig:1d}a). Making use of the simple structure of the fixed QS protocol, we are able to obtain an approximate analytic expression for its classification error $\mathcal{E}_{\rm QS}$ for this task (see Appendix~\ref{app:analyticGauss}). For the special case where each region has an equal extent $\delta = \frac{\pi/4}{2R}$, we find that an approximate analytic expression can be obtained for the classification error, given by $\mathcal{E}_{\rm QS} \simeq \frac{2}{\pi/4}\cdot\frac{2R-1}{2}\left[ 2\delta \left( {\rm erf}(2\delta\sqrt{2S}) - {\rm erf}(\delta\sqrt{2S}) \right) + \frac{1}{\sqrt{2\pi S}}\left(1 + e^{-8S\delta^2} - 2e^{-2S\delta^2} \right) \right]$, in the limit where $\sqrt{S} \gtrsim R$ (recall that $N=S$ here). For $\sqrt{S} \gg R$, this expression reduces to $\mathcal{E}_{\rm QS} \approx \frac{2}{\pi/4}\cdot\frac{2R-1}{2}\frac{1}{\sqrt{2\pi S}}$; as the classification error increases with $R$ for fixed sensing resources, it is clear that increasing $R$ increases the task difficulty. For $N=S=2^6$, $\mathcal{E}_{\rm QS} \approx 0.063(2R-1)$.

By comparing the performance of both QS and QCS protocols as a function of $R$, we are therefore able to identify the dependence on QCSA on the complexity of this task. The performance of the QCS protocols depends on the number of processing layers $L$, and even for fixed $L$ on the specific $\Uqcs$ learned via training; therefore a general analytic expression is more nontrivial to obtain. Empirically, we find that the error of QCS protocols for tasks with equal $\delta$ and for $N=L=2^6$ scales as $\mathcal{E}_{\rm QCS} \approx 0.015(2R-1)$. As a result, the extent of QCSA (the difference between $\mathcal{E}_{\rm QS}$ and $\mathcal{E}_{\rm QCS}$) clearly increases with $R$. We also compare the performance of both QS and QCS protocols as a function of $R$ for more general tasks where the extent of regions $R$ are unequal as in Fig.~\ref{fig:1d}a (see Appendix~\ref{app:complexity}). 

Our results confirm that QCSA can depend strongly on the difficulty of the task under consideration, and encourage the exploration of more complex tasks for the observation of larger advantages. Furthermore, the expression for $\mathcal{E}_{\rm QS}$ for the selection of tasks we have introduced can serve as a useful benchmark to compare the performance of other QCS protocols, providing an approximate quantitative metric for the extent of QCSA, and also enabling a comparison  of different QCS protocols against each other.

\subsection{QCS using multiple qubits: binary and multi-class discrimination of multi-dimensional signals}
\label{subsec:2q}

We next show that quantum computational sensing can naturally be extended to more complex computational tasks characterized by high-dimensional (or multi-parameter) inputs. The sensing of multiple inputs also often requires multiple sensors, for example to measure magnetic fields at differing spatial locations. Increasing the scale of the sensing system can also have other advantages. For example, the dimension of the space of measured features $\bm{x}$ defined by a complete set of commuting observables is constrained by the number of quantum modes, given by $2^M$ for an $M$-qubit sensor, for example. This measurement space dimension determines the expressive capacity of a physical system~\cite{dambre_information_2012, wright_capacity_2019, hu_tackling_2023}; in short, increasing $M$ enables a system to perform more varied and complex tasks.

A related but more subtle factor emerges in the special case of single-shot sensing. Specifically, a single shot from a single qubit sensor can have only two possible outcomes, or equivalently can provide a single bit of information. Successfully performing a classification task over $C$ classes requires obtaining $\log_2 C$ bits of information. As a result, while a single shot from a quantum-computational sensor comprising a single qubit can in-principle be sufficient for binary classification, it does not provide enough information to distinguish $C>2$ classes. This is true regardless of the number of sensing periods $N$ or the complexity of $\Uqcs$, which limits the utility of QCS resources; the only solution is to collect more than a single shot to eventually extract enough information for the task to be performed. In contrast, the use of a suitable $M$-qubit quantum computational sensor (specifically, where $M \geq \log_2 C$) can provide the requisite information in a single shot. These potential applications to multi-class discrimination over multivariate signals motivate our exploration of QCSA using multi-qubit sensors in this section.


\begin{figure}[t]
    \centering
    \includegraphics{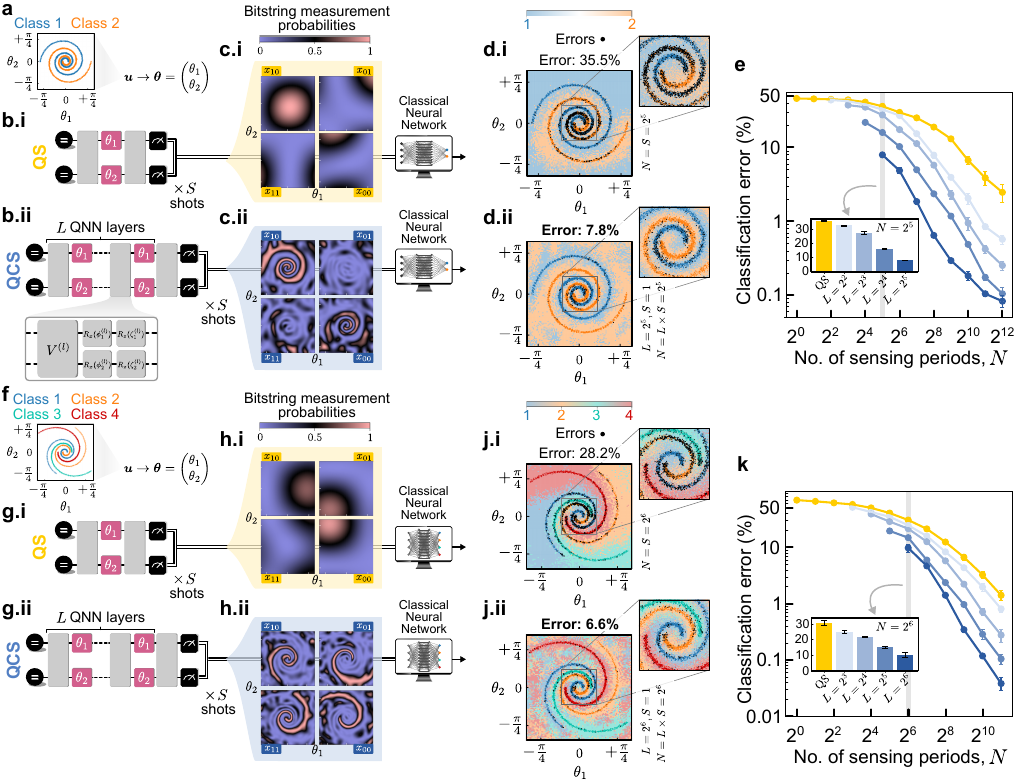}
    \caption{\textbf{Quantum computational sensing using multiple qubits for multivariate classification tasks.} \textbf{a}, Binary classification task for two-variable phases, sampled from the depicted dataset, here an instance of the \logspirals{} task. \textbf{b.i}, Conventional QS scheme using a two-qubit quantum sensor. \textbf{b.ii}, QCS scheme using $L$ coherent processing layers are used to process multiple sensing signals, followed by measurement in the computational basis; the resulting protocol bears similarity to a QNN architecture. \textbf{c.i}, Expected measurement outcomes for the QS scheme with $N=S=2^5$. \textbf{c.ii}, Expected measurement outcomes for the QCS scheme with $N=L\times S = 2^5$, $L=2^5, S=1$. \textbf{d.i}, Output of QS protocol and classical postprocessing as shown by the colorbar, over a fixed domain of $\bm{\theta}$ values. Dots are instances of $\bm{\theta}$ sampled from the testing dataset in \textbf{a}: where shown in color, these are correctly classified by the QS protocol and classical postprocessing. Dots shown in black are incorrectly classified. Inset shows zoom-in of marked square region. \textbf{d.ii}, Same as \textbf{d.i}, but for the QCS protocol followed by classical postprocessing. \textbf{e}, Classification error for binary \logspirals{} task using QS and QCS schemes. Error bars indicate performance differences due to variation in training of QCS protocols; where not visible, they are smaller than the height of plot markers. Inset: classification error for fixed $N=2^5$. The error \% difference between the QS protocol and the single-shot ($L=2^5$) QCS protocol is $35.5\%-7.8\%=27.7$ percentage points. \textbf{f}, 4-class discrimination task for two-variable phases, sampled from the depicted dataset, here an instance of a modified \logspirals{} task. \textbf{g.i}, Conventional QS scheme using a two-qubit quantum sensor, same architecture as \textbf{b.i}. \textbf{g.ii}, QCS scheme using a two-qubit quantum sensor, same architecture as \textbf{b.ii}. \textbf{h.i}, Expected measurement outcomes for the QS scheme with $N=S=2^6$. \textbf{h.ii}, Expected measurement outcomes for the QCS scheme with $N=L\times S = 2^6$, $L=2^6, S=1$. \textbf{j.i}, Output after QS protocol and classical postprocessing, as in \textbf{d.i}. \textbf{j.ii}, Same as \textbf{j.i}, but for the QCS protocol followed by classical postprocessing. \textbf{k}, Classification error for 4-class \logspirals{} task using QS and QCS schemes; error bars again show variation in training of QCS protocols. Inset: classification error for fixed $N=2^6$. The error \% difference between the QS protocol and the single-shot ($L=2^6$) QCS protocol is $28.2\%-6.6\%=22.2$ percentage points. }
    \label{fig:2d}
\end{figure}
\clearpage


The first task we analyze is a two-variable binary classification task; each pair of phases $\bm{\theta} = \begin{psmallmatrix} \theta_1 \\ \theta_2 \end{psmallmatrix}$ is sampled from one of the two classes depicted in the scatter plot in Fig.~\ref{fig:2d}a, here an instance of the \logspirals{} dataset. Such a task requires nonlinear processing: any linear decision boundary drawn on in Fig.~\ref{fig:2d}a is unable to fully separate the two sets of colored points. As before, we begin by discussing the conventional QS approach (see Fig.~\ref{fig:2d}b.i). We consider a two-qubit quantum sensor where the phase $\theta_j$ is sensed by qubit-$j$ only. The natural extension of the QS protocol from Fig.~\ref{fig:1d}b would consider two qubits used as Ramsey interferometers, to individually sense each phase $\theta_j$. Here we additionally allow the QS scheme access to two-qubit gates to construct the $\Uenc$ and $\Udec$ operations, allowing the quantum sensor to harness entanglement, if useful for this task. In Appendix~\ref{app:mqramsey}, we show that uncoupled Ramsey interferometers achieve a similar or marginally worse performance for the tasks considered in this section.

The resulting expected measurement results $\xqs= (x_{10},x_{01},x_{00},x_{11})$---now the probabilities of measuring each of $2^2$ bitstrings---are shown in Fig.~\ref{fig:2d}c.i. These are sinusoidal functions of $\theta_1,\theta_2$, with different offsets learned via training. Following measurement, stochastic measurement outcomes $\Xbqs$ are constructed using $S$ shots as usual. An important distinction from the previous single-variable task is that the classical postprocessing step applied to $\Xbqs$ is no longer a Bayes classifier; such a classifier becomes increasingly computationally expensive to construct when the dimension of output variables $\bm{\bar{X}}$ increases. We therefore consider classical postprocessing using a multi-layer perceptron (MLP) comprising two layers and ReLU activation, trained alongside $\Uqcs$ in an end-to-end framework. We have verified that using a more complex classical postprocessing step does not improve the performance of either the QS or the QCS approaches, ensuring that postprocessing is not the source of any observed performance differences~(see Appendix~\ref{app:postp} for full details).

Classical postprocessing of the output of the conventional quantum sensor should predict the underlying class label. The output of the trained protocol is plotted in $(\theta_1,\theta_2)$ space in the top panel of Fig.~\ref{fig:2d}d.i, for $N=S=2^5$ sensing periods (per qubit); therefore $S=2^5$ shots are used to construct $\Xbqs$ that are subsequently passed to the classical postprocessor. We note that the computation has many errors: while larger $\theta$ values where the underlying classes are well-separated are accurately classified, smaller $\theta$ values where points belonging to different classes are much closer leads to many incorrect classification results. We emphasize that this is again a limitation enforced by the sampling noise in measurement. Increasing the complexity of classical postprocessing makes no difference; on the other hand, obtaining more shots (over more sensing periods) reduces the variance of $\Xbqs$ and enables a reduction in classification error, as shown in Fig.~\ref{fig:2d}e.

Our QCS protocol again uses the coherent processing structure of Eq.~(\ref{eq:cohprocessing}) with $L>1$ sensing operations interleaved with $L$ coherent processing operations, the latter taking the form of arbitrary, trainable two-qubit unitary operations $V^{(l)}$ as well as single-qubit rotations (see Fig.~\ref{fig:2d}b.ii). The QCS protocol is concluded by a computational basis measurement of both qubits. The total number of sensing periods (per qubit) is again $N=L\times S$. The overall scheme therefore resembles the framework of Quantum Neural Networks (QNNs)~\cite{farhi_classification_2018}, here adapted for use with sensing protocols. The trainable parameters of $\Uqcs$ are once more optimized via supervised learning, for finite shots and using a training dataset. 

The expected probabilities $\xqcs = (x_{10},x_{01},x_{00},x_{11})$ shown in the lower panel of Fig.~\ref{fig:2d}c.ii indicate the impact of QCS: the quantum-computational sensor can be engineered to yield a highly-nonlinear map between $\xqcs$ and $\bm{\theta}$. Here for $L=2^5$, we see that the trained QCS protocol is such that for $\bm{\theta}$ values that lie in Class~1, to a good approximation $x_{10} \to 1$, while for $\bm{\theta}$ values in Class~2, $x_{10} \to 0$ instead. As in the single-variable case, measurement probabilities $\bm{x}$ close to $0,1$ simultaneously lead to a reduction in quantum sampling noise; therefore, such a QCS protocol should also provide good single-shot performance.

This expectation is borne out by the MLP output following the quantum computational sensor shown in the lower panel of Fig.~\ref{fig:2d}d.ii for $N=2^5$; for the QCS protocol with $L=2^5$, this means that $S=1$. In particular, note that for $\theta$ values near the origin, the QCS protocol commits substantially fewer errors when compared to the QS scheme. The reason is simple: the QCS protocol nonlinearity is able to map $x_{10} \to 0,1$ depending on the class label even in this domain, directly predicting the class label with reduced sampling noise, and effectively allowing these finer features to be resolved even with a single measurement shot. We therefore achieve a classification error of $7.8\%$, substantially lower than the $35.5\%$ achievable using conventional QS protocol for the same $N$. The classification error reduction can also be observed for smaller $L$, as shown in Fig.~\ref{fig:2d}e, emphasizing that even limited QCS resources can provide an advantage over the conventional QS scheme.

We note that performing the present \logspirals{} classification task should in-principle be possible even using a single-qubit sensor in the single-shot regime, given that it demands only binary discrimination. In Appendix~\ref{app:mvbc} we provide a generalized single-qubit QCS protocol, which adapts the multivariable-QSP protocol~\cite{rossi_multivariable_2022} and is able to perform this task with high accuracy. However, as mentioned earlier, single-shot multi-class discrimination benefits from the larger measurement space enabled by sensors comprising multiple qubits. To this end, we now consider a $C=4$ class modified \logspirals{} discrimination task, depicted in Fig.~\ref{fig:2d}f. Here, each pair of phases is now indexed by one of four class labels. Both the QS and QCS protocols we employ are identical to those used for the binary \logspirals{} task, as the dimension of the input $\bm{\theta}$ is unchanged; these are depicted in Figs.~\ref{fig:2d}g.i and \ref{fig:2d}g.ii respectively. The QCS protocol is again trained via supervised learning.

The expected measurement results $\xqs$ for the QS protocol are shown in Fig.~\ref{fig:2d}h.i, which again are sinusoidal functions of $\theta_1,\theta_2$. The final MLP output to predict the class label is presented in Fig.~\ref{fig:2d}j.i for $N=S=2^6$. Analogously to the previous task, larger values of $\theta$ where the four classes are more separated can be classified correctly. However, the QS output is too noisy to distinguish $\bm{\theta}$ values near the origin, where the classes are much closer together, regardless of the complexity of classical postprocessing resources.

For a trained QCS protocol with $L=2^6$, the obtained bitstring probabilities for the same task are shown in Fig.~\ref{fig:2d}h.ii. We now find that each bitstring probability (approximately) approaches unity for inputs $\bm{\theta}$ belonging to a given class; more precisely, $x_{10},x_{11},x_{01},x_{00} \to 1$ for Class~1,~2,~3,~4 respectively. This was precisely the intuition behind our consideration of a quantum computational sensor comprising $M= \log_2 C$ qubits. This mapping also implies a reduction in sampling noise, and again suggests that such a quantum computational sensor should enable single-shot classification with high accuracy. This is observed in the MLP output shown in Fig.~\ref{fig:2d}j.ii, for $N=L=2^6$ so that $S=1$. This QCS protocol is able to achieve a classification error of $6.6\%$, much lower than the QS error rate of $28.2\%$; we note that the precise value of the relative improvement will depend on both the specific task and the optimization of $\Uqcs$. The classification error for different $L$ in Fig.~\ref{fig:2d}k follows a familiar theme: QCS with increasing $L$ provides a lower classification error than the conventional QS approach for the same $N$.

\subsection{QCS of spatiotemporal signals using multiple qubits}
\label{sec:meg}

We have thus far analyzed quantum computational sensing of multi-parameter but static signals. In many settings, however, the physical signal to be sensed may vary over time, $\bm{u}(t)$, so that repeated sensing operations are sensitive to a different signal. It is natural to ask whether quantum computational sensing protocols can be devised for this sensing paradigm.

\begin{figure}[t]
    \centering
    \includegraphics{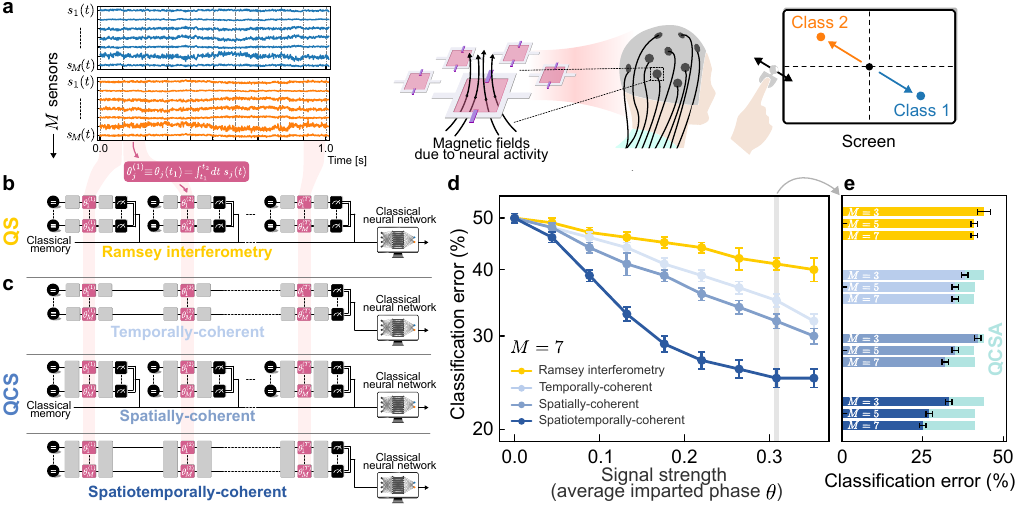}
    \caption{\textbf{Quantum computational sensing of spatiotemporal signals.} \textbf{a}, Motor imagery classification tasks using magnetoencephalography (MEG). Neural activity of a subject is mapped using MEG while the subject is asked to perform one of $C=2$ physical actions, here moving their hand in the direction of a cursor that can take a position on the bottom-right or upper-left of a screen. Left panel shows an example of the recorded spatiotemporal signals, obtained from an actual motorimagery experiment in Ref.~\cite{yeom2023magnetoencephalography}. \textbf{b}, Protocol for conventional QS of spatiotemporal signals. \textbf{c}, Protocols for QCS of spatiotemporal signals. \textbf{d}, Classification error as a function of signal strength for the conventional QS scheme and three depicted QCS schemes. \textbf{e}, Classification error as a function of number of sensing qubits used at a fixed signal strength indicated by the shaded gray line in \textbf{d}, for the conventional QS scheme and three depicted QCS schemes. Green shaded region marks reduction in classification error, and hence achieved QCSA, of each QCS protocol in comparison to the conventional QS protocol using the same number of qubits $M$.}
    \label{fig:meg}
\end{figure}

The specific problem we consider is from the field of cognitive neuroscience, and for which magnetic field sensing is required in practice: the analysis of the brain waves of a patient performing a specific physical task. This neural activity of human subjects can be recorded using non-invasive techniques such as magnetoencephalography (MEG), namely the detection of magnetic fields produced by electric currents in the brain~\cite{gross_magnetoencephalography_2019}. Present sensing platforms for this task already use cryogenic SQUID-based magnetometers; it is therefore feasible that such tasks could present a possible future application of superconducting qubit-based quantum sensors~\cite{aslam2023quantum}. We use this motivation to theoretically analyze quantum computational sensing using an actual measured MEG dataset for motor imagery analysis from Ref.~\cite{yeom2023magnetoencephalography}. In this study, MEG was used to map neural activity of human subjects during the movement of the subject's hand in the direction of a presented visual stimulus, as depicted in the right panel of Fig.~\ref{fig:meg}a. The motor imagery analysis task then required processing the recorded neural activity to correctly deduce the movement direction among two choices as depicted here, therefore defining a binary classification task. 


Examples of typical MEG signals from the dataset of Ref.~\cite{yeom2023magnetoencephalography} are depicted in Fig.~\ref{fig:meg}a for each of the two movement directions that define the two classes. The index $j$ marks spatial locations on the head from which the corresponding signal $s_j(t)$ is sensed, and $t\in [0,\mathcal{T}]$ for a total signal length $\mathcal{T}$. We first detail our model for how a qubit-based sensor receives such time-dependent signals. We consider signals at spatial location $j$ to only be sensed by qubit-$j$, as before. Secondly we introduce $\tau$ as the sensing duration for a single sensing operation. The resulting measured phase is given by the integral of the signal over the duration $\tau$, $\theta_j(t_l) = \int_{t_l}^{t_l+\tau} dt~s_j(t)$. The time index $t_l$ then represents the coarse-grained signal that the sensor receives in the $l$th sensing step. The total number of discrete sensing steps is therefore $T = \lfloor \frac{\mathcal{T}}{\tau} \rfloor$. 

Before detailing the quantum sensor and quantum computational sensor architectures and their relative performance, we discuss some features that are common across both approaches. First, the entire time-varying signal is allowed to be received by either sensor only once before a prediction for the movement direction must be made; more precisely, the signal cannot be repeatedly sensed to collect more information and boost the classification fidelity. This is informed by practical constraints of sensing time-varying transient signals, which cannot be assumed to be reproducible. Secondly, as with all other tasks considered so far, we allow measurement results $\bm{X}$ of either sensor to be passed through a classical postprocessing step, here a neural network with an MLP architecture with ReLU activation. Full details of the end-to-end training procedure for the QCS protocols and the following MLP can be found in Appendix~\ref{app:meg}.

The conventional QS approach for such signals is to use a Ramsey interferometer comprising $M$ qubits, with each qubit used to sense the time-varying signal at a specified spatial location (see Fig.~\ref{fig:meg}b). A single shot of the quantum sensor consists of a single sensing operation, followed by measurement. As a result, following the first sensing step during which the sensor is sensitive to the phase $\theta_j(t_1)$, a measurement is performed and the sensor is reset. This prepares the quantum sensor for the next sensing operation, during which it receives $\theta_j(t_2)$; the protocol continues until the entire time-varying signal has been sensed. 

For QCS we analyze three distinct protocols, informed by the protocols analyzed earlier for static signals, all still within the general structure of Eq.~(\ref{eq:cohprocessing}); these schemes are depicted in Fig.~\ref{fig:meg}c. The first $M$-qubit quantum-computational sensor adopts the coherent processing layers used by the protocol in Sec.~\ref{subsec:1q}, although now subsequent sensing operations are sensitive to a time-varying signal instead of a static one (recall that $\bm{\theta}^{(l)}$ was independent of $l$ in Sec.~\ref{subsec:1q}). In particular, we set $L=T$, so the entire signal is sensed and coherently processed before a single measurement is performed in the $M$-qubit computational space. However, no multi-qubit operations are performed; we therefore refer to this scheme as a \textit{temporally-coherent} scheme, since it generates no entanglement between the sensing qubits. From our prior analysis, such a protocol should be able to coherently process the sensed signal to identify temporal correlations of the phase $\theta_j$ sensed by qubit-$j$ at different times, but will not be able to compute correlations across different qubits: any such spatial correlations will have to be extracted from the measurement results by the classical neural network in postprocessing.

The second QCS protocol instead allows for general, trainable $M$-qubit unitary operations across the sensing qubits similar to Sec.~\ref{subsec:2q}, but after each sensing operation a measurement of the sensor is performed and the result stored in classical memory. The sensor is then reset before sensing the signal for the next sensing step. As the result, this \textit{spatially-coherent} scheme utilizes coupling across $M$-qubits and can process spatial correlations of the signals via the dynamics of the quantum computational sensor itself. On the other hand, this architecture is unable to coherently process temporal correlations, due to the projective nature of the measurement (which destroys entanglement between the qubits). Again, any temporal correlations in signals must then be extracted via the classical postprocessing layer.

Lastly, we consider the natural combination of these two protocols: a fully \textit{spatiotemporally-coherent} scheme. Similar to the temporally-coherent protocol, this scheme utilizes coherent processing layers, sensing phases $\bm{\theta}^{(l)}$ corresponding to different times $t_l$ before a single computational-basis measurement is performed; however, it also incorporates multi-qubit entangling operations like the spatially-coherent scheme. We expect such a scheme to be able to coherently process and compute complex functions across the full spatio-temporal space of the sensed signals. Of course, this comes at the expense of requiring both temporal coherence for the duration of the sensed signal, and entanglement across $M$ qubits.

Having introduced the QS and QCS protocols and their distinct architectures, we note that the total number of measurements performed differs across them: in particular the QS and spatially-coherent QCS scheme perform $M\times T$ measurements, while the temporally-coherent and spatiotemporally-coherent QCS schemes require $M$ measurements. The total number of sensing periods (per qubit), however, remains fixed at $N=T$ across all protocols. 

Fig.~\ref{fig:meg}d shows the performance of the conventional QS protocol and the various QCS protocols, as a function of the strength of the received signal (here given by the RMS imparted phase $\theta_{\rm RMS}$, see Appendix~\ref{app:meg}) for sensors with $M=7$ qubits. All of the QCS schemes provide a substantial reduction in the classification error in comparison to the QS protocol. The fully spatiotemporally-coherent QCS scheme provides the lowest classification error, with the spatially-coherent scheme providing the next best performance. Fig.~\ref{fig:meg}e shows the performance comparison for a fixed signal strength but now as a function of $M$. While all schemes show an improvement with increasing $M$, we see again that the QCS protocols outperform the conventional QS protocol. 

Our results emphasize that QCS protocols can be applied to spatiotemporally varying signals, and further that such protocols can be trained using supervised learning to achieve a QCSA. Comparisons of the various QCS protocols also provide some insights into the origin of QCSA for this task. For example, the spatially-coherent QCS protocol marginally outperforming the temporally-coherent protocol indicates that spatial correlations across sensors carry more useful information to discriminate the two classes in comparison to the temporal correlations of signals received by a single sensor. Since both outperform the QS approach, it is reasonable to expect the fully spatiotemporally-coherent QCS scheme to be the best option, as is duly found. 

We note of course that if the time-varying signals could be repeatedly sensed and an arbitrary number of measurements could be performed, a classical postprocessing step could compute any required function of the signal (e.g. computing spatial and or temporal correlations) to a sufficiently-high precision, suppressing the classification error even in the QS protocol. However, as was seen with the simpler static signal examples, given constrained sensing resources and finite measurements, sampling noise can strongly limit the amount of useful information extractable for complex tasks. QCS provides a means to compute such correlations via the dynamics of the quantum computational sensor prior to measurement, enabling a substantially-reduced classification error.

\subsection{Summary of QCS using qubit-only sensors}

By starting our analysis with the simplest case of sensing static, few-parameter signals, we have been able to investigate several key aspects of quantum computational sensing in quantitative detail. Low-dimensional tasks are highly-interpretable: the function $\Ft$ that is the target for the QCS protocol can even be visualized, and therefore the ability of increasingly-complex QCS protocols (for the qubit-based architecture, quantified by the number of processing layers $L$) to infer $\Ft$ from data can be clearly observed. We emphasize here that the reduced dimensionality of sensed parameters does not restrict the complexity of the target function $\Ft$ within this reduced space. In particular, we consider classification tasks for which $\Ft$ can be highly nonlinear functions of sensed inputs, and even explore QCSA as a function of the complexity of $\Ft$ in a controlled fashion for single-variable tasks.

We also see how certain target functions also minimize sampling noise, so the QCS protocols can achieve a low classification error even using a single shot. Perhaps most importantly, the QCS protocols can be applied to general tasks where the target function $\Ft$ is not \textit{a priori} known; we show that using supervised learning schemes, QCS protocols can be successfully trained for a variety of tasks even using finitely-sampled measurement results. This general architecture of QCS protocols using multiple sensing and processing operations is then applied to binary classification of more complex, spatiotemporally-varying signals from a real dataset. We find that qubit-based quantum computational sensors can also provide a QCSA for these more complex sensing tasks.

Overall, our results emphasize that qubit-based quantum computational sensors can be used to achieve a QCSA for general tasks. The scale of the advantage can vary, and will in general be dependent both on the specific task, as well as on the specific QCS protocols. The latter dependence includes both macroscopic properties such as the general structure of $\Ucoh{l}$ in Eq.~(\ref{eq:cohprocessing}) and the number of processing layers $L$, as well as more microscopic quantities such as the specific parameters optimized via supervised learning. In the same vein, we note that alternative QCS protocols may also exist, especially if one also allows control over the sensing operations (by controlling the individual sensing times, for example). The exploration of optimal QCS protocols provides rich avenues for future work.

\section{Quantum Computational Sensing using bosonic sensors}
\label{sec:bosonic}

We have thus far explored quantum computational sensing using qubit-based systems. However, quantum computational sensing protocols can equivalently be designed for other sensing platforms, as we now demonstrate. 

An example of sensing using modes other than qubits is the sensing of electromagnetic fields using bosonic modes. An electromagnetic signal---such as that reflecting off an aircraft and detected via radar---can cause a bosonic receiver that it is incident on to experience a displacement $\bm{u} \to \alpha$. Several physical platforms can realize this sensing interaction, ranging from microwave microstrip antennae to optomechanical sensors, to optical Rydberg receivers, to name just a few (see Fig.~\ref{fig:bosonic}a). In general, the displacement $\alpha = \alpha_x + i\alpha_y$ can be complex-valued (where $\alpha_x,\alpha_y$ are real), with the real and imaginary parts describing the in-phase and quadrature components of an incoming field. As before, we will consider tasks where the objective is not to estimate $\alpha$, but instead to extract more relevant information, such as whether the received signal was reflected off of a friendly or a hostile aircraft.

In the present section, we will demonstrate a QCS protocol for such signals, using quantum computational sensors that are composed of bosonic modes only. The QCS protocol follows the general architecture we have proposed in Eq.~(\ref{eq:cohprocessing}). However, in contrast to the qubit-based quantum computational sensors analyzed thus far, this all-bosonic QCS protocol requires only a single processing layer ($L=1$), and therefore only a single sensing operation, foregoing the need for repeated sensing. We will further assume that the sole processing operation---which is equivalent to a probe-state-preparation step, $\Ucoh{1} = \mathcal{U}_{\rm probe}$---is just the identity operation; as a result, both QS and QCS protocols rely on the nontrivial $\Udec$ operation to achieve their respective goals. Finally, we note that while bosonic sensors have been considered for QCS before, the resulting protocols are typically designed to estimate arbitrary linear combinations of sensed displacements~\cite{zhuang_physical-layer_2019}; such schemes can be used as a sub-routine for arbitrary function approximation, but further require the aid of adaptive protocols~\cite{bringewatt_optimal_2024, ge_heisenberg-limited_2024}. The QCS scheme we present here is non-adaptive, and uses an engineered nonlinear Hamiltonian on bosonic modes to compute arbitrary nonlinear functions via the dynamics of the quantum computational sensor itself.

\subsection{Function-approximation tasks}

We consider the very general task of approximating an arbitrary nonlinear, real-valued polynomial $\Ft$ of the complex-valued displacement $\alpha = \alpha_x + i \alpha_y$ over a domain $\mathcal{R}$, defined as:
\begin{align}
    \Ft = \sum_{n,m=0}^D \mathcal{W}_{nm} \alpha^{*n}\alpha^m,
    \label{eq:ftbosonic}
\end{align}
where $\mathcal{W}_{nm} = \mathcal{W}_{mn}^*$ to ensure that $\Ft$ is a real-valued function. The (arbitrary) integer $D$ determines the nonlinearity of the target function; in particular, $2D$ is the largest degree of any term in $\Ft$. 

Before discussing the QS and QCS approaches to this task, we must define a fidelity metric that will allow a comparison of the two approaches, analogous to classification error metrics we have considered for discrimination tasks. We define $\Fqs$ and $\Fqcs$ as the estimates of the target function $\Ft$ computed using the QS and QCS approaches respectively. While the construction of these estimates will be fundamentally different -- as we will detail shortly -- both are constructed using finite measurements from a quantum system: as a result, such an estimate will be stochastic. The fidelity of such a noisy estimate of $\Ft$ can be quantified using an \textit{expected} mean-squared-error (MSE) over repeated measurements, 
\begin{align}
    \mathbb{E}[\MSE]_j = \frac{1}{f}\int_{\mathcal{R}} d\bm{u}~\mathbb{E}[( \Ft- \mathcal{F}_j )^2]~~\stackrel{\mathbb{E}[ \mathcal{F}_j ] = \Ft}{=}~~ \frac{1}{f}\int_{\mathcal{R}} d\bm{u}~{\rm Var}[\mathcal{F}_j]
    \label{eq:emse}
\end{align}
where $j \in \{{\rm QS},{\rm QCS}\}$, and $f=\int_{\mathcal{R}} d\bm{u}~(\Ft)^2$ is a normalization constant. The second equality follows in the special case where we are able to construct an \textit{unbiased} estimator of $\Ft$, namely one whose mean value equals the desired target, $\mathbb{E}[ \mathcal{F}_j ] = \Ft$. In this case, the expected MSE depends entirely on the variance of the estimator, which will be determined by quantum sampling noise. Using a variety of function approximation tasks, we will show that our proposed QCS protocol can achieve a lower expected MSE than a comparable QS protocol, $\mathbb{E}[\MSE]_{\rm QCS} < \mathbb{E}[\MSE]_{\rm QS}$, leading to a QCSA.

\subsection{QS approach: linear phase-preserving amplifier}

\begin{figure}[t]
    \centering
    \includegraphics{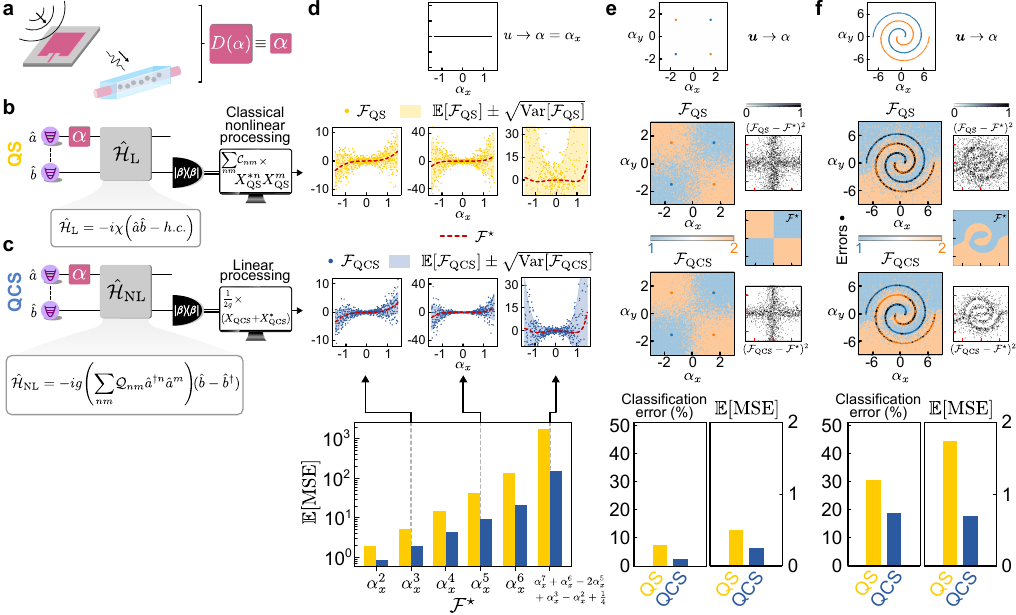}
    \caption{\textbf{Quantum computational sensing using bosonic modes.} \textbf{a}, An RF signal is received by a bosonic mode, leading to the sensing of a displacement via a physical interaction that can be realized in a variety of quantum sensing platforms; some examples depicted here include microwave microstrip antennae and optical Rydberg receivers. \textbf{b}, Conventional QS approach using a phase-preserving quantum linear amplifier to sense a complex-valued displacement, followed by classical nonlinear postprocessing to compute arbitrary polynomials of $\alpha$. \textbf{c}, QCS approach using a quantum nonlinear amplifier realized using the depicted quantum non-demolition Hamiltonian. Measurement of the nonlinear amplifier directly computes nonlinear polynomials of $\alpha$, engineered using the choice of coefficients $\mathcal{Q}_{nm}$ in the Hamiltonian. \textbf{d}, Single-variable nonlinear function approximation using QS and QCS approaches. Lower panel shows the expected mean-squared-error for a selection of polynomials of increasing degree; examples of some specific polynomials are shown in the center panel. \textbf{e}, Two-variable function approximation for the \XOR{} task, which requires $D=2$. Center panel shows a single instance of the estimators $\Fqs$ and $\Fqcs$ (passed through a thresholding function for classification), with the target function $\F$ and the squared error shown in the right column. The lower panel shows classification error for the \XOR{} task and the expected mean-squared-error over the input dataset. \textbf{f}, Same as \textbf{e}, now for a \spirals{} classification task which requires $D=4$. }
    \label{fig:bosonic}
\end{figure}

We begin by describing the conventional QS approach to sensing a displacement $\alpha$. A bosonic mode $\hat{a}$ is defined as the receiver, which undergoes the displacement $\alpha$ after sensing the incoming electromagnetic signal, as depicted in the top panel of Fig.~\ref{fig:bosonic}b. The receiver mode can then be directly measured to extract information about the signal, for example via heterodyne measurement. If the received signal is extremely weak, however, interfacing the receiver with a noisy classical measurement setup has the potential to swamp any extractable information. As a result, weak quantum signals are typically first amplified using a quantum amplifier, raising the total power and making it robust to noise~\cite{castellanos-beltran_amplification_2008, castellanos-beltran_development_2010}.

Since we are interested in measuring both quadratures of $\hat{a}$, the requisite quantum amplifier must provide phase-preserving amplification, as we now describe. Quantum phase-preserving amplification requires coupling the receiver $\hat{a}$ to a second bosonic mode $\hat{b}$ via the Hamiltonian $\hat{\mathcal{H}}_{\rm L} = -i \chi (\hat{a}\hat{b} - h.c.)$, as depicted in the top panel of Fig.~\ref{fig:bosonic}b. As a result of the amplifier interaction, the final state of mode $\hat{b}$ become sensitive to displacement $\alpha$ amplified by a gain factor $\mathcal{G}$ (which depends on $\chi$). When $\alpha$ is complex-valued, extracting all information about the sensed displacement requires heterodyne measurement of mode $\hat{b}$. In this case the outcome of heterodyne measurement of the quantum sensor is best described as a scalar but complex-valued stochastic variable, $X_{\rm QS} = \beta$. It is straightforward to show that $\mathbb{E}[X_{\rm QS}] = \sqrt{\mathcal{G}-1}~\alpha^*$, so that $X_{\rm QS}$ can be used to construct an unbiased estimator of $\alpha$ in the conventional QS protocol. Phase-preserving amplification comes at a cost, however; the variance of this estimate  ${\rm Var}[X_{\rm QS}] = \mathcal{G}$, instead of $\frac{\mathcal{G}}{2}$ that would be expected from vacuum fluctuations amplified by $\mathcal{G}$; this is typically referred to as the ``half a photon'' of noise added at the input by phase-preserving amplification~\cite{caves_quantum_1982, clerk_introduction_2010}. 

Of course we are not interested in estimating $\alpha$, but in computing unbiased estimators of polynomials of $\alpha$ as defined by Eq.~(\ref{eq:ftbosonic}); this necessitates classical nonlinear postprocessing of the measurement results $X_{\rm QS}$ obtained from the linear amplifier. In Appendix~\ref{app:bosonicFA}, we show that for $\Ft$ of degree $D$, an unbiased estimator can be constructed using an analogous nonlinear construction, $\Fqs = \sum_{nm}^D \mathcal{C}_{nm} X_{\rm QS}^{*n}X_{\rm QS}^m$. We further provide an analytic method to compute $\mathcal{C}_{nm}$ for arbitrary $\mathcal{W}_{nm}$. The nonlinear postprocessing of stochastic measurement results required for the QS protocol can worsen the effect of the amplifier added noise, thereby limiting the fidelity of function approximation, as we will see.


\subsection{QCS approach: nonlinear amplifier}

Our proposed QCS approach, on the other hand, completely eliminates the need for nonlinear classical postprocessing, by instead performing the required nonlinear computation via the dynamics of the quantum computational sensor itself, prior to measurement. We show that this is possible using a class of bosonic nonlinear amplifiers~\cite{epstein_quantum_2021}. In particular, following the sensing operation on mode $\hat{a}$, a nonlinear quantum non-demolition interaction $\hat{\mathcal{H}}_{\rm NL}$ is used to couple the sensing mode and the readout mode, as depicted in Fig.~\ref{fig:bosonic}c. This interaction takes the specific form $\hat{\mathcal{H}}_{\rm NL} = \sqrt{2} g \hat{f} \hat{P}_b$ where $\hat{P}_b = -\frac{i}{\sqrt{2}}(\hat{b}-\hat{b}^{\dagger})$ is the mode $\hat{b}$ canonical momentum operator, and $\hat{f} = \hat{f}(\hat{a},\hat{a}^{\dagger})$ is a Hermitian operator that can be \textit{nonlinear} in the mode $\hat{a}$ creation and annihilation operators. 

This specially-chosen quantum non-demolition interaction means that dynamics under $\hat{\mathcal{H}}_{\rm NL}$ effectively lead to the readout mode $\hat{b}$ being driven by the operator $\hat{f}$; consequently, a heterodyne measurement of the readout mode yields measurement outcomes $X_{\rm QCS}$ that satisfy $\mathbb{E}[X_{\rm QCS}] = g \avg{\hat{f}}$, where $\avg{\hat{f}} \equiv \bra{\psi_0}\Usense^{\dagger}(\alpha)\hat{f}~\Usense(\alpha)\ket{\psi_0}$. Remarkably, this is true for arbitrary Hermitian $\hat{f}$; in particular, for the choice $\hat{f} = \sum^D_{nm} \mathcal{Q}_{nm}\hat{a}^{\dagger n}\hat{a}^m$ (with the constraint that $\mathcal{Q}_{nm} = \mathcal{Q}_{mn}^*$ to enforce Hermiticity)---which describes a nonlinear interaction of mode $\hat{a}$---we can compute $\avg{\hat{f}}$ and show that $\mathbb{E}[X_{\rm QCS}] = g \sum_{nm}^D \mathcal{Q}_{nm}\alpha^{* n}\alpha^m$. Therefore a simple heterodyne measurement of the readout mode $\hat{b}$ directly yields an estimate of a nonlinear polynomial of $\alpha$. If we now define $\mathcal{Q}_{nm} = \mathcal{W}_{nm}$ and the real-valued estimator constructed using the QCS approach as $\Fqcs = \frac{1}{2g}(X_{\rm QCS}+X_{\rm QCS}^*)$, we immediately see that $\mathbb{E}[\Fqcs] = \Ft$. This has the following interpretation: the entire nonlinearity required for computation has been incorporated into the quantum dynamics via $\hat{\mathcal{H}}_{\rm NL}$. This essentially eliminates the need for any additional complex classical postprocessing (at most, a simple linear operation may be further required). Furthermore, the interaction strengths $\mathcal{Q}_{nm}$ required are simply set by the polynomial coefficients $\mathcal{W}_{nm}$ we wish to build an estimator for. Whether this approach provides a \QCSA{} depends on the error in estimating $\Ft$, which is given by ${\rm Var}[\Fqcs]$ via Eq.~(\ref{eq:emse}). Our specific choice of $\hat{f}$ simplifies this calculation for the QCS approach, allowing us to determine the error of function approximation for arbitrary $\Ft$ of the form of Eq.~(\ref{eq:ftbosonic})~(see Appendix~\ref{app:bosonicFA}).

\subsection{Quantum computational-sensing advantage}

We now compare the conventional QS approach against the QCS scheme, here for a variety of function-approximation tasks; we will discuss shortly how function approximation can enable classification tasks. The target coefficients $\mathcal{W}_{nm}$, and the required $\mathcal{C}_{nm}, \mathcal{Q}_{nm}$ for the various tasks we consider are included in Appendix~\ref{app:bosonicFA}. We start with the simple case where $\alpha = \alpha_x$ and the task is the approximation of degree-$D$ polynomials of this real-valued displacement.
For a selection of polynomials $\Ft$ of degree up to $D=7$, we calculate the expected MSE of unbiased estimators $\Fqs$ and $\Fqcs$ for the QS and QCS approaches respectively; this is straightforward using our analytic methods to obtain ${\rm Var}[\Fqs]$ and ${\rm Var}[\Fqcs]$, and the results are shown in Fig.~\ref{fig:bosonic}d. We observe a general feature that while the expected MSE increases with $D$ for both schemes, the QS approach exhibits a larger increase in error. For illustration, we show the estimator mean and variance for select polynomials over the domain of $\alpha_x$, together with individual stochastic instances of $\Fqs$ and $\Fqcs$; the latter are sampled assuming a Gaussian distribution for simplicity, which does not change the estimator variance or expected MSE. The increase in variance for the QS approach relative to the QCS approach with increasing $D$ is apparent here.

The true use of phase-preserving amplifiers is when both quadratures of the sensed signal carry useful information and hence must both be amplified and measured. We now consider this more general case in the context of one of the most prototypical nonlinear classification tasks: \XOR{} classification. More precisely, we consider the sensing of the complex-valued displacement $\alpha = \alpha_x + i \alpha_y$, which can only take one of four possible values, as shown in the top panel of Fig.~\ref{fig:bosonic}e: $\alpha_x = -\alpha_y = \pm \eta$, both of which are indexed as Class~1, and $\alpha_x = +\alpha_y = \pm \eta$, both of which are indexed as Class~2, and $\eta$ sets the displacement scale. The task is to determine the true class label, as usual. Information from a single quadrature (i.e. sensing only $\alpha_x$ or $\alpha_y$) yields no information about the true class label. However, an extremely simple nonlinear processing step can be used to distinguish the two classes: we note that ${\rm sign}\{\alpha_x\alpha_y\} = -1$ for Class~1 and ${\rm sign}\{\alpha_x\alpha_y\} = +1$ for Class~2. For our purposes, this quantity can be estimated by computing (an approximation of) the quadratic polynomial $\Ft =i\alpha^{*2} - i\alpha^2  = 4\alpha_x\alpha_y$, providing another example of how classification tasks can be cast in the language of function approximation. Using our analytic results, we again construct unbiased estimators $\Fqs$ and $\Fqcs$ of $\Ft$ (for a worked example, see Appendix~\ref{app:xor}); a single instance of each is shown in Fig.~\ref{fig:bosonic}e. The squared error in the approximation is also shown over the $\alpha_x,\alpha_y$ values that define the \XOR{} task, indicating that the QCS approach provides an estimate with lower error. The result of the function approximation can directly be passed through a simple thresholding function (determining its sign) to output the predicted class label. This leads to an improved classification accuracy using the QCS approach.

Our analytic results can be used to carry out the above analysis for arbitrarily-complex function approximation and classification tasks. For more general tasks involving signals belonging to more complex datasets, the required target function $\Ft$ may not be as simple to infer as in the \XOR{} task. However in such cases we can efficiently determine $\Ft$ via supervised learning approaches, namely learning the target coefficients $\mathcal{W}_{nm}$ using a training dataset. Subsequently, our analytic method for determining the required $\mathcal{C}_{nm}$ and $\mathcal{Q}_{nm}$ coefficients for QS and QCS approaches respectively can be applied. We use this method to consider highly-nonlinear tasks such as the \spirals{} classification task considered in Fig.~\ref{fig:bosonic}f. For the resulting target function $\Ft$---which requires the maximum degree $D=4$---we are again able to determine the expected MSE for unbiased estimators $\Fqs$ and $\Fqcs$. Relative to the \XOR{} task, the expected MSE increases for both QS and QCS approaches, consistent with the increased complexity of this target function. Importantly, the separation between the two approaches also increases; this  reiterates an observation we have now made in several examples, that the extent of QCSA can increase with task complexity.

\section{Quantum Computational Sensing using hybrid sensors}
\label{sec:hybrid}

In this final section, we present a protocol for quantum computational sensing using hybrid systems comprising qubits and bosonic modes. The use of hybrid architectures has been of interest for various quantum computational sensing protocols, for example to equip otherwise linear quantum systems with non-Gaussian resources~\cite{sinanan-singh_single-shot_2024, liao_quantum-enhanced_2024}. 

\begin{figure}
    \centering
    \includegraphics{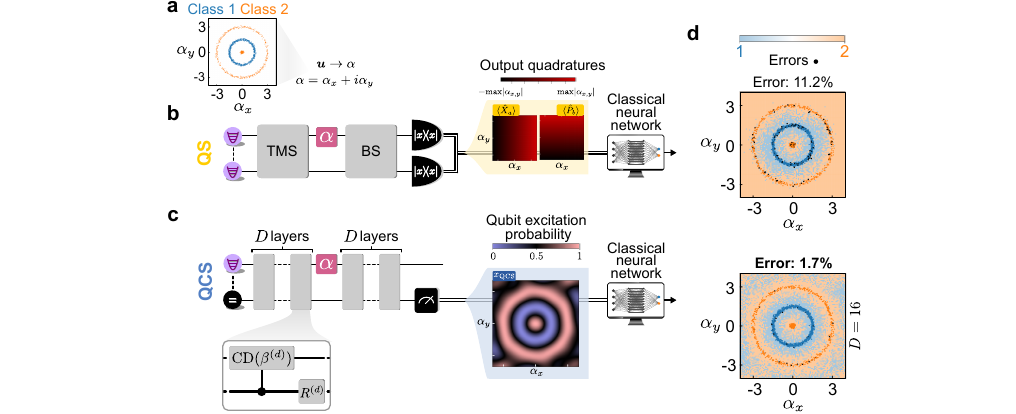}
    \caption{\textbf{Quantum computational sensing using a hybrid qubit-bosonic mode sensor.} \textbf{a}, Example of the binary classification tasks analyzed, here an instance of the \circles{} dataset. \textbf{b}, Conventional bosonic quantum sensor for sensing a complex-valued displacement $\alpha$, where the sensing bosonic mode is coupled to a second mode with which it is entangled.  \textbf{c}, Quantum computational sensor realized by coupling the sensing bosonic mode to a single qubit, whose state is read out and passed through a classical postprocessor, here a neural network. \textbf{d}, Output of QS protocol followed by classical postprocessing (top panel), and QCS protocol followed by classical postprocessing (bottom panel) according to the colorbar, over a fixed domain of $\alpha$ values, similar to Fig.~\ref{fig:2d}. Dots are instances of $\alpha$ sampled from the testing dataset in \textbf{a}: where shown in color, these are correctly classified, while black dots are errors. The resulting classification errors are also provided. }
    \label{fig:bqsp}
\end{figure}

We also return to classification tasks where a static signal received by a sensor belongs to one of two possible classes; a specific instance we consider is from a \circles{} dataset shown in Fig.~\ref{fig:bqsp}a. Like the previous section, we consider displacement sensing; only a bosonic mode is used to sense the complex-valued displacement $\alpha$. The conventional QS protocol we consider is inspired by a standard entanglement-enhanced protocol for the measurement of complex displacements~\cite{braunstein_dense_2000}, and is depicted in Fig.~\ref{fig:bqsp}b. The protocol starts with a probe state preparation step $\Uenc$ consisting of a two-mode squeezing operation (TMS) to entangle the sensing mode with an auxiliary bosonic mode prior to the sensing operation. Following sensing, $\Udec$ performs a balanced beam splitter (BS). Finally, homodyne measurements are performed on both modes, to measure the quadratures $\hat{X}_a = \frac{1}{\sqrt{2}}(\hat{a}+\hat{a}^{\dagger})$ and $\hat{P}_b =\frac{-i}{\sqrt{2}}(\hat{b}-\hat{b}^{\dagger})$. Defining the average measurement results $\xqs = (\avg{\hat{X}_a},\avg{\hat{P}_b})$, we find that $\avg{\hat{X}_a} = \alpha_x$, and $\avg{\hat{P}_b} = \alpha_y$, as depicted in the right panel of Fig.~\ref{fig:bqsp}b; as a result, these measurements can directly estimate the real and imaginary parts of the complex displacement $\alpha$~\cite{braunstein_dense_2000}. 


Unlike the QS protocol based on a phase-preserving linear amplifier, however, the variance of these estimates can now be reduced below the vacuum variance, by virtue of the TMS operation that enables the measured $\hat{X}_a$ and $\hat{P}_b$ quadratures to be squeezed; this directly improves the accuracy of estimating $\alpha$. As will be discussed shortly, the QCS protocol we consider for hybrid sensors will involve a probe state preparation step $\Uenc$ (unlike QCS using bosonic nonlinear amplifiers); therefore, allowing the QS protocol to also prepare a probe state allows for a fair performance comparison. The QS protocol is completed by passing the stochastic measurement results $\Xqs$ from the quantum sensor through a classical neural network, which is trained to predict the class label using supervised learning.

Now, we can define the hybrid quantum computational sensors we analyze, which consist of a single bosonic mode coupled to a single qubit; only the bosonic mode is used to sense the displacement $\alpha$, and only the qubit is used for readout. We note that other versions of hybrid sensors could of course be considered, for example sensors using the qubit for sensing instead, sensors allowing for joint sensing and readout using the same mode, as well as sensors utilizing multiple modes. The protocols here are described by Eq.~(\ref{eq:cohprocessing}) with a single sensing operation ($L=1$), so that $\Ucoh{1} = \Uenc$. To engineer a useful QCS protocol, these trainable computing operations are based on a computing ansatz that can equip the quantum computational sensor to process the sensed displacement prior to measurement. Here, this ansatz is inspired by a generalization of the bosonic quantum signal processing~\cite{sinanan-singh_single-shot_2024} algorithm, as depicted in Fig.~\ref{fig:bqsp}b. Here, $\Udec = \Uenc^{\dagger}$, and $\Uenc$ is decomposed into $D$ layers: each layer interleaves a qubit-state-conditioned displacement ${\rm CD}(\beta^{(d)})$ of the bosonic mode by the complex amplitude $\beta^{(d)}$ with a general single qubit rotation $R^{(d)}$, where $d=1,\ldots,D$ indexes the ansatz layer (for full details, see Appendix~\ref{app:bqsp}). A larger value of $D$ can increase the complexity of the functions that can be approximated by the resulting QCS protocols~\cite{sinanan-singh_single-shot_2024}. We note that such conditional displacement gates have been experimentally demonstrated in various platforms~\cite{campagne-ibarcq_quantum_2020, eickbusch_fast_2022}.

We then use our supervised learning approach to train this hybrid QCS architecture to perform binary classification tasks. Unlike the case of qubit-based sensors, the simulation of a bosonic mode with a formally infinite Hilbert space requires careful management to ensure our results are not influenced by truncation effects. We use a specific loss function to ensure such effects can be avoided, and further verify during testing that the trained QCS protocols are unchanged for larger Hilbert space sizes. More details on our approach can be found in Appendix~\ref{app:bqsp}. As in all other instances of supervised learning in this paper, we train the hybrid sensor using a training dataset of finitely-sampled measurement results, here with just $S=1$. 

Using supervised learning, the QCS protocol can be trained to engineer $x_{\rm QCS}$ over the domain of complex-valued displacements sensed by the bosonic mode that the qubit is coupled to, for a given task. For the \circles{} task using this hybrid sensor, the qubit excitation probability $x_{\rm QCS}$ of our trained QCS protocol with $D=16$ is shown in the right panel of Fig.~\ref{fig:bqsp}c. Note how this protocol allows $x_{\rm QCS} \to 0$ for Class 1, while $x_{\rm QCS} \to 1$ for Class 2, much like QCS protocols using qubit-only quantum computational sensors. 

We can finally compare the performance of the QS and QCS protocols for the \circles{} task. For a fair comparison, we ensure that the average photon number in the bosonic modes following $\Uenc$ is identical in both the QS and QCS protocols. The output of the classical neural network is shown in Fig.~\ref{fig:bqsp}d, together with the classification error, for both protocols. The QS protocol achieves an error or 11.2\%. The QCS protocol, on the other hand, achieves a substantially lower error of 1.7\%, leading to a QCSA.

These results demonstrate that hybrid qubit-bosonic mode sensors can also be used for QCS. We note that QCS protocols using hybrid sensors have recently been proposed for both real-valued~\cite{sinanan-singh_single-shot_2024} and complex-valued displacement sensing~\cite{liao_quantum-enhanced_2024}, for constructing thresholding (`step') functions for binary decision problems. Our work analyzes binary decision tasks where more general nonlinear decision functions must be constructed, and demonstrates that hybrid QCS protocols to do so can be efficiently trained via supervised learning. Finally, while the conventional QS protocol used here makes use of $\Uenc$, simpler protocols are often used for practical convenience, such as those based on phase-preserving amplifiers. In Appendix~\ref{app:bqsp} we compare our QCS protocol against this simpler QS protocol, finding an even larger QCSA across a variety of binary classification tasks.

\section{Discussion and Outlook}
\label{sec:outlook}

\subsection{Overview of our work}

In this paper we have presented a variety of quantum computational sensing (QCS) protocols covering qubit-based sensors, bosonic sensors, and hybrid qubit--bosonic sensors; different numbers of sensing qubits/modes; and a range of tasks: binary and multiclass discrimination of static signals, arbitrary nonlinear function approximation, and binary discrimination of time-varying signals. For each QCS protocol, we identify a conventional quantum sensing (QS) protocol as a baseline to compare against: the conventional QS protocol first performs parameter estimation using a standard quantum sensor, followed by classical postprocessing to complete the task. We show that each of the QCS protocols we present outperforms its respective baseline QS protocol for a fixed number $N$ of sensing operations per sensing mode, which determines the total sensing time, thereby achieving a quantum computational-sensing advantage (QCSA). The task definitions, datasets, and quantum models used are publicly available (see Data and Code Availability) and we hope will find use as open benchmarks for future proposals and studies of QCS protocols.

Our analysis of this seemingly-disparate collection of QCS protocols is unified by a common protocol structure, defined by Eq.~(\ref{eq:cohprocessing}): $L$ sensing operations are performed and interleaved with $L$ computing operations, to coherently process the received signals prior to measurement. This structure applies regardless of whether a protocol is for static or for time-varying signals, and regardless of the number of modes $M$ comprising the quantum computational sensor. The complexity of the QCS protocol generally depends on the choice of individual processing operations as well as the number of coherent operations $L$. We show in several cases that increasing complexity enables more powerful QCS protocols that achieve a larger QCSA.

Choosing the free parameters in the quantum operations of a QCS protocol is an important step in engineering and specializing it to a particular task. Depending on the task and QCS protocol, we either give a closed-form solution or give a numerical optimization procedure. When the target function $\Ft$ is known \textit{a priori}, it is sometimes possible to design the QCS protocol entirely analytically. For example in our analysis of bosonic sensors for the task of computing bivariate polynomials of a complex-valued displacement, we presented an analytic method to determine the required parameters of a nonlinear amplifier. However, in many cases a task is defined without having an explicit definition of the target function $\Ft$. A setting where this typically arises is in tasks where the objective is defined by a training dataset and is most naturally solved using machine learning. We show that supervised learning can be used to efficiently optimize QCS protocols across a variety of tasks and sensors. 

In supervised learning, a labeled input is sampled from the training dataset and sent to the quantum system as synthetic parameters to sense; the output of the quantum computational sensor is used to calculate a cost function, which is minimized. Several prior works using supervised learning for QCS assume this output to be the measurement probabilities, expectation values of observables, or some other function of exact density matrix elements of the quantum system. Having these exact probabilities or expectation values as outputs requires either a quantum simulation model of the system that is quantitatively accurate and can directly provide these outputs---but such a model can be difficult to construct and calibrate for many current noisy quantum systems---or the estimation of these quantities from many repeated experimental shots---formally requiring infinitely many measurement samples, and in practice requiring enough that training becomes unacceptably slow. We instead use finitely-sampled measurement results as the outputs of a quantum system for training, and show that QCS protocols can be trained using no more samples to compute the cost function in training than one would use during inference. For single-shot inference, this means we are able to train protocols even using single-shot measurement results.

\subsection{Experimental prospects and future research directions}

There are two major open questions for experimental realization of the protocols we have studied: how to train them for noisy hardware, and how much advantage can be retained in the presence of noise.

For supervised-learning QCS protocols, it is desirable to be able to train the quantum circuit without using a simulation of it.\footnote{Even for small hardware systems, quantitatively accurate simulation models are often difficult to develop. Furthermore, as quantum computational sensors become larger, using more and more qubits and/or bosonic modes, simulating them classically will also in general become more expensive.} Our training approach relies only on finite measurement samples to compute the cost function for a current choice of circuit parameters, as opposed to exact probabilities or expectation values. In an experimental hardware demonstration, these finite measurement samples could be obtained directly by running the experiment multiple times, without any need for a simulation---and in our work, we have quantified how many samples are needed. However, in our simulation studies we have performed the parameter optimization using gradient descent where it has been assumed that we have access to the exact gradient. An important final step to realize our training method on experimental hardware, without relying on quantitatively accurate simulations, is to apply the techniques from variational optimization in quantum computing---for example, the parameter-shift rule~\cite{cerezo2021variational} to obtain gradient information, or the use of surrogate models~\cite{lerch2024efficient}---to the QCS circuits we have proposed. We conjecture that even with noisy gradients, the QCS protocols we have presented could be efficiently trained.

Our analyses in this paper used unitary models of quantum evolution (up to measurement). The impact of experimental imperfections---including finite coherence times, imperfect gate fidelities, finite photon loss, etc.---on the QCS protocols we have presented has yet to be evaluated. It is an open question, for each protocol, to what extent the advantage of QCS over conventional QS can be maintained under realistic experimental conditions. However, there are two reasons to be optimistic on this front. First, we have shown that QCS with even a limited number of computing operations can provide an improvement over conventional QS. Therefore, even if practical constraints limit the maximum complexity of computing operations that can be performed with reasonable fidelity, one can expect an observation of a (smaller) QCSA with limited circuit depth and width. Second, we have observed empirically that by reducing the difficulty of a task, the required resources to achieve a QCSA also reduce. It seems likely that it will be possible in the near term to experimentally observe QCSA with at least some of the QCS protocols we have proposed, for some of the tasks we have considered. For example, trapped-ion platforms can have high coherence times and gate fidelities, and are often used to demonstrate concepts in quantum sensing~\cite{marciniak_optimal_2022}, so seem promising. 

We have shown advantages from QCS for a variety of tasks, including one---binary classification of spatiotemporal signals from magnetoencephalography (Fig.~\ref{fig:meg})---based on a practical application, and evaluated using realistic data. An open area for study is to investigate how sophisticated classification tasks can be and still have quantum computational sensors be able to achieve an advantage in performing them. For example, what advantage could a QCS protocol achieve on the magnetoencephalography use case we explored if the task were extended beyond binary classification, and beyond just 7 sensors (towards the hundreds used in practical magnetoencephalography experiments)? What other spatiotemporal-signal classification tasks could QCS give an advantage for? As the number of sensors and the complexity of the QCS protocols are increased, how can one best avoid or mitigate barren plateaus~\cite{cerezo2022challenges} to enable efficient training?

While there is much to learn from `toy' problems like the ones we have studied in this paper, another important direction for future work will be in identifying applications for which QCS protocol can give a practical advantage. The most direct route to achieving a practical advantage is probably to consider the applications where quantum sensors are used in the practical state-of-the-art solution, to then select those applications that have an end goal that is not just collecting and storing the raw sensor data, and finally to study what QCS protocols can be implemented on the quantum platforms used for sensing for those applications.

In addition to developing case studies of QCS protocols for ever-more sophisticated and realistic tasks, there remains much to be explored theoretically in general. While the complexity of functions that a quantum system can produce in the infinite-sampling limit has been explored, for example by Schuld et al.~\cite{schuld_circuit-centric_2020} in quantum machine learning, understanding what functions can be computed with high signal-to-noise ratio when there are finite shots and the sampling noise is nontrivial~\cite{hu_tackling_2023} is not as well understood. We can also ask: what are the limits to the advantages that can be obtained by QCS? How can one design optimal protocols that saturate these limits? The answers to these questions will likely depend on the definition of the task being addressed, but it would be useful to have a theory that can be applied to analyze how much advantage is possible for a given task, perhaps also taking into account constraints on how large the protocol's quantum circuit can be. Finally, one can also ask the inverse question: given a specific sensing platform and computing resources, can tasks be identified that maximize the QCSA? The answers to this question, in addition to being of abstract theoretical interest, might also provide insight into how to design or frame practical tasks in a way that is most amenable to getting a benefit from quantum computational sensing.

\section*{Data and code availability}

All data generated and code used in this work is available at \url{https://doi.org/10.5281/zenodo.15692099}.

\section*{Author contributions}

S.A.K. and S.P. carried out the research, performing the numerical simulations and analysis of the results, with S.A.K leading on quantum computational sensing of static signals and S.P. leading on quantum computational sensing of time-varying magnetoencephalography signals. S.A.K., S.P. and P.L.M. wrote the manuscript, with input from L.G.W.. P.L.M. and L.G.W. initiated the project. P.L.M. supervised the research.

\section*{Acknowledgements}

We would like to thank Richard~Allen, Valla~Fatemi, June~Sic~Kim, Vladimir~Kremenetski, Jérémie~Laydevant, Shi-Yuan~Ma, Benjamin~Malia, Tatsuhiro~Onodera, Mathieu~Ouellet, Mandar~Sohoni, Hakan~T\"ureci, Fan~Wu and Ryotatsu~Yanagimoto for helpful discussions.

We gratefully acknowledge financial support from the Air Force Office of Scientific Research under award number FA9550-22-1-0203. We thank NTT Research for their financial and technical support. P.L.M. acknowledges membership in the CIFAR Quantum Information Science Program as an Azrieli Global Scholar.

\stoptoc

\bibliographystyle{mcmahonlab}
\bibliography{references}

\begin{thebibliography}{10}

\bibitem{perspective}
S.~A. Khan, S.~Prabhu, L.~G. Wright, and P.~L. McMahon, Quantum Computational-Sensing Advantage.
\newblock {\em arXiv, to appear} (2025).

\bibitem{eldredge2018optimal}
Z.~Eldredge, M.~Foss-Feig, J.~A. Gross, S.~L. Rolston, and A.~V. Gorshkov, Optimal and secure measurement protocols for quantum sensor networks.
\newblock {\em \href{https://doi.org/10.1103/PhysRevA.97.042337}{Physical Review A}} \href{https://doi.org/10.1103/PhysRevA.97.042337}{{\bfseries 97}, 042337} (2018).

\bibitem{zhuang_physical-layer_2019}
Q.~Zhuang and Z.~Zhang, Physical-Layer Supervised Learning Assisted by an Entangled Sensor Network.
\newblock {\em \href{https://link.aps.org/doi/10.1103/PhysRevX.9.041023}{Physical Review X}} \href{https://link.aps.org/doi/10.1103/PhysRevX.9.041023}{{\bfseries 9}, 041023} (2019).

\bibitem{banchi_quantum-enhanced_2020}
L.~Banchi, Q.~Zhuang, and S.~Pirandola, Quantum-Enhanced Barcode Decoding and Pattern Recognition.
\newblock {\em \href{https://link.aps.org/doi/10.1103/PhysRevApplied.14.064026}{Physical Review Applied}} \href{https://link.aps.org/doi/10.1103/PhysRevApplied.14.064026}{{\bfseries 14}, 064026} (2020).

\bibitem{sinanan-singh_single-shot_2024}
J.~Sinanan-Singh, G.~L. Mintzer, I.~L. Chuang, and Y.~Liu, Single-shot {Quantum} {Signal} {Processing} {Interferometry}.
\newblock {\em \href{https://quantum-journal.org/papers/q-2024-07-30-1427/}{Quantum}} \href{https://quantum-journal.org/papers/q-2024-07-30-1427/}{{\bfseries 8}, 1427} (2024).

\bibitem{allen_quantum_2025}
R.~R. Allen, F.~Machado, I.~L. Chuang, H.-Y. Huang, and S.~Choi, Quantum Computing Enhanced Sensing.
\newblock {\em \href{http://arxiv.org/abs/2501.07625}{arXiv:2501.07625}} (2025).

\bibitem{qian_heisenberg-scaling_2019}
K.~Qian, Z.~Eldredge, W.~Ge, G.~Pagano, C.~Monroe, J.~V. Porto, and A.~V. Gorshkov, Heisenberg-scaling measurement protocol for analytic functions with quantum sensor networks.
\newblock {\em \href{https://link.aps.org/doi/10.1103/PhysRevA.100.042304}{Physical Review A}} \href{https://link.aps.org/doi/10.1103/PhysRevA.100.042304}{{\bfseries 100}, 042304} (2019).

\bibitem{bringewatt_optimal_2024}
J.~Bringewatt, A.~Ehrenberg, T.~Goel, and A.~V. Gorshkov, Optimal function estimation with photonic quantum sensor networks.
\newblock {\em \href{https://link.aps.org/doi/10.1103/PhysRevResearch.6.013246}{Physical Review Research}} \href{https://link.aps.org/doi/10.1103/PhysRevResearch.6.013246}{{\bfseries 6}, 013246} (2024).

\bibitem{liao_quantum-enhanced_2024}
P.~Liao, B.~Zhang, and Q.~Zhuang, Quantum-enhanced learning with a controllable bosonic variational sensor network.
\newblock {\em \href{https://dx.doi.org/10.1088/2058-9565/ad752d}{Quantum Science and Technology}} \href{https://dx.doi.org/10.1088/2058-9565/ad752d}{{\bfseries 9}, 045040} (2024).

\bibitem{cerezo2021variational}
M.~Cerezo, A.~Arrasmith, R.~Babbush, S.~C. Benjamin, S.~Endo, K.~Fujii, J.~R. McClean, K.~Mitarai, X.~Yuan, L.~Cincio et~al. Variational quantum algorithms.
\newblock {\em \href{https://doi.org/10.1038/s42254-021-00348-9}{Nature Reviews Physics}} \href{https://doi.org/10.1038/s42254-021-00348-9}{{\bfseries 3}, 625--644} (2021).

\bibitem{martyn2021grand}
J.~M. Martyn, Z.~M. Rossi, A.~K. Tan, and I.~L. Chuang, Grand unification of quantum algorithms.
\newblock {\em \href{https://doi.org/10.1103/PRXQuantum.2.040203}{PRX Quantum}} \href{https://doi.org/10.1103/PRXQuantum.2.040203}{{\bfseries 2}, 040203} (2021).

\bibitem{degen2017quantum}
C.~L. Degen, F.~Reinhard, and P.~Cappellaro, Quantum sensing.
\newblock {\em \href{http://dx.doi.org/10.1103/RevModPhys.89.035002}{Reviews of Modern Physics}} \href{http://dx.doi.org/10.1103/RevModPhys.89.035002}{{\bfseries 89}, 035002} (2017).

\bibitem{devroye_bayes_1996}
L.~Devroye, L.~Györfi, and G.~Lugosi, (1996) The {Bayes} {Error}. In {\em A {Probabilistic} {Theory} of {Pattern} {Recognition}}.
\newblock (Springer, New York, NY), pp. 9--20.

\bibitem{dambre_information_2012}
J.~Dambre, D.~Verstraeten, B.~Schrauwen, and S.~Massar, Information {Processing} {Capacity} of {Dynamical} {Systems}.
\newblock {\em \href{https://www.nature.com/articles/srep00514}{Scientific Reports}} \href{https://www.nature.com/articles/srep00514}{{\bfseries 2}, 514} (2012).

\bibitem{wright_capacity_2019}
L.~G. Wright and P.~L. McMahon, The Capacity of Quantum Neural Networks.
\newblock {\em \href{https://doi.org/10.48550/arXiv.1908.01364}{arXiv:1908.01364}} (2019).

\bibitem{hu_tackling_2023}
F.~Hu, G.~Angelatos, S.~A. Khan, M.~Vives, E.~Türeci, L.~Bello, G.~E. Rowlands, G.~J. Ribeill, and H.~E. Türeci, Tackling {Sampling} {Noise} in {Physical} {Systems} for {Machine} {Learning} {Applications}: {Fundamental} {Limits} and {Eigentasks}.
\newblock {\em \href{https://link.aps.org/doi/10.1103/PhysRevX.13.041020}{Physical Review X}} \href{https://link.aps.org/doi/10.1103/PhysRevX.13.041020}{{\bfseries 13}, 041020} (2023).

\bibitem{farhi_classification_2018}
E.~Farhi and H.~Neven, Classification with {Quantum} {Neural} {Networks} on {Near} {Term} {Processors}.
\newblock {\em \href{http://arxiv.org/abs/1802.06002}{arXiv:1802.06002}} (2018).

\bibitem{rossi_multivariable_2022}
Z.~M. Rossi and I.~L. Chuang, Multivariable quantum signal processing (M-QSP): prophecies of the two-headed oracle.
\newblock {\em \href{https://quantum-journal.org/papers/q-2022-09-20-811/}{Quantum}} \href{https://quantum-journal.org/papers/q-2022-09-20-811/}{{\bfseries 6}, 811} (2022).

\bibitem{gross_magnetoencephalography_2019}
J.~Gross, Magnetoencephalography in {Cognitive} {Neuroscience}: {A} {Primer}.
\newblock {\em \href{https://www.sciencedirect.com/science/article/pii/S0896627319305999}{Neuron}} \href{https://www.sciencedirect.com/science/article/pii/S0896627319305999}{{\bfseries 104}, 189--204} (2019).

\bibitem{aslam2023quantum}
N.~Aslam, H.~Zhou, E.~K. Urbach, M.~J. Turner, R.~L. Walsworth, M.~D. Lukin, and H.~Park, Quantum sensors for biomedical applications.
\newblock {\em \href{https://doi.org/10.1038/s42254-023-00558-3}{Nature Reviews Physics}} \href{https://doi.org/10.1038/s42254-023-00558-3}{{\bfseries 5}, 157--169} (2023).

\bibitem{yeom2023magnetoencephalography}
H.~G. Yeom, J.~S. Kim, and C.~K. Chung, A magnetoencephalography dataset during three-dimensional reaching movements for brain-computer interfaces.
\newblock {\em \href{https://doi.org/10.1038/s41597-023-02454-y}{Scientific Data}} \href{https://doi.org/10.1038/s41597-023-02454-y}{{\bfseries 10}, 552} (2023).

\bibitem{ge_heisenberg-limited_2024}
W.~Ge and K.~Jacobs, Heisenberg-limited continuous-variable distributed quantum metrology with arbitrary weights.
\newblock {\em \href{http://arxiv.org/abs/2412.01074}{arXiv:2412.01074}} (2024).

\bibitem{castellanos-beltran_amplification_2008}
M.~A. Castellanos-Beltran, K.~D. Irwin, G.~C. Hilton, L.~R. Vale, and K.~W. Lehnert, Amplification and squeezing of quantum noise with a tunable {Josephson} metamaterial.
\newblock {\em \href{https://www.nature.com/articles/nphys1090}{Nature Physics}} \href{https://www.nature.com/articles/nphys1090}{{\bfseries 4}, 929--931} (2008).

\bibitem{castellanos-beltran_development_2010}
M.~Castellanos-Beltran, (2010) {PhD} {Thesis} (University of Colorado Boulder, Boulder).

\bibitem{caves_quantum_1982}
C.~M. Caves, Quantum limits on noise in linear amplifiers.
\newblock {\em \href{https://link.aps.org/doi/10.1103/PhysRevD.26.1817}{Physical Review D}} \href{https://link.aps.org/doi/10.1103/PhysRevD.26.1817}{{\bfseries 26}, 1817--1839} (1982).

\bibitem{clerk_introduction_2010}
A.~A. Clerk, M.~H. Devoret, S.~M. Girvin, F.~Marquardt, and R.~J. Schoelkopf, Introduction to quantum noise, measurement, and amplification.
\newblock {\em \href{http://link.aps.org/doi/10.1103/RevModPhys.82.1155}{Reviews of Modern Physics}} \href{http://link.aps.org/doi/10.1103/RevModPhys.82.1155}{{\bfseries 82}, 1155--1208} (2010).

\bibitem{epstein_quantum_2021}
J.~M. Epstein, K.~B. Whaley, and J.~Combes, Quantum limits on noise for a class of nonlinear amplifiers.
\newblock {\em \href{https://link.aps.org/doi/10.1103/PhysRevA.103.052415}{Physical Review A}} \href{https://link.aps.org/doi/10.1103/PhysRevA.103.052415}{{\bfseries 103}, 052415} (2021).

\bibitem{braunstein_dense_2000}
S.~L. Braunstein and H.~J. Kimble, Dense coding for continuous variables.
\newblock {\em \href{https://link.aps.org/doi/10.1103/PhysRevA.61.042302}{Physical Review A}} \href{https://link.aps.org/doi/10.1103/PhysRevA.61.042302}{{\bfseries 61}, 042302} (2000).

\bibitem{campagne-ibarcq_quantum_2020}
P.~Campagne-Ibarcq, A.~Eickbusch, S.~Touzard, E.~Zalys-Geller, N.~E. Frattini, V.~V. Sivak, P.~Reinhold, S.~Puri, S.~Shankar, R.~J. Schoelkopf, L.~Frunzio, M.~Mirrahimi, and M.~H. Devoret, Quantum error correction of a qubit encoded in grid states of an oscillator.
\newblock {\em \href{https://www.nature.com/articles/s41586-020-2603-3}{Nature}} \href{https://www.nature.com/articles/s41586-020-2603-3}{{\bfseries 584}, 368--372} (2020).

\bibitem{eickbusch_fast_2022}
A.~Eickbusch, V.~Sivak, A.~Z. Ding, S.~S. Elder, S.~R. Jha, J.~Venkatraman, B.~Royer, S.~M. Girvin, R.~J. Schoelkopf, and M.~H. Devoret, Fast universal control of an oscillator with weak dispersive coupling to a qubit.
\newblock {\em \href{https://www.nature.com/articles/s41567-022-01776-9}{Nature Physics}} \href{https://www.nature.com/articles/s41567-022-01776-9}{{\bfseries 18}, 1464--1469} (2022).

\bibitem{lerch2024efficient}
S.~Lerch, R.~Puig, M.~S. Rudolph, A.~Angrisani, T.~Jones, M.~Cerezo, S.~Thanasilp, and Z.~Holmes, Efficient quantum-enhanced classical simulation for patches of quantum landscapes.
\newblock {\em \href{https://doi.org/10.48550/arXiv.2411.19896}{arXiv:2411.19896}} (2024).

\bibitem{marciniak_optimal_2022}
C.~D. Marciniak, T.~Feldker, I.~Pogorelov, R.~Kaubruegger, D.~V. Vasilyev, R.~van Bijnen, P.~Schindler, P.~Zoller, R.~Blatt, and T.~Monz, Optimal metrology with programmable quantum sensors.
\newblock {\em \href{https://www.nature.com/articles/s41586-022-04435-4}{Nature}} \href{https://www.nature.com/articles/s41586-022-04435-4}{{\bfseries 603}, 604--609} (2022).

\bibitem{cerezo2022challenges}
M.~Cerezo, G.~Verdon, H.-Y. Huang, L.~Cincio, and P.~J. Coles, Challenges and opportunities in quantum machine learning.
\newblock {\em \href{https://doi.org/10.1038/s43588-022-00311-3}{Nature Computational Science}} \href{https://doi.org/10.1038/s43588-022-00311-3}{{\bfseries 2}, 567--576} (2022).

\bibitem{schuld_circuit-centric_2020}
M.~Schuld, A.~Bocharov, K.~M. Svore, and N.~Wiebe, Circuit-centric quantum classifiers.
\newblock {\em \href{https://link.aps.org/doi/10.1103/PhysRevA.101.032308}{Physical Review A}} \href{https://link.aps.org/doi/10.1103/PhysRevA.101.032308}{{\bfseries 101}, 032308} (2020).

\bibitem{kaubruegger_optimal_2023}
R.~Kaubruegger, A.~Shankar, D.~V. Vasilyev, and P.~Zoller, Optimal and {Variational} {Multiparameter} {Quantum} {Metrology} and {Vector}-{Field} {Sensing}.
\newblock {\em \href{https://link.aps.org/doi/10.1103/PRXQuantum.4.020333}{PRX Quantum}} \href{https://link.aps.org/doi/10.1103/PRXQuantum.4.020333}{{\bfseries 4}, 020333} (2023).

\bibitem{fagaly2006superconducting}
R.~Fagaly, Superconducting quantum interference device instruments and applications.
\newblock {\em \href{https://doi.org/10.1063/1.2354545}{Review of Scientific Instruments}} \href{https://doi.org/10.1063/1.2354545}{{\bfseries 77}} (2006).

\bibitem{paszke2019pytorchimperativestylehighperformance}
A.~Paszke, S.~Gross, F.~Massa, A.~Lerer, J.~Bradbury, G.~Chanan, T.~Killeen, Z.~Lin, N.~Gimelshein, L.~Antiga, A.~Desmaison, A.~Köpf, E.~Yang, Z.~DeVito, M.~Raison, A.~Tejani, S.~Chilamkurthy, B.~Steiner, L.~Fang, J.~Bai, and S.~Chintala, PyTorch: An Imperative Style, High-Performance Deep Learning Library.
\newblock {\em \href{https://doi.org/10.48550/arXiv.1912.01703}{arXiv:1912.01703}} (2019).

\bibitem{Schollw_ck_2011}
U.~Schollwöck, The density-matrix renormalization group in the age of matrix product states.
\newblock {\em \href{http://dx.doi.org/10.1016/j.aop.2010.09.012}{Annals of Physics}} \href{http://dx.doi.org/10.1016/j.aop.2010.09.012}{{\bfseries 326}, 96–192} (2011).

\bibitem{carmichael_statistical_2002}
H.~J. Carmichael, (2002) {\em Statistical {Methods} in {Quantum} {Optics} 1 - {Master} {Equations} and {Fokker}-{Planck} {Equations} {\textbar} {Springer}}.

\bibitem{deepak_general_2023}
Deepak and A.~Chatterjee, General expansion of natural power of linear combination of {Bosonic} operators in normal order.
\newblock {\em \href{http://arxiv.org/abs/2305.18113}{arXiv:2305.18113}} (2023).

\bibitem{johansson_qutip_2013}
J.~R. Johansson, P.~D. Nation, and F.~Nori, {QuTiP} 2: {A} {Python} framework for the dynamics of open quantum systems.
\newblock {\em \href{http://www.sciencedirect.com/science/article/pii/S0010465512003955}{Computer Physics Communications}} \href{http://www.sciencedirect.com/science/article/pii/S0010465512003955}{{\bfseries 184}, 1234--1240} (2013).

\end{thebibliography}

\resumetoc

\appendix


\startcontents[Appendices]
\addtocontents{toc}{\protect\setcounter{tocdepth}{0}} 
\section*{Appendices} 
\thispagestyle{empty} 
\addtocontents{toc}{\protect\setcounter{tocdepth}{2}} 
\printcontents[Appendices]{l}{1}{}

\setcounter{page}{1}

\makeatletter
\let\toc@pre\relax
\let\toc@post\relax
\makeatother

\newpage

\section{Classical postprocessing and error scaling with nonlinearity: unbiased estimation of nonlinear polynomials of a classical Gaussian random variable}
\label{app:gauss}

In the main text, we note that amount of useful information extractable from measurements of a quantum system can depend nontrivially on the complexity of the task, even if the quantum system and its measurement properties, and hence sampling noise, remain fixed (see Sec.~\ref{subsec:taskcomp}). In other words, postprocessing of stochastic measurement results from a quantum system can degrade the signal-to-noise ratio by an amount that depends on the complexity of postprocessing required.

In this Appendix section, we show that this degradation of signal-to-noise ratio is not special to the results of quantum measurements: it affects stochastic variables exhibiting classical statistical properties as well, and is therefore a general phenomenon. We quantify the difficulty associated with classical nonlinear postprocessing by analyzing the task of estimating nonlinear polynomials of increasing degree using samples of a classical, Gaussian random variable. We show that even for a Gaussian random variable with a fixed variance, the error in the estimation of polynomials can increase with the polynomial degree, and hence with the complexity of the nonlinear postprocessing required. This dependence has been implicitly studied for a variety of quantum systems in the main text; Appendix~\ref{app:complexity} explores this relationship more quantitatively for measurements from a qubit-based quantum system (which exhibit multinomial, instead of Gaussian statistics).

Consider a Gaussian random variable $Z$ that follows a Gaussian (or Normal) distribution, $Z \sim \mathcal{N}(z,\sigma^2)$, with mean $z$ and variance $\sigma^2$. The variable $Z$ could describe measurements from a noisy classical dynamical system just as we have used $X$ to describe measurements from a quantum system (although such an association is not necessary for our analysis; furthermore, we note that a general, noisy classical system can exhibit more complex statistics than simply Gaussian statistics). Without loss of generality, we can write the stochastic variable $Z$ in terms of its mean value and a zero-mean stochastic process,
\begin{align}
    Z = z + \xi,
\end{align}
so that the stochastic process $\xi$ follows the distribution $\mathcal{N}(0,\sigma^2)$. Analogously to the quantum case, we introduce $\bar{Z} = \frac{1}{S}\sum Z$ as the average of $S$ samples of the stochastic variable $Z$, which has the same mean $z$ but a suitably-reduced variance $\frac{\sigma^2}{S}$.

The task we consider is: given an instance of the stochastic variable $\bar{Z}$, construct an \textit{unbiased} estimator of degree $m$ polynomials of the mean value $z$, $\Ft_m = z^m$, for $m \geq 1$, for $z$ over a fixed one-dimensional domain $\mathcal{R}$. The fidelity of the estimation can again be quantified using the expected mean-squared-error, defined in the same way as in Sec.~\ref{sec:bosonic} of the main text,
\begin{align}
    \mathbb{E}[{\rm MSE}]_m &= \frac{1}{f_m}\int_{\mathcal{R}} dz~\mathbb{E}[( \Ft_m- \mathcal{F}_m)^2]~~\stackrel{\mathbb{E}[ \mathcal{F}_m ] = \Ft_m}{=}~~ \frac{1}{f_m}\int_{\mathcal{R}} dz~{\rm Var}[\mathcal{F}_m]
    \label{appeq:emse}
\end{align}
where $f_m=\int_{\mathcal{R}} dz~(\Ft_m)^2$ is a normalization constant as before. The domain over which we wish to approximate $\Ft_m$ is $\mathcal{R}: z \in [-1,+1]$. The second line follows in the special case where $\mathcal{F}_m$ is an unbiased estimator of $\Ft_m$.

The required estimators $\mathcal{F}_m$ are constructed by computing functions of the random variable $\bar{Z}$. Note that any linear processing of $\bar{Z}$ can only produce estimators of linear functions of $z$, since $\mathbb{E}[\mathcal{C}_0 + \mathcal{C}_1 Z] = C_0 + C_1 z$ (where $\mathcal{C}_0,\mathcal{C}_1$ are constants). Therefore, constructing estimators for the target functions $\Ft_m$ for $m \geq 2$ will require nonlinear operations to be performed on the stochastic variable $\bar{Z}$. We define the general estimator $\mathcal{F}_m$ for the $m$th-order polynomial as:
\begin{align}
    \mathcal{F}_m = \sum_{n=0}^m \mathcal{C}^{(m)}_n \bar{Z}^n.
    \label{appeq:gaussub}
\end{align}
For completeness we now show how the required unbiased estimator $\mathcal{F}_m$ can be constructed for arbitrary $\Ft_m$. We first write the target function as $\Ft_m = \sum_{j=0}^m \mathcal{W}_j^{(m)} z^j$; this includes the case $\Ft_m = z^m$ for which $\mathcal{W}_n^{(m)} = \delta_{nm}$. Now, for $\mathcal{F}_m$ to be an unbiased estimator for $\Ft_m$, we require $\mathbb{E}[\mathcal{F}_m] = \sum_{n=0}^m \mathcal{C}^{(m)}_n \mathbb{E}[\bar{Z}^n] = \Ft_m$. This requirement places constraints on the coefficients $\mathcal{C}^{(m)}_n$, and ultimately defines their unique values for specific $\Ft_m$. To evaluate $\mathbb{E}[\mathcal{F}_m]$, we require arbitrary $n$th order moments of the Gaussian random variable $\bar{Z}$; these are given by:
\begin{align}
    \mathbb{E}[\bar{Z}^n] = \sum_{j=0}^n 
    \begin{pmatrix}
        n \\ 
        j
    \end{pmatrix}
    \mathcal{I}_{n-j}
    \left(\frac{\sigma^2}{S} \right)^{n-j} z^{j}
    \label{appeq:gaussmoments}
\end{align}
where:
\begin{align}
    \mathcal{I}_{j} = 
    \begin{cases}
        \frac{j!}{2^{j/2}(j/2)!}, &j=0,2,4,\ldots \\
        0, &{\rm otherwise}.
    \end{cases}
\end{align}
Using Eq.~(\ref{appeq:gaussmoments}), the condition $\mathbb{E}[\mathcal{F}_m] = \Ft_m $ can be rewritten as:
\begin{align}
    \sum_{n=0}^m \mathcal{C}^{(m)}_n \sum_{j=0}^n 
    \begin{pmatrix}
        n \\ 
        j
    \end{pmatrix}
    \mathcal{I}_{n-j}
    \left(\frac{\sigma^2}{S} \right)^{n-j} z^{j} =  \sum_{j=0}^m \mathcal{W}_j^{(m)} z^j
\end{align}
The sum over $j$
 depends on $n$ in the sense that we can never have $j>n$. We instead enforce this constraint using the step function $\Theta_{j\leq n}$ which is equal to $1$ if $j \leq n$ and $0$ otherwise. With the introduction of $\Theta_{j\leq n}$, the above can be rewritten as:
 \begin{align}
    \sum_{j=0}^m \left( \sum_{n=0}^m   \left[\Theta_{j\leq n} 
    \begin{pmatrix}
        n \\ 
        j
    \end{pmatrix}
    \mathcal{I}_{n-j}
    \left(\frac{\sigma^2}{S} \right)^{n-j}\mathcal{C}^{(m)}_n\right] \right)z^{j} =  \sum_{j=0}^m \left( \mathcal{W}_j^{(m)} \right) z^j
\end{align}
Equating coefficients of the powers $z^j$ on both sides for each $j$ (the terms in round brackets) yields a matrix system for the coefficients $\mathcal{C}^{(m)}_n$. By solving this system of equations, we can obtain the coefficients $\mathcal{C}^{(m)}_n$ that ensure the construction of an unbiased estimator. For $m \in \{1,\ldots,6\}$, the coefficients obtained by solving the above system are summarized in Table~\ref{tab:gaussian}.


\begin{table}[h]
    \centering
    \begin{tabular}{c|c||c|c|c|c|c|c|c}
         \multicolumn{1}{c}{} & & \multicolumn{7}{c}{$\mathcal{C}_{n}^{(m)}$} \\
         \cline{3-9}
         \multicolumn{1}{c}{} & & $n\!=\!0$ & $n\!=\!1$ & $n\!=\!2$ & $n\!=\!3$ & $n\!=\!4$ & $n\!=\!5$ & $n\!=\!6$    \\
         \hline 
         \hline
         \multirow{8.5}{*}{$\Ft$} & $z^1$ & 0 & 1 & 0 & 0 & 0 & 0 & 0 \\ [0.3em]
         \cline{2-9}
        & $z^2$ & $-\frac{\sigma^2}{S}$ & 0 & 1 & 0 & 0 & 0 & 0 \\  [0.3em]
        \cline{2-9}
        & $z^3$ & 0 & $-\frac{3\sigma^2}{S}$ & 0 & 1 & 0 & 0 & 0 \\ [0.3em]
        \cline{2-9}
        & $z^4$ & $\frac{3\sigma^4}{S^2}$ & 0 & $-\frac{6\sigma^2}{S}$ & 0 & 1 & 0 & 0 \\ [0.3em] 
         \cline{2-9}
        & $z^5$ & 0  & $\frac{15\sigma^4}{S^2}$ & 0 & $-\frac{10\sigma^2}{S}$ & 0 & 1 & 0 \\ [0.3em]
         \cline{2-9}
        & $z^6$ & $-\frac{15\sigma^6}{S^3}$ & 0 & $\frac{45\sigma^4}{S^2}$ & 0 & $-\frac{15\sigma^2}{S}$ & 0 & 1 \\ 
    \end{tabular}
    \caption{Coefficients $\mathcal{C}_{n}^{(m)}$ for construction of unbiased estimators via Eq.~(\ref{appeq:gaussub}) of the corresponding target polynomials $\Ft_m = z^m$. }
    \label{tab:gaussian}
\end{table}

\begin{figure}[t]
    \centering
    \includegraphics{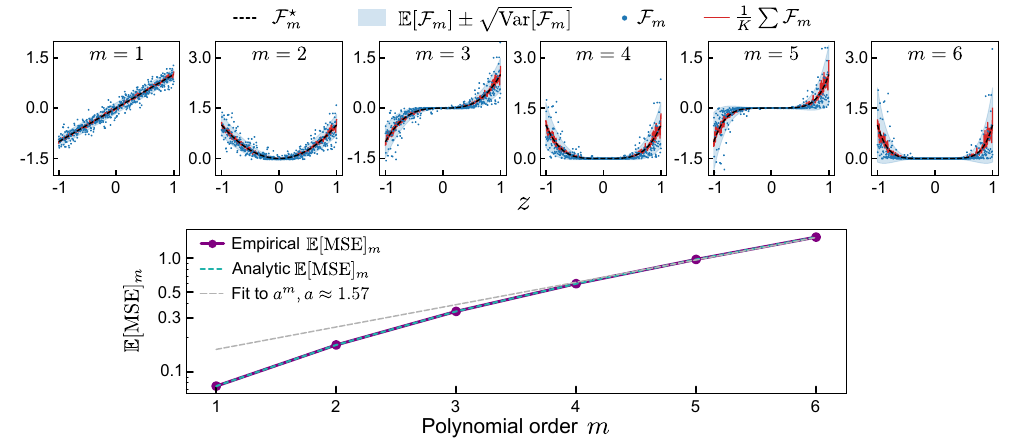}
    \caption{\textbf{Scaling of expected MSE in the estimation of degree-$m$ polynomials using samples of a Gaussian random variable.} Plot: $\mathbb{E}[{\rm MSE}]_m$ against degree $m$ of target functions $\Ft_m$ for $m\in \{1,\ldots,6\}$. Both the empirically-computed values and the analytic values (calculated using the analytic variance, Eq.~(\ref{appeq:fmvar}), of the unbiased estimator) are shown. Asymptotic exponential fit is also shown, indicating the exponential increase in expected MSE with $m$. Top row: Target functions $\Ft_m = z^m$ for $m\in \{1,\ldots,6\}$. Blue dots are values of the corresponding unbiased estimators $\mathcal{F}_m$. Blue shaded regions show the analytically-computed window of one standard deviation around the expected value of $\mathcal{F}_m$, while the red curve is the average of $K=25$ stochastic instances of $\mathcal{F}_m$.  }
    \label{appfig:gaussian}
\end{figure}

Once the required coefficients $\mathcal{C}_n^{(m)}$ are known, we can construct $\mathcal{F}_m$ using stochastic samples of $\bar{Z}$. Examples of $\mathcal{F}_m$ for the target polynomials $\Ft_m = z^m$, as a function of $m$ for $m \in \{1,\ldots,6\}$, are shown in the top panel of Fig.~\ref{appfig:gaussian} (blue dots), alongside the corresponding target $\Ft_m$ (dashed lines). Visually, the stochastic estimates appear to be distributed around the desired target functions. To verify that the constructed $\mathcal{F}_m$ are indeed unbiased estimators of the corresponding $\Ft_m$, we must estimate $\mathbb{E}[\mathcal{F}_m]$. To do so empirically, we average $K=25$ independent instances of the unbiased estimators $\mathcal{F}_m$ (red lines), which show convergence to the target function. 

Our analytic approach to computing the unbiased estimator also allows us to compute the variance in this unbiased estimator. In particular, the variance is given by ${\rm Var}[\mathcal{F}_m] = \mathbb{E}[\mathcal{F}_m^2] - (\mathbb{E}[\mathcal{F}_m])^2$, which can be explicitly written in the form:
\begin{align}
    {\rm Var}[\mathcal{F}_m] = \sum_{p=0}^m\sum_{n=0}^m \mathcal{C}_n^{(m)}\mathcal{C}_p^{(m)} \left( \mathbb{E}[\bar{Z}^{n+p}] - \mathbb{E}[\bar{Z}^n]\mathbb{E}[\bar{Z}^{p}] \right)
    \label{appeq:fmvar}
\end{align}
The known coefficients $\mathcal{C}_n^{(m)}$, together which Eq.~(\ref{appeq:gaussmoments}) that allows computing arbitrary moments of the Gaussian random variables $\bar{Z}$, we can compute ${\rm Var}[\mathcal{F}_m]$ analytically using the above equation (as opposed to only estimating the variance empirically).

To quantify the fidelity of constructing an unbiased estimate of the target functions, we can finally compute the expected MSE via Eq.~(\ref{appeq:emse}). Using the stochastic samples of $\mathcal{F}_m$, the expected MSE can be directly estimated using the first term in Eq.~(\ref{appeq:emse}); this result is plotted in the lower panel of Fig.~\ref{appfig:gaussian}, here using $K=10^4$ samples to ensure convergence. We see the expected MSE increases with $m$.  Furthermore, since we have access to the variance in the unbiased estimators analytically, we can estimate the expected MSE directly using the second line of Eq.~(\ref{appeq:emse}); this result is shown in dashed green in Fig.~\ref{appfig:gaussian}, and shows excellent agreement with its empirically computed counterpart, as expected. Furthermore, the growth in expected MSE is asymptotically exponential in $m$ (note that the vertical axis is shown in logscale). The precise scaling of the expected MSE will depend on the nature of the functions $\Ft$ that are being approximated. 

The results of this section show how nonlinear operations on stochastic data can lead to a degradation of the signal-to-noise ratio, and subsequently to an increase in the error associated with function approximation. Furthermore, the error can scale with the complexity of the function to be approximated, here the polynomial degree. By computing required functions prior to measurement, quantum computational sensing aims to avoid this nonlinear processing of stochastic measurement results.


\clearpage


\newpage

\section{QCS of static signals using qubit-based quantum computational sensors}
\label{app:svbc}

In this Appendix section, we include additional details of QCS using single-qubit sensors. We begin by presenting the optimal Bayes classifier used for the single-variable binary classification task in Sec.~\ref{subsec:1q} of the main text. We also include a heuristic discussion of the role of sampling noise in QCSA, and explore how QCSA depends on the complexity of the single-variable task. We finally present results of a QCS protocol that enables the use of a single qubit sensor for binary classification tasks over multi-variate signals.

\subsection{Architectures for qubit-based quantum computational sensors of static signals}

\begin{figure}[h!]
    \centering
    \includegraphics{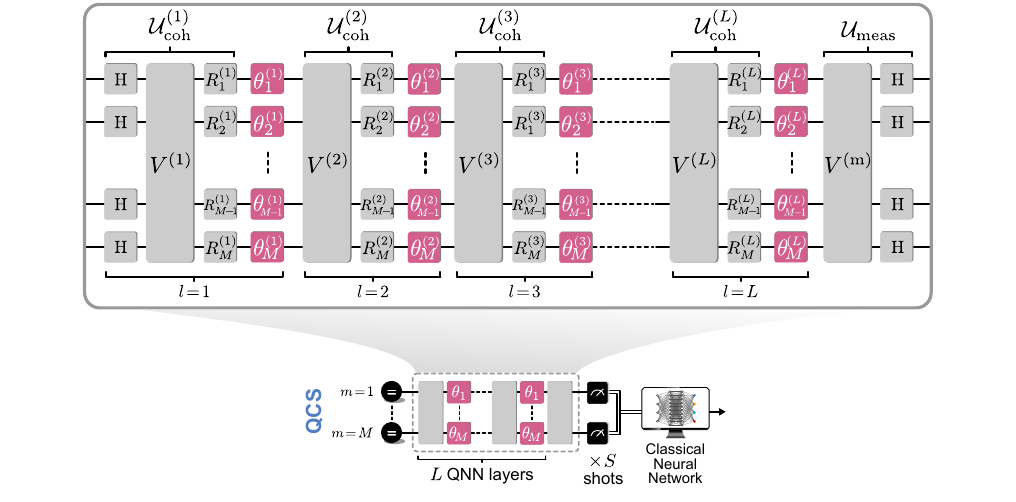}
    \caption{\textbf{Multi-qubit quantum computational sensor architecture.} All gray blocks describe operations that are trained using supervised learning, apart from the Hadamard gate $\text{H}$. Precise definitions of the gates and trainable parameters are included in Eqs.~(\ref{appeq:H})-(\ref{appeq:Munitary}).}
    \label{appfig:qubits}
\end{figure}

In Fig.~\ref{appfig:qubits} we show the structure of the protocol used for QCS for the single and two-qubit simulations in the main text, namely Sec.~\ref{subsec:1q} and Sec.~\ref{subsec:2q}. We now provide details of the specific computing operations employed by these qubit-based quantum computational sensors. However, before doing so, we first introduce the physical sensing operation across $M$ sensing qubits for phase sensing. For $\bm{u} \to \bm{\theta} = (\theta_1,\ldots,\theta_M)$, this is given by:
\begin{align}
    \Usense(\bm{\theta}^{(l)}) = \otimes_m^M \exp \left\{ -i \theta^{(l)}_m \hat{\sigma}^z_m  \right\}
\end{align}


We can now introduce the constituent gates that comprise the full protocol, starting with the simplest: ${\rm H}$, which is the single-qubit Hadamard gate, and is given by:
\begin{align}
    {\rm H} = \exp \left\{ -i \frac{\pi}{2} \left( \frac{1}{\sqrt{2}}\hat{\sigma}^z + \frac{1}{\sqrt{2}}\hat{\sigma}^x \right) \right\} = -\frac{i}{\sqrt{2}}
    \begin{bmatrix}
    1 & 1 \\
    1 & -1 
    \end{bmatrix}
    \label{appeq:H}
\end{align}
where we have left out the qubit indices for simplicity, and the second equality presents the matrix representation. Note that the above definition differs from the usual Hadamard gate by an overall phase of $-i$; this has no influence on measured observables.

Secondly, we introduce  the trainable single qubit operations $R^{(l)}_{m}$, given by:
\begin{align}
    R^{(l)}_{m} = \exp \left\{ -i \zeta_m^{(l)}\hat{\sigma}_m^z \right\}\exp \left\{ -i \frac{\phi_m^{(l)}}{2}\hat{\sigma}_m^x \right\}
    \label{appeq:Sunitary}
\end{align}

Finally, we introduce the trainable $M$-qubit unitary operations ${\rm V}^{(l)}$. For maximum generality we first introduce the trainable Hermitian generator $\hat{G}^{(l)}$, with $4^M$ trainable parameters $c_j^{(l)}$ per layer $l$:
\begin{align}
    \hat{G}^{(l)} = \frac{1}{2}\left( \sum_{j=1}^{4^M} c^{(l)}_j \hat{P}_j + h.c. \right).
    \label{appeq:HGen}
\end{align}
We have decomposed this generator in terms of the operators $\hat{P}_j$, which can be constructed from states defining the Hilbert space of the quantum computational sensor. In particular, for a sensor comprising $M$ qubits, the Hilbert space can be defined using states $\ket{\bm{s}_k}$, where $\bm{s}_k$ is the $k$th bitsring of length $M$; there are $2^M$ such bitstrings in total. For $M=2$, for example, $\bm{s}_k \in \{00,01,10,11\}$. Then, $\hat{P}_j \in \{ \ket{\bm{s}_n}\!\bra{\bm{s}_m} \}_{n,m = 1,\ldots 2^M} $; there are hence $4^M$ such $\hat{P}_j$'s.

The trainable Hermitian generator from Eq.~(\ref{appeq:HGen}) allows to construct the trainable $M$-qubit unitary operation:
\begin{align}
    V^{(l)} = \Big( \otimes_{m}^M {\rm H} \Big) \exp \left\{ -i \hat{G}^{(l)} \right\} \Big( \otimes_{m}^M {\rm H}^{\dagger} \Big)
    \label{appeq:Munitary}
\end{align}

Having defined all the constituent gates, we now explain how the composite operations that make up the qubit-based quantum computational sensing protocols are constructed. The protocols all follow the structure given by Eq.~(\ref{eq:cohprocessing}) of the main text. From Fig.~\ref{appfig:qubits}, the first coherent processing operation is distinct from the others:
\begin{align}
    \Ucoh{l} = 
    \begin{cases}
        \Big( \otimes_{m}^M R_m^{({1})} \Big) V^{({1})} \Big( \otimes_{m}^M {\rm H} \Big), & l = 1, \\
        \Big( \otimes_{m}^M R_m^{(l)} \Big) V^{(l)}, &l=2,\ldots,L.
    \end{cases}
\end{align}
Finally, the operation $\Udec$ applied just prior to measurement in the computational basis is composed of a final $M$-qubit trainable unitary operation, followed by an $M$-qubit Hadamard rotation:
\begin{align}
    \Udec =   \Big( \otimes_{m}^M {\rm H} \Big)  V^{({\rm m})}
\end{align}

The specific structure in Fig.~\ref{appfig:qubits} could have been consolidated to reduce the total number of trainable parameters, for example by absorbing single-qubit operations $R_m^{(l)}$ as part of the general $M$-qubit unitaries ${V}^{(l)}$ wherever they appear consecutively. The reason for our choice of this apparently over-parameterized general structure is that is allows us to more easily compare a variety of protocols. As just one example, toggling between using $M$-qubit operations or not allows us to compare entangled versus unentangled protocols; in the latter case, the use of separate, trainable single-qubit operations ensures the quantum computational sensing protocol can still perform some computation on the sensed signals. If one wanted to analyze a given quantum system with fixed architecture, a more purpose-built simulation tool could be designed, to have a greater economy of trainable parameters.

\subsection{Bayes (optimal) classifier for binary classification using a single qubit}
\label{app:bayes}

In this Appendix subsection, we provide details of the Bayes classifier used for binary classification tasks using single qubit systems in Sec.~\ref{subsec:1q} of the main text. It will prove useful to define the variable $Y = S \bar{X}$
where $\bar{X}$ is the averaged measurement result over $S$ shots, for both the QS and QCS protocols, namely $\bar{X} \in \{\bar{X}_{\rm QS},\bar{X}_{\rm QCS}\}$. Therefore, the random variable $Y$ defines the number of times the qubit is measured in the excited state in $S$ measurement shots; formally, in $S$ shots, the possible values of $y$ of $Y$ are enumerated as $y \in \{0,1,\ldots,S\}$. The probability of each outcome is given by a binomial distribution, $Y \sim {\rm Bin}(S,x(\theta))$, where $x(\theta)$ is the qubit excitation probability; the distribution again holds for both the QS and QCS approaches, with $x \in \{x_{\rm QS},x_{\rm QCS}\}$ respectively. For measurements of a single qubit, the single variable $Y$ contains all information that can be used to perform any task using this quantum system. This means the dimension of the feature space is therefore low (one-dimensional), enabling a Bayes classifier to be efficiently constructed. We now explain how this construction proceeds.

Consider the task of distinguishing two classes, indexed $j=1,2$ using the value of a random variable $Y$. We will assume the two classes are equally likely to occur, namely the prior probabilities $P(C=1) = P(C=2) = 0.5$. The optimal Bayes classifier $\mathcal{C}_{\rm Bayes}(Y)$ takes any value $y$ of the variable $Y$ and returns the predicted class label according to the principle:
\begin{align}
    \mathcal{C}_{\rm Bayes}(y) = \underset{j \in \{1,2\}}{{\rm argmax}}~P(C=j| Y=y ),
\end{align}
namely the Bayes classifier returns the class label that has the largest probability of having occurred given the specific measurement result $Y=y$ was obtained. 

For binary classification tasks in particular, the ${\rm argmax}$ operation in the Bayes classifier reduces to a simple inequality,
\begin{align}
    \mathcal{C}_{\rm Bayes}(y) = 
    \begin{cases}
        1,~{\rm if~} P(C=1| Y=y ) > P(C=2| Y=y ), \\
        2,~{\rm otherwise}.
        \label{appeq:bayes}
    \end{cases}
\end{align}
The probability distribution we actually have access to via measurements of the quantum system is $P(Y=y|C=j)$, as will be shown shortly. This quantity can be related to the `reversed' conditional probability that appears in the Bayes classifier via Bayes' theorem:
\begin{align}
    P(C=j| Y=y ) = \frac{P(Y=y|C=j)P(C=j)}{P(Y=y)}
    \label{appeq:bayesthm}
\end{align}
The difference between the two conditional probabilities is the factor of $P(C=j)/P(Y=y)$; for equal prior probabilities, this is just a class label-independent scaling factor. For binary classification via Eq.~(\ref{appeq:bayes}), this overall scaling factor will cancel in the comparison of conditional probabilities, and therefore does \textit{not} affect the prediction of the Bayes classifier.

It will therefore be sufficient to compute $P(Y=y|C=j)$. Since we know $Y$ follows a binomial distribution parameterized by $\theta$, we have access to $P(Y=y|\theta)$, which is simply given by the probability mass function of the binomial distribution:
\begin{align}
    P(Y=y|\theta) = 
    \begin{pmatrix}
    S \\ 
    y
    \end{pmatrix} (x(\theta))^y(1-x(\theta))^{n-y},
\end{align}
where $\begin{psmallmatrix}
    S \\ 
    y
    \end{psmallmatrix}$ is the standard binomial coefficient.


\begin{figure}
    \centering
    \includegraphics{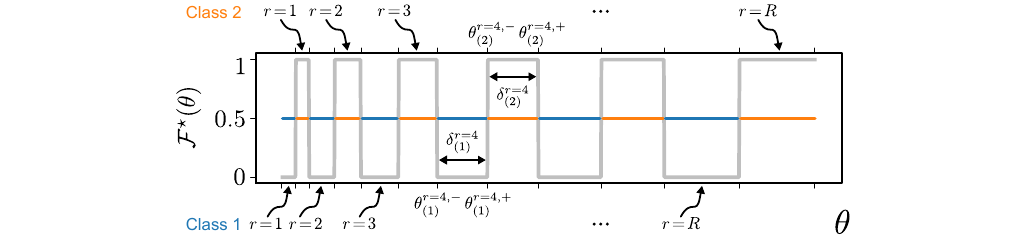}
    \caption{\textbf{Definition of single-variable binary classification task.} The domain of the sensed parameter $\theta$ is divided into $R$ noncontiguous regions for each class, depicted in blue for Class~1 and orange for Class~2; here $R=6$. The `length' or spacing of each region can vary, and for the $r$th region belonging to the $j$th class is given by $\delta_{(j)}^r = \theta_{(j)}^{r,+}-\theta_{(j)}^{r,-}$ . Grey curve shows the target function $\Ft(\theta)$ for the specific task instance shown.  }
    \label{appfig:taskdef}
\end{figure}


We can now discuss the specifics of the binary classification task defined over the phases $\theta$, which is shown in complete generality in Fig.~\ref{appfig:taskdef}. We consider $\theta$ to be sampled from a fixed domain $\theta \in [0,\frac{\pi}{4}]$. This domain is equally divided into two macroscopic regions $\mathcal{R}_j$ for $j=1,2$ corresponding to each of two classes, which therefore must have equal total `lengths' $\frac{1}{2}\left(\frac{\pi}{4}\right) = \frac{\pi}{8}$. Importantly, these regions are not required to be continuously connected: each region $\mathcal{R}_j$ is further divided into $R$ sub-regions. Each sub-region, indexed by $r=1,\ldots,R$, is delimited by phase values $[\theta_{(j)}^{r,-},\theta_{(j)}^{r,+}]$, with the subscript $(j)$ indexing the class label. Formally, the macroscopic regions $\mathcal{R}_j$ are compactly defined as:
\begin{align}
    \mathcal{R}_j : \bigcup_{r=1}^R~  [\theta_{(j)}^{r,-},\theta_{(j)}^{r,+}]
\end{align}
Within each region, the distribution of $\theta$ values is uniformly random, so that the probability density of $\theta$ belonging to class $j$ is given by:
\begin{align}
    P(\theta| C=j) = 
    \begin{cases}
        \frac{1}{\pi/8},~& \theta \in \mathcal{R}_j, \\
        0,~&{\rm otherwise}.
    \end{cases}
\end{align}

Then we can write for the conditional probability density of measurement outcomes: 
\begin{align}
    P(Y=y\cap C=j) =  \int_{\mathcal{R}_j} d\theta~P(Y=y\cap \Theta = \theta) = \int_{\mathcal{R}_j} d\theta~P(Y=y|\Theta = \theta)P(\Theta=\theta)
    \label{appeq:pygj}
\end{align}

Using Eq.~(\ref{appeq:pygj}) we can use the Bayes classifier $\mathcal{C}_{\rm Bayes}$ of Eq.~(\ref{appeq:bayes}) to determine the predicted class label for any value $y$ of the measurement outcome. Of course we are ultimately interested in the classification error of the Bayes classifier $\mathcal{C}_{\rm Bayes}$, which is simply given by the probability that the outcome of the classifier is incorrect. Intuitively, the Bayes classifier associates a label with any measurement outcome $y$. However, if there is a finite nonzero probability that the same outcome $y$ could have been measured but for $\theta$ belonging to the class other than that predicted by the Bayes classifier, a finite error probability exists. Formally, we can write this error probability $\mathcal{E}_{\rm Bayes}$ as:
\begin{align}
    \mathcal{E}_{\rm Bayes} &= P( \mathcal{C}_{\rm Bayes}(y) = 1 \cap C = 2 ) + P( \mathcal{C}_{\rm Bayes}(y) = 2 \cap C = 1 )  \nonumber \\
    &=\sum_{y | \mathcal{C}_{\rm Bayes}(y)=1 } \!\!\!\!\!\!\!\!\!\!\! P(Y=y \cap C=2) + \sum_{y | \mathcal{C}_{\rm Bayes}(y)=2 } \!\!\!\!\!\!\!\!\!\!\ P(Y=y \cap C=1)
    \label{appeq:perrbayes}
\end{align}
The above expression can be explained as follows. The two sums together run over all possible measurement outcomes $y$. For this binary classification task, the measurement outcomes can be split into those for which the Bayes classifier predicts the output label as Class 1 (first term), and those for which the output label is predicted to be Class 2 (second term). Then, the probability that the prediction for any $y$ is incorrect is equal to the conditional probability that $y$ was measured given an input $\theta$ from the opposite class was sensed. We use Eq.~(\ref{appeq:perrbayes}) to compute the classification error for binary classification using single-qubit systems both in the main text and in this Appendix section.

\subsection{Approximate Gaussian classifier for binary classification using a single-qubit conventional quantum sensor}

While $\mathcal{E}_{\rm Bayes}$ in Eq.~(\ref{appeq:perrbayes}) is straightforward to calculate numerically, it does not generally yield a simple semi-analytic expression. However in certain special cases an approximate expression for the error probability can be obtained. One such case is the limit of a large number of shots $S$, where the binomial distribution of the scaled random variable $\frac{1}{S}Y = \bar{X}$ can be approximated to be a Gaussian distribution, with mean $x(\theta)$ and variance $\sigma^2(\theta) = \frac{1}{S}x(\theta)(1-x(\theta))$. Under this approximation, $\bar{X}$ is now taken to be a continuous random variable, and its probability density is given by:
\begin{align}
    P_{\rm G}(\bar{X}|\theta) = \frac{1}{\sqrt{2\pi \sigma^2(\theta)} } \exp \left\{ - \frac{(\bar{X}-x(\theta))^2}{2\sigma^2(\theta)} \right\}
\end{align}
We can therefore define the probability $P(\bar{X} \cap C=j)$ as:
\begin{align}
    P_{\rm G}(\bar{X} \cap C=j) = \int d\theta~P_{\rm G}(\bar{X}|\theta)P(\theta|C=j)P(C=j) = \frac{1}{\pi/4} \int_{\mathcal{R}_j} d\theta~P_{\rm G}(\bar{X}|\theta) 
\end{align}
where we have used as before $P(C=j) = \frac{1}{2}~\forall~j$. For later use, it will prove useful to write the integral over the disconnected domain $\mathcal{R}_j$ as a sum over integrals:
\begin{align}
    P_{\rm G}(\bar{X} \cap C=j) = \frac{1}{\pi/4} \sum_{r=1}^R \int_{\theta_{(j)}^{r,-} }^{\theta_{(j)}^{r,+}} d\theta~P_{\rm G}(\bar{X}|\theta) 
\end{align}

\begin{figure}
    \centering
    \includegraphics{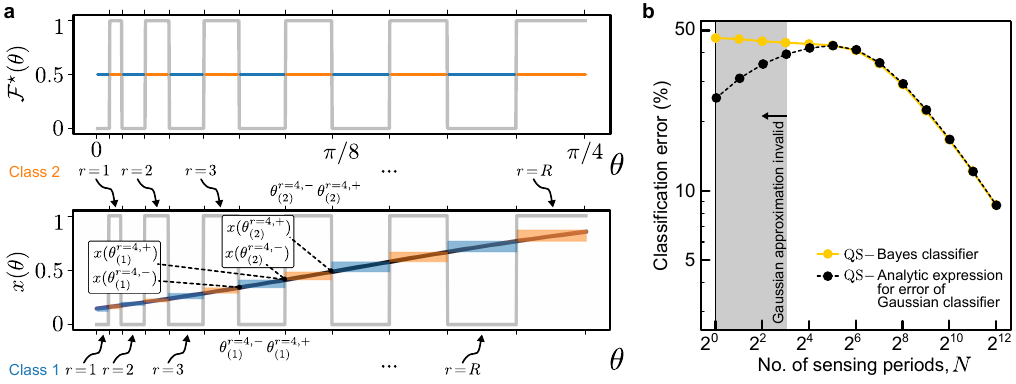}
    \caption{\textbf{Gaussian classifier for conventional QS protocol.} \textbf{a}, Definition of single-variable binary classification task considered in Sec.~\ref{subsec:1q} of the main text, from Fig.~\ref{appfig:taskdef}, is shown for reference. Working principle of approximate Gaussian classifier. \textbf{c}, Performance of exact Bayes classifier and analytic formula for performance of Gaussian classifier (given by Eq.~(\ref{appeq:gausscerr})) for the conventional QS protocol, as a function of the number of sensing periods $N$. }
    \label{appfig:gaussclassifier}
\end{figure}

We will now define an approximate classifier that is valid for the conventional QS case in a regime where a Gaussian approximation for $\bar{X}$ holds. In particular, we define the Gaussian classifier as:
\begin{align}
    \mathcal{C}_{\rm G}(\bar{X}) = 
    \begin{cases}
        1,~{\rm if}~\bar{X} \in \mathcal{X}_1, \\
        2,~{\rm if}~\bar{X} \in \mathcal{X}_2.
    \end{cases}
    \label{appeq:gaussclassifier}
\end{align}
where we have defined:
\begin{align}
    \mathcal{X}_j : \bigcup_{r=1}^R~  [x(\theta_{(j)}^{r,-}),x(\theta_{(j)}^{r,+})].
    \label{appeq:gaussregion}
\end{align}
The working of the Gaussian classifier defined by Eq.~(\ref{appeq:gaussclassifier}) is easily explained visually. In the lower panel of Fig.~\ref{appfig:gaussclassifier}a, the various shaded regions for Class $j$ define the regions $[\theta_{(j)}^{r,-},\theta_{(j)}^{r,+}]$ along the $\theta$ axis, and $[x(\theta_{(j)}^{r,-}),x(\theta_{(j)}^{r,+})]$ along the $x(\theta)$ axis. Note that $j=1$ is given by the blue regions while $j=2$ is given by the orange regions. There are six such noncontiguous regions here, as $r \in \{1,\ldots,6\}$. The region $\mathcal{X}_j$ introduced in Eq.~\ref{appeq:gaussregion} then defines the union of these regions along the $x(\theta)$ axis. The classifier operation is then as follows: if the measurement result $\bar{X}$ falls into any of the blue shaded regions, it is classified as being from Class 1. If it instead falls into any of the orange regions, it is classified as being from Class 2 (this is in words what Eq.~(\ref{appeq:gaussclassifier}) implements).


As before, we are interested in the error probability of this classifier, namely the probability that the classifier output is wrong. We must again sum over all possible values of the random variable $\bar{X}$, and determine the probability that the outcome was incorrect. Again, the outcomes can be divided into regions that are classified as $C=1$ or $C=2$, the only two possibilities. Furthermore, since $\bar{X}$ is now assumed to be a continuous random variable, this sum becomes an integral:
\begin{align}
    \mathcal{E}_{\rm G} = \int_{\bar{X}|\mathcal{C}_{\rm G}=1} \!\!\!\!\!\! d\bar{X}~P_{\rm G}(\bar{X} \cap C=2) + \int_{\bar{X}|\mathcal{C}_{\rm G}=2} \!\!\!\!\!\! d\bar{X}~P_{\rm G}(\bar{X} \cap C=1)
\end{align}
The classified regions are a set of non-contiguous regions, so each integral reduces to a sum over integrals. In particular we have:
\begin{align}
    \int_{\bar{X}|\mathcal{C}_{\rm G}=1} \!\!\!\!\!\! d\bar{X}~P_{\rm G}(\bar{X} \cap C=2) &= \sum_{r'=1}^R \int_{x(\theta_1^{r',-})}^{x(\theta_1^{r',+})}  d\bar{X}~P_{\rm G}(\bar{X} \cap C=2) 
\end{align}
Finally,
\begin{align}
    \int_{\bar{X}|\mathcal{C}_{\rm G}=1} \!\!\!\!\!\! d\bar{X}~P_{\rm G}(\bar{X} \cap C=2) = \frac{1}{\pi/4}\sum_{r'=1}^R \sum_{r=1}^R \int_{x(\theta_{(1)}^{r',-})}^{x(\theta_{(1)}^{r',+})} d\bar{X} \int_{\theta_{(2)}^{r,-} }^{\theta_{(2)}^{r,+}} d\theta~P_{\rm G}(\bar{X}|\theta) 
\end{align}
The classification error of the approximate Gaussian classifier is therefore given by:
\begin{align}
    \mathcal{E}_{\rm G} = \frac{1}{\pi/4}\sum_{r'=1}^R \sum_{r=1}^R \int_{x(\theta_{(1)}^{r',-})}^{x(\theta_{(1)}^{r',+})} d\bar{X} \int_{\theta_{(2)}^{r,-} }^{\theta_{(2)}^{r,+}} d\theta~P_{\rm G}(\bar{X}|\theta)  + \frac{1}{\pi/4}\sum_{r'=1}^R \sum_{r=1}^R \int_{x(\theta_{(2)}^{r',-})}^{x(\theta_{(2)}^{r',+})} d\bar{X} \int_{\theta_{(1)}^{r,-} }^{\theta_{(1)}^{r,+}} d\theta~P_{\rm G}(\bar{X}|\theta) 
    \label{appeq:gausscerr}
\end{align}
Eq.~(\ref{appeq:gausscerr}) describes the \textit{analytically-calculated} error of the Gaussian classifier; note that this will not always equal the performance that would be achieved if the Gaussian classifier as defined by Eq.~(\ref{appeq:gaussclassifier}) was used to classify stochastic measurement results $\bar{X}_{\rm QS}$. This is because the measurement results $\bar{X}_{\rm QS}$ in practice obey binomial statistics, whereas Eq.~(\ref{appeq:gausscerr}) is derived under the assumption that the measurement results exactly obey Gaussian statistics.

We can now compare this analytically-computed error of the Gaussian classifier against the actual performance of the Bayes classifier in Fig.~\ref{appfig:gaussclassifier}b. Note that for small $S$ there is a disagreement between the two, as could be expected. In particular, we emphasize that for low $S$, the predicted lower classification error than the exact Bayes classifier is simply an artifact that arises since the Gaussian assumption that Eq.~(\ref{appeq:gausscerr}) is centered on is not valid in this regime. If the Gaussian classifier of Eq.~(\ref{appeq:gaussclassifier}) was used to classify actual stochastic measurement results $\bar{X}_{\rm QS}$ (which obey binomial, not approximate Gaussian statistics for low $S$), the error would always be larger than that achieved by the Bayes classifier (not shown here, although we have verified this). For larger $S$, however, the analytically-calculated error of the Gaussian classifier shows excellent agreement with the performance of the exact Bayes classifier. Typically, even values of $S$ as low as $S \sim 2^4$ can be sufficient for good agreement.

\subsection{Role of sampling noise in QCSA using a single-variable example}

In the main text we have seen several examples of how quantum sampling noise and its reduction can play an important role in achieving a QCSA. For the simple case of a single-variable binary classification task using a single-qubit sensor, the sampling noise characteristics for both QS and QCS protocols can be clearly visualized. In this appendix subsection, we will analyze these sampling noise characteristics to  further explain how QCS enables a QCSA.


\begin{figure}
    \centering
    \includegraphics{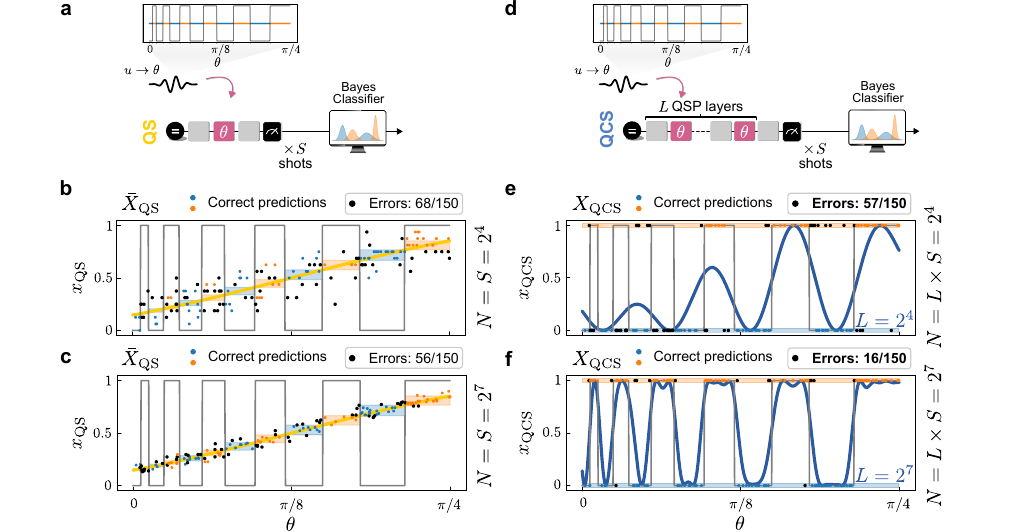}
    \caption{\textbf{Sampling noise characteristics: QS vs. QCS.} \textbf{a}, Conventional quantum sensor used for the binary classification task shown. \textbf{b}, Output of the conventional quantum sensing protocol as a function of phase $\theta$, for $N=S=2^4$. \textbf{c}, Same as \textbf{b}, for $N=S=2^7$. In both plots the solid curve is the expected measurement result $x_{\rm QS}$ (which is independent of $S$), while dots are finitely-sampled outputs $\bar{X}_{\rm QS}$. The shaded regions are used to determine the output of a Gaussian classifier (see text). Increasing number of shots $S$ reduces sampling noise and reduces the number of errors in classification. \textbf{d}, Quantum computational sensor used for the binary classification task shown. \textbf{e}, Output of the quantum computational sensing protocol as a function of phase $\theta$, for $N=2^4$ as for the QS protocol in \textbf{b}; for the QCS protocol $L=2^4$ and hence $S=1$. \textbf{f}, Output of the quantum computational sensing protocol as a function of phase $\theta$, for $N=2^7$ as for the QS protocol in \textbf{c}; for the QCS protocol $L=2^7$ and hence $S=1$. For the QCS protocol with a single shot, the shaded regions are used to determine the output of a Bayes classifier. }
    \label{appfig:sampling}
\end{figure}


In Fig.~\ref{appfig:sampling}, we consider the binary classification task from Sec.~\ref{subsec:1q} of the main text, starting with the conventional QS protocol in Fig.~\ref{appfig:sampling}a. As in the main text, we will analyze the measurement output of the conventional quantum sensor as a function of the sensed phase $\theta$, and for varying $N$ values. Note that the expected measurement result $x_{\rm QS}$ for the conventional QS strategy is fixed, and therefore does not vary with $N$; this is seen in the plots of $x_{\rm QS}$ in Fig.~\ref{appfig:sampling}b for $N=2^4$, and in Fig.~\ref{appfig:sampling}c for $N=2^7$.

Now, however, in addition to $x_{\rm QS}$, we also plot instances of the finite-sampled measurement results $\bar{X}_{\rm QS}$ for 150 uniformly-sampled $\theta$ values in the sensing domain, for both $N$ values considered; these are shown by the dots. These dots are also marked by colors that encode the result of a Gaussian classifier applied to the measurement results, defined by Eq.~(\ref{appeq:gaussclassifier}), as we will explain shortly. We use a Gaussian classifier for the conventional QS here since its operation is more interpretable in general than the Bayes classifier, and will allow us to better visualize classification performance. Note that, as seen in Fig.~\ref{appfig:gaussclassifier}c, for $S\gtrsim 2^4$ the predicted accuracy of this Gaussian classifier is very close to that of the exact Bayes classifier, so the performance of the QS protocol is not underestimated by our use of the Gaussian classifier in this analysis.

The shaded regions in Figs.~\ref{appfig:sampling}b,~\ref{appfig:sampling}c are the regions $\mathcal{X}_1$ (blue) and $\mathcal{X}_2$ (orange) that define the Gaussian classifier for Class 1 and 2 respectively, as defined by Eq.~(\ref{appeq:gaussregion}). For $\theta$ values in a blue region ($\mathcal{R}_1$, i.e. belonging to Class 1), if the corresponding measurement result $\bar{X}_{\rm QS}(\theta)$ also falls into a blue region, the Gaussian classifier result is Class 1, and is therefore correct; such values of $\bar{X}_{\rm QS}$ are colored blue. Similarly, for $\theta$ values in an orange region ($\mathcal{R}_1$) that hence belong to Class 2, if the corresponding measurement result also falls into an orange region, the classifier result is correct again, and such values of $\bar{X}_{\rm QS}$ are colored orange. The remaining option is if $\theta$ and the corresponding $\bar{X}_{\rm QS}(\theta)$ fall in differently-colored regions; here the classifier result will be incorrect, and these errors are shown as black dots. 

We can now analyze how the sampling noise of finitely-sampled measurement results changes with increasing $N$, and more importantly how this influences the classification error. Recall that for the QS, increasing $N$ increases the number of shots $S$ used to construct $\bar{X}_{\rm QS}$; this reduces sampling noise by a factor of $\frac{1}{S}$. This reduction is seen in going from Fig.~\ref{appfig:sampling}b to Fig.~\ref{appfig:sampling}c: the measurement results are distributed around their mean---the infinite-sampling result $x_{\rm QS}(\theta)$---but their variance visibly decreases with $N=S$. The reduction in variance means that for any $\theta$, the stochastic measurement results are distributed more closely around $x_{\rm QS}(\theta)$, and are less likely to fall outside the shaded region corresponding to the given $\theta$. This leads to fewer errors with increasing $S$, visualized in Fig.~\ref{appfig:sampling}c, and observed in Fig.~\ref{fig:1d}e of the main text.

Next, we analyze the sampling noise characteristics of QCS protocols (Fig.~\ref{appfig:sampling}d). To do so, we first show the infinite-sampling measurement result $x_{\rm QCS}(\theta)$ for QCS protocols with $N=L=2^4$ in Fig.~\ref{appfig:sampling}e, and $N=L=2^7$ in Fig.~\ref{appfig:sampling}f. Here the trained QCS protocol varies with $L$ and therefore $x_{\rm QCS}(\theta)$ is different for the two protocols shown. Since we have set $N=L$, only a single measurement shot can be obtained from the quantum computational sensor (since $N=L\times S$ and $N=L \implies S = 1$). As a result the measurement results $X_{\rm QCS}$ can only yield the value 0 or 1. The Bayes classifier simply assigns the most likely class to each of these two measurement results: if $X_{\rm QCS} = 0$ the predicted class is Class 1 (blue), while if $X_{\rm QCS}=1$ the predicted class is Class 2 (orange). 

We can now analyze how classification proceeds in the single shot regime. For any given measurement shot, the measurement result $X_{\rm QCS}$ is again a stochastic variable, exhibiting quantum sampling noise. Crucially, the quantum sampling noise depends on the underlying value of $x_{\rm QCS}(\theta)$: for values $0,1$, the sampling noise is zero! Unlike the conventional quantum sensor for which $x_{\rm QS}$ is around $0.5$ (the linear response regime), a quantum computational sensor is not restricted to operate in such a regime. As we see in Figs.~\ref{appfig:sampling}e, f, QCS protocols can be engineered so that the value of $x_{\rm QCS}$ approaches $0$ for $\theta$ in Class 1, and approaches $1$ for $\theta$ in Class 2. Then, for such $\theta$ values as inputs, the impact of sampling noise in the measurement of the quantum computational sensor is strongly suppressed. Here, even a single measurement shot provides sufficient information to perform classification. Errors can still occur, at the transition regions between $x_{\rm QCS}\to 0$ and $x_{\rm QCS}\to 1$. However these are much fewer in number than the errors committed by the conventional QS protocol at the same $N$, leading to a QCSA.

This visualization of classification allows another observation. For a sensing phase $\theta$ over a fixed domain, the more noncontiguous regions $R$ the domain is divided into for each class, the more transition regions exist where errors can be made. As a result, increasing $R$ should increase the difficulty of the classification task, for both protocols. This dependence of performance of both the QS and QCS protocols on task complexity is analyzed in the next two subsections.

\subsection{QCSA vs. task complexity for simplified tasks: approximate analytic error probability for a single-qubit conventional quantum sensor}
\label{app:analyticGauss}

The Gaussian classifier for the QS protocol is not just useful for the visualization of classification. In particular, our efforts in deriving an analytic form for the error of this classifier will allow us to arrive at an \textit{approximate} analytic form for the classification error of the QS protocol for specific binary classification tasks. We will derive such an expression in the present appendix section, and use it to explore the dependence of QS performance on task complexity, which we will also quantify precisely. We will finally also empirically determine QCS performance for the same set of tasks, ultimately obtaining the dependence of QCSA on task complexity.

To obtain an approximate analytic form of Eq.~(\ref{appeq:gausscerr}), we first focus on simplifying the double integral. Considering the first term (the second can be obtained from the first by swapping indices $1 \leftrightarrow 2$), we first rewrite the integral by swapping the order of integration:
\begin{align}
    \int_{x(\theta_{(1)}^{r',-})}^{x(\theta_{(1)}^{r',+})} d\bar{X} \int_{\theta_{(2)}^{r,-} }^{\theta_{(2)}^{r,+}} d\theta~P_{\rm G}(\bar{X}|\theta) = \int_{\theta_{(2)}^{r,-} }^{\theta_{(2)}^{r,+}} d\theta \int_{x(\theta_{(1)}^{r',-})}^{x(\theta_{(1)}^{r',+})} d\bar{X}~P_{\rm G}(\bar{X}|\theta).
    \label{appeq:doubleI}
\end{align}
Since $P_{\rm G}(\bar{X}|\theta)$ is a Gaussian probability density function, a particularly simple form exists for its integral any interval, in terms of error function:
\begin{align}
    \int_{x(\theta_{(1)}^{r',-})}^{x(\theta_{(1)}^{r',+})} d\bar{X}~P_{\rm G}(\bar{X}|\theta) = \frac{1}{2} \left[ {\rm erf}\left(\frac{x(\theta_{(1)}^{r',+})-x(\theta)}{\sqrt{2\sigma^2(\theta)}}\right) - {\rm erf}\left(\frac{x(\theta_{(1)}^{r',-})-x(\theta)}{\sqrt{2\sigma^2(\theta)}}\right)  \right]
\end{align}
To proceed, we now make a simplifying approximations. In particular, since we will still be required to perform an integral over the phase $\theta$, it will be very helpful if the $\theta$-dependence in the argument of the error function can be simplified. To this end, we will consider the small $\theta$ approximation of the mean $x(\theta)$ and the variance $\sigma^2(\theta)$ for the single-qubit conventional quantum sensor:
\begin{subequations}
\begin{align}
    x(\theta) &= 1-\sin^2 \left( \theta-\frac{3\pi}{8} \right) \stackrel{\theta\to 0}{\approx} \frac{1}{2} + \left( \theta- \frac{\pi}{8} \right),  \\
    \sigma^2(\theta) &= \frac{1}{S}x(\theta)(1-x(\theta)) = \frac{1}{S}\left[1-\sin^2 \left( \theta-\frac{3\pi}{8} \right) \right] \cos^2 \left( \theta-\frac{3\pi}{8} \right) \stackrel{\theta\to 0}{\approx} \frac{1}{S}\left[ \frac{1}{4}-\left(\theta-\frac{\pi}{8} \right)^2  \right] \approx \frac{1}{4S}.
\end{align} 
\end{subequations}
The validity of these (and all other) approximations will be assessed self-consistently. Under the above, $x(\theta)$ is now linear in $\theta$; as a result, we have:
\begin{align}
    x(\theta_{(j)}^{r',\pm})-x(\theta) = \theta_{(j)}^{r',\pm} - \theta
\end{align}
The required integral therefore simplifies to:
\begin{align}
    \int_{x(\theta_{(1)}^{r',-})}^{x(\theta_{(1)}^{r',+})} d\bar{X}~P_{\rm G}(\bar{X}|\theta) \approx \frac{1}{2} \left[ {\rm erf} \sqrt{2S}\left(\theta_{(1)}^{r',+} - \theta \right) - {\rm erf} \sqrt{2S}\left(\theta_{(1)}^{r',-} - \theta \right)  \right]
\end{align}
Returning to our double integral, we have:
\begin{align}
    \int_{x(\theta_{(1)}^{r',-})}^{x(\theta_{(1)}^{r',+})} d\bar{X} \int_{\theta_{(2)}^{r,-} }^{\theta_{(2)}^{r,+}} d\theta~P_{\rm G}(\bar{X}|\theta) \approx \frac{1}{2} \int_{\theta_{(2)}^{r,-} }^{\theta_{(2)}^{r,+}} d\theta \left[ {\rm erf} \sqrt{2S}\left(\theta_{(1)}^{r',+} - \theta \right) - {\rm erf} \sqrt{2S}\left(\theta_{(1)}^{r',-} - \theta \right)  \right]
\end{align}
We are now left with the integral over $\theta$, which requires performing a definite integral over the error function. To evaluate this, we use the standard result for the indefinite integral:
\begin{align}
    \int d\theta'~{\rm erf}\left(a\theta'+b \right) = \frac{1}{a}\left(a\theta + b \right){\rm erf}\left(a\theta+b \right) + \frac{1}{\sqrt{\pi}a} e^{-(a\theta+b )^2} + {\rm constant}.
\end{align}
Note that the constant of integration will be canceled out in the calculation of the definite integral. Therefore, the double integral of Eq.~(\ref{appeq:doubleI}) takes the form:
\begin{align}
    \int_{x(\theta_{(1)}^{r',-})}^{x(\theta_{(1)}^{r',+})} d\bar{X} \int_{\theta_{(2)}^{r,-} }^{\theta_{(2)}^{r,+}} d\theta~P_{\rm G}(\bar{X}|\theta) \approx +\frac{1}{2}\Bigg\{ -&\frac{1}{\sqrt{2S}}\left[ \sqrt{2S}\left(\theta_{(1)}^{r',+} - \theta_{(2)}^{r,+} \right){\rm erf}\sqrt{2S}\left(\theta_{(1)}^{r',+} - \theta_{(2)}^{r,+} \right) + \frac{1}{\sqrt{\pi}} e^{-2S\left(\theta_{(1)}^{r',+} - \theta_{(2)}^{r,+} \right)^2}  \right] \nonumber \\
    +&\frac{1}{\sqrt{2S}}\left[ \sqrt{2S}\left(\theta_{(1)}^{r',+} - \theta_{(2)}^{r,-} \right){\rm erf}\sqrt{2S}\left(\theta_{(1)}^{r',+} - \theta_{(2)}^{r,-} \right) + \frac{1}{\sqrt{\pi}} e^{-2S\left(\theta_{(1)}^{r',+} - \theta_{(2)}^{r,-} \right)^2}  \right]
    \Bigg\} \nonumber \\
    -\frac{1}{2}\Bigg\{ -&\frac{1}{\sqrt{2S}}\left[ \sqrt{2S}\left(\theta_{(1)}^{r',-} - \theta_{(2)}^{r,+} \right){\rm erf}\sqrt{2S}\left(\theta_{(1)}^{r',-} - \theta_{(2)}^{r,+} \right) + \frac{1}{\sqrt{\pi}} e^{-2S\left(\theta_{(1)}^{r',-} - \theta_{(2)}^{r,+} \right)^2}  \right] \nonumber \\
    +&\frac{1}{\sqrt{2S}}\left[ \sqrt{2S}\left(\theta_{(1)}^{r',-} - \theta_{(2)}^{r,-} \right){\rm erf}\sqrt{2S}\left(\theta_{(1)}^{r',-} - \theta_{(2)}^{r,-} \right) + \frac{1}{\sqrt{\pi}} e^{-2S\left(\theta_{(1)}^{r',-} - \theta_{(2)}^{r,-} \right)^2}  \right]
    \Bigg\} 
    \label{appeq:intpg1}
\end{align}


\begin{figure}
    \centering
    \includegraphics{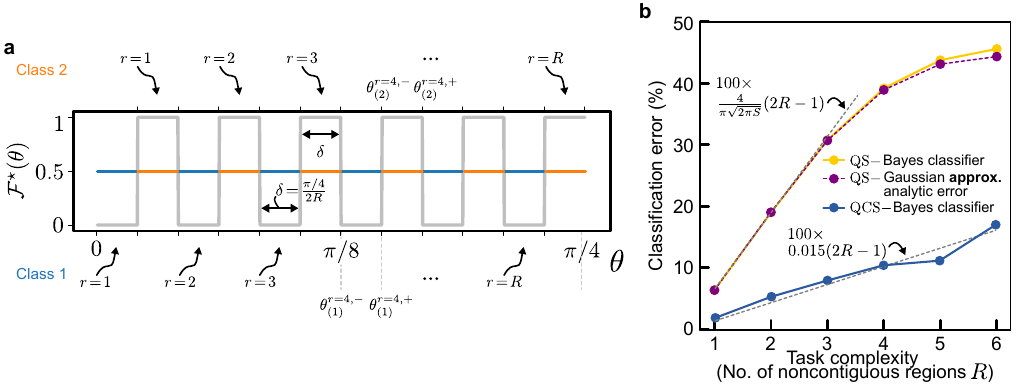}
    \caption{\textbf{Classification error of QS and QCS protocols for a simple binary classification task with increasing task difficulty} \textbf{a}, Binary classification task definition, which is a special case of the task considered in Fig.~\ref{appfig:taskdef}a where the spacings $\delta$ defining each of the $R$ noncontiguous regions are now equal. \textbf{b}, Classification error for fixed number of sensing periods $N=2^6$ as a function of task complexity, as determined by number of noncontiguous regions $R$ in \textbf{a}, comparing QS protocol (with $N=S=2^6$) and QCS protocol (with $N=L=2^6$ and $S=1$). Dashed gray lines show linear fits valid in the regime $N \gg R$. }
    \label{appfig:simplecomplexity}
\end{figure}


Under the assumption of a linear response between the quantum sensor output $x(\theta)$ and the phase $\theta$, and further that the stochastic measurement results $\bar{X}_{\rm QS}$ follow a Gaussian distribution, Eq.~(\ref{appeq:intpg1}) holds for \textit{any} single-variable binary classification task of the form of Fig.~\ref{appfig:gaussclassifier}a, and is therefore very general. Beyond this point, however, it will be difficult to obtain a compact form of the above integral, and therefore of $\mathcal{E}_{\rm G}$, for completely arbitrary tasks. However, for certain binary classification tasks, the above integral can be simplified further. In particular, we consider the case where the classification task has equally-spaced regions $\delta = \frac{\pi/4}{2R}$ for each class, as depicted in Fig.~\ref{appfig:simplecomplexity}a. In this case we have:
\begin{align}
    \theta_{(1)}^{r',+} - \theta_{(2)}^{r,+} &= \left[2(r'-r)-1\right]\delta \nonumber \\
    \theta_{(1)}^{r',+} - \theta_{(2)}^{r,-} &= \left[2(r'-r)\right]\delta \nonumber\\
    \theta_{(1)}^{r',-} - \theta_{(2)}^{r,+} &= \left[2(r'-r)-2\right]\delta \nonumber \\
    \theta_{(1)}^{r',-} - \theta_{(2)}^{r,-} &= \left[2(r'-r)-1\right]\delta
\end{align}
Substituting the above expressions into Eq.~(\ref{appeq:intpg1}) and simplifying, we arrive at:
\begin{align}
    &\sum_{r'=1}^R\sum_{r=1}^R\int_{x(\theta_{(1)}^{r',-})}^{x(\theta_{(1)}^{r',+})} d\bar{X} \int_{\theta_{(2)}^{r,-} }^{\theta_{(2)}^{r,+}} d\theta~P_{\rm G}(\bar{X}|\theta) \approx \nonumber \\
    \frac{1}{2}&\sum_{r'=1}^R\sum_{r=1}^R \Big\{  2(r'-r)\delta~{\rm erf}(2(r'-r)\sqrt{2S}\delta) + \left[2(r'-r)-2\right]\delta~{\rm erf}(\left[2(r'-r)-2\right]\sqrt{2S}\delta) \nonumber \\
    &-2\left[2(r'-r)-2\right]\delta~{\rm erf}(\left[2(r'-r)-2\right]\sqrt{2S}\delta) +\frac{1}{\sqrt{2\pi S}}\left( e^{-2S\delta^2\left[2(r'-r)\right]^2 } + e^{-2S\delta^2\left[2(r'-r)-2\right]^2 } -2 e^{-2S\delta^2\left[2(r'-r)-1\right]^2 }  \right) \Big\}
\end{align}
The remaining double sum is in-principle to be performed over $R^2$ terms. However, we can now make a second simplifying approximation. We limit the inner sum to neighboring indices only, $r'-r = 0,1$. This assumes that errors are only caused by from neighboring regions, and is valid in the regime where $S$ is large so that the variance of measurement results can be assumed to be small enough that errors from beyond neighboring regions are rare. The restriction to $r'-r=0,1$ simplifies the sum over $r$:
\begin{align}
    \sum_{r'=1}^R\sum_{r=1}^R\int_{x(\theta_{(1)}^{r',-})}^{x(\theta_{(1)}^{r',+})} d\bar{X} \int_{\theta_{(2)}^{r,-} }^{\theta_{(2)}^{r,+}} d\theta~P_{\rm G}(\bar{X}|\theta) \approx \frac{1}{2}&\sum_{r'=1}^R\Big\{  2\delta~{\rm erf}(2\sqrt{2S}\delta) -2\delta~{\rm erf}(\sqrt{2S}\delta) +\frac{1}{\sqrt{2\pi S}}\left( 1+e^{-8S\delta^2} -2 e^{-2S\delta^2} \right) \Big\} \nonumber \\
    +\frac{1}{2}&\sum_{r'=1}^{R-1}\Big\{2\delta~{\rm erf}(2\sqrt{2S}\delta) -2\delta~{\rm erf}(\sqrt{2S}\delta) +\frac{1}{\sqrt{2\pi S}}\left( 1+e^{-8S\delta^2} -2 e^{-2S\delta^2} \right) \Big\}
\end{align}
where we have also used ${\rm erf}(-z) = -{\rm erf} z$. The terms on the first line arise from setting $r'-r=0$; there will be $R$ such terms. The terms on the second line, on the other hand, arise from the condition $r'-r=1$; however note that there are only $R-1$ such terms (in particular, we cannot set $r=R$, as then $r'=R+1$). Note further that the summands are equal, and further have no dependence on $r'$; as a result the summation will simply count the number of terms in each term, and can be simplified to the compact form:
\begin{align}
    \sum_{r'=1}^R\sum_{r=1}^R\int_{x(\theta_{(1)}^{r',-})}^{x(\theta_{(1)}^{r',+})} d\bar{X} \int_{\theta_{(2)}^{r,-} }^{\theta_{(2)}^{r,+}} d\theta~P_{\rm G}(\bar{X}|\theta) \approx \frac{2R-1}{2}\left[  2\delta~{\rm erf}(2\sqrt{2S}\delta) -2\delta~{\rm erf}(\sqrt{2S}\delta) +\frac{1}{\sqrt{2\pi S}}\left( 1+e^{-8S\delta^2} -2 e^{-2S\delta^2} \right) \right]
    \label{appeq:intpg2}
\end{align}
The result has no dependence on class labels any more; this means that the second integral in Eq.~(\ref{appeq:gausscerr}) will have the same form as the above. Therefore, substituting Eq.~(\ref{appeq:intpg2}) into Eq.~(\ref{appeq:gausscerr}) provides the approximate Gaussian classifier error $\mathcal{E}_{\rm G}$ (for tasks with equal spacing $\delta$):
\begin{align}
    \mathcal{E}_{\rm G} \approx \frac{2}{\pi/4}\cdot\frac{2R-1}{2}\left[ 2\delta \left( {\rm erf}(2\delta\sqrt{2S}) - {\rm erf}(\delta\sqrt{2S}) \right) + \frac{1}{\sqrt{2\pi S}}\left(1 + e^{-8S\delta^2} - 2e^{-2S\delta^2} \right) \right],
    \label{appeq:gausserrorapprox}
\end{align}
where recall that $\delta = \frac{\pi/4}{2R}$. This expression is presented in Sec.~\ref{subsec:taskcomp} of the main text as the approximate classification error of the QS protocol, $\mathcal{E}_{\rm QS} = \mathcal{E}_{\rm G}$.

In Fig.~\ref{appfig:simplecomplexity}b we compare the performance of the conventional QS protocol against QCS protocols for fixed $N=2^6$. For the conventional QS case we include both the exact Bayes classifier and the approximate result of Eq.~(\ref{appeq:gausserrorapprox}). The two show excellent agreement, especially for small $R$, emphasizing that the approximations we have made in arriving at this analytic form are valid for the considered values of $R$ and $S$. For the QCS protocols, we set $N=L=2^6$ and hence $S=1$, and train protocols using supervised learning as in the main text. The classification error obtained by applying a Bayes classifier to single-shot measurement results of the trained QCS protocols is plotted in Fig.~\ref{appfig:simplecomplexity}b. 

\subsection{QCSA vs. task complexity for more general tasks}
\label{app:complexity}

To emphasize that the performance difference between QS and QCS protocols for increasing task difficulty is not special to the tasks considered in Appendix~\ref{app:analyticGauss}, we now compare their performance for more general tasks. 


\begin{figure}[h]
    \centering
    \includegraphics{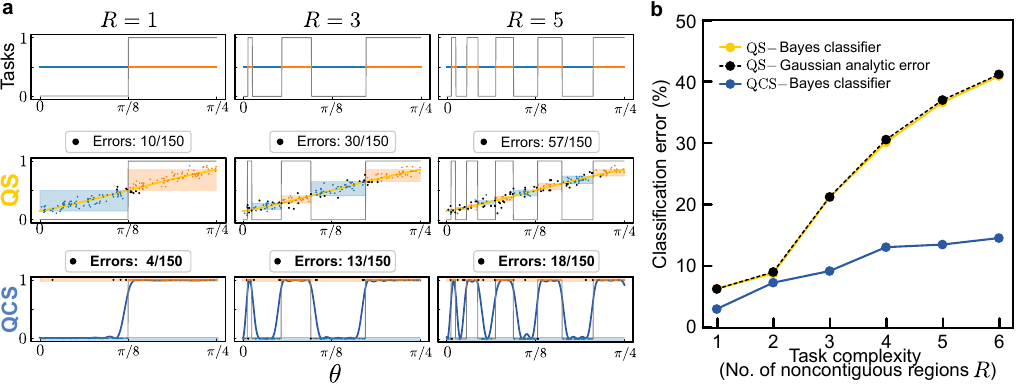}
    \caption{\textbf{Increasing QCSA with increasing task difficulty.} \textbf{a}, Top panel: tasks for $R=1,3,5$ from left to right. Both QS and QCS protocols are allowed $N=2^6$ sensing periods. Center panel: Conventional quantum sensor measurement outputs and visualization of QS scheme performance for the considered tasks.  For conventional QS, $N=S=2^6$.  Bottom panel: Quantum computational sensor measurement outputs and visualization of QCS scheme performance for the considered tasks. For QCS, we set $N=L=2^6$, which forces $S=1$ as $N=L\times S$ for the QCS protocol. For both QS and QCS protocols, colored dots show correctly classified points, while black dots indicate errors (shown enlarged in comparison to colored dots to aid visualization). A total of 150 points are depicted, and the corresponding errors for each protocol are indicated in the plot headers. \textbf{b}, Classification error of QS and QCS approaches for single-variable binary classification tasks of increasing complexity, quantified by the number of noncontiguous regions $R$ defining each class; here $R\in\{1,\ldots,6\}$. }
    \label{appfig:complexity}
\end{figure}


In particular, while we still consider our general binary classification tasks with $R$ noncontiguous regions depicted in Fig.~\ref{appfig:taskdef}, now consider tasks with unequal spacings $\delta^r$ unlike the previous section. We still vary the number of regions $R\in\{1,\ldots,6\}$ to controllably increase the task difficulty. Some instances of the tasks we consider are shown in the top panel of Fig.~\ref{appfig:complexity}a for $R=1,3,5$; here $R=6$ is the task analyzed in Sec.~\ref{subsec:1q} of the main text. 

The relative performance of the QS and QCS protocols for a fixed number of sensing periods---here $N=2^6$---can again be visualized as a function of the number of regions $R$. In the center panel of Fig.~\ref{appfig:complexity}a, we show the output of the conventional QS protocol ($N=S=2^6$) and classification results, here using the Gaussian classifier to aid visualization. For the QCS protocol, we set $N=L=2^6$, so that only a single measurement shot from the quantum computational sensor must be used for classification. We clearly see that while the error of both protocols increases with task difficulty, the error of the QCS protocol grows much more slowly. As a result, the extent of QCSA increases with task difficulty, as shown in Fig.~\ref{appfig:complexity}b.

\newpage
\subsection{Multi-variate sensing: comparing quantum computational sensors against multiple Ramsey interferometers}
\label{app:mqramsey}

\begin{figure}
    \centering
    \includegraphics{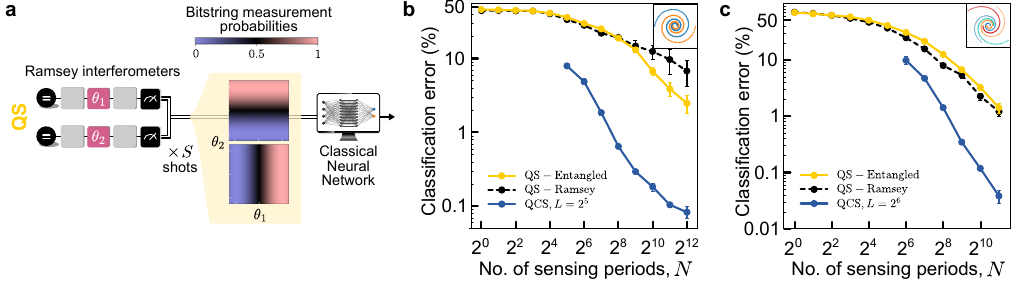}
    \caption{\textbf{Multiple Ramsey interferometers for multi-variate sensing.} \textbf{a}, Using a two-qubit system as two uncoupled Ramsey interferometers, each sensing one of the phases $\theta_j$. Color plots show excitation probabilities of each of the individual qubits over the domain of $\bm{\theta}$ relevant for the classification tasks under consideration. \textbf{b}, Classification error against number of sensing periods $N$ for the binary \logspirals{} classification task studied in Fig.~\ref{fig:2d}a of the main text (depicted in the top-right corner). \textbf{c}, Classification error against number of sensing periods $N$ for the multiclass \logspirals{} classification task studied in Fig.~\ref{fig:2d}f of the main text (depicted in the top-right corner). }
    \label{appfig:mqramsey}
\end{figure}

In Sec.~\ref{subsec:2q} of the main text, for the conventional QS benchmark we consider a protocol for which $\Uenc$ and $\Udec$ are entangling operations and $L=0$, namely with a structure similar to quantum sensor networks~\cite{eldredge2018optimal}. In particular, this quantum sensor is not engineered specifically to operate in a linear response regime where measurements from it can directly yield an estimate of the individual phases $\theta_j$. In this subsection, we instead deploy the two-qubit system as two uncoupled Ramsey interferometers, each used to sense one of the phases $\theta_1$ and $\theta_2$ respectively, as depicted in Fig.~\ref{appfig:mqramsey}a.

Instead of probabilities for measuring the four possible bitstrings across two-qubit measurements, we show the excitation probabilities $x_1 = x_{00} + x_{10}$ and $x_2 = x_{00} + x_{01}$ for the first and second qubit respectively over the input domain of $\bm{\theta}$. It is clear that $x_1$, $x_2$ vary monotonically with $\theta_1$, $\theta_2$ respectively; as a result, measurements of each qubit can be used to estimate the sensed phases directly, as desired. The measurement results given access to $N$ sensing periods (recalling $N=S$  shots for the QS protocol) are passed through a trainable classical neural network and the classification error computed for the task under consideration.

Our results are shown in Fig.~\ref{appfig:mqramsey}b and Fig.~\ref{appfig:mqramsey}c for the two classification tasks considered in Sec.~\ref{subsec:2q}. For reference, we also show the performance of the conventional QS benchmark and the QCS protocol with largest $L$ used in the main text. Our results show that using two Ramsey interferometers supplemented by classical postprocessing yields a performance that is at most equal to, and can be worse, than the entangled QS protocols. As a result, our conclusions about the achievable QCSA using the developed QCS protocol still hold.


\newpage
\subsection{QCSA for multi-variable binary discrimination tasks using a single qubit}
\label{app:mvbc}

In Sec.~\ref{subsec:2q} of the main text, we showed how binary multi-variable classification tasks can be performed using quantum computational sensors comprising two qubits. However, in principle a single-shot measurement of even a single qubit can provide sufficient information for binary classification. Therefore one may ask whether a single-qubit quantum computational sensor can be engineered to perform binary but multi-variable classification tasks. In this Appendix section we show that this is indeed possible, by providing a QCS protocol that achieves this objective.

\begin{figure}
    \centering
    \includegraphics{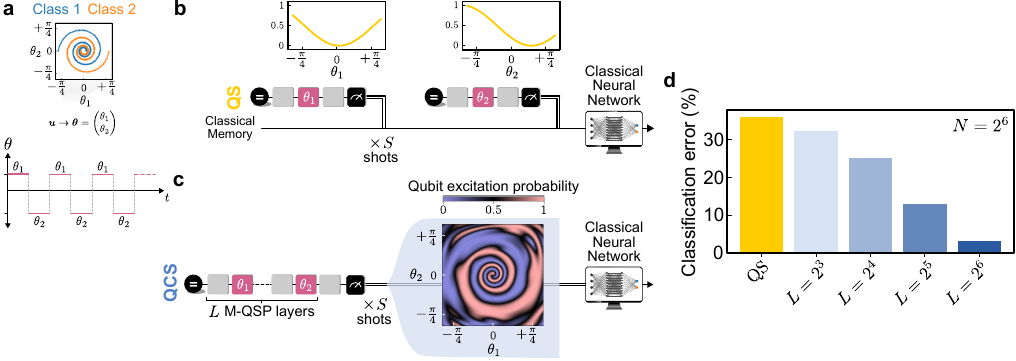}
    \caption{\textbf{Quantum computational sensing of bivariate signals using a single qubit.} \textbf{a}, Time-varying bivariate signal to be sensed using single-qubit sensors. \textbf{b}, The conventional QS protocol is given by a Ramsey interferometer. \textbf{c}, The QCS protocol used here bears similarity to the M-QSP protocol; for more details, see text. \textbf{d}, Classification error for a fixed number of sensing periods $N=2^6$, for the QS protocol and the QCS protocol with varying $L$. }
    \label{appfig:mqsp}
\end{figure}

We consider the binary classification task over bivariate signals shown in Fig.~\ref{appfig:mqsp}a, the same instance of the \logspirals{} classification task that was considered in Sec.~\ref{subsec:2q}. We consider performing this task using a single qubit and the sensing operation we have considered so far, which is sensitive to a scalar phase. Note that other approaches to sensing multivariate signals using a single qubit have also been considered~\cite{kaubruegger_optimal_2023}; QCS protocols using such sensing operations could potentially also be considered.

To sense multivariate signals using a single qubit and this scalar sensing operation, we consider the situation where the multivariate phases are received at different times; namely we consider the sensing of the time-varying signal depicted in Fig.~\ref{appfig:mqsp}a. While this signal is no longer static in time, we have already seen in Sec.~\ref{sec:meg} of the main text that QCS protocols can be designed for the processing of time-varying signals. We do note, however, that the variation of this signal is not entirely arbitrary, as phases repeat at every other time step; this situation could describe, for example, signals transmitted using a preset communication protocol.

The conventional QS benchmark uses a single-qubit quantum sensor to perform a single sensing operation before measurement of the sensor; operations before and after the sensing operation are trainable. Unlike the static case, the input phase now alternates between $\theta_1$ and $\theta_2$ in successive sensing operations. Since this repetition structure is known (defined for example by the communication protocol being used to transmit the signal), we allow the trainable operations to have different learned parameters for the sensing of $\theta_1$ and $\theta_2$, essentially assuming that the sensor can be deterministically prepared for the sensing of $\theta_1$ or $\theta_2$. We have observed that using a standard Ramsey interferometer with fixed operations provides a qualitatively similar or slightly worse performance. 

We now describe the QCS protocol, which shares the standard QCS architecture we have introduced in Eq.~(\ref{eq:cohprocessing}) of the main text, and is depicted in Fig.~\ref{appfig:mqsp}b. The time-varying signal is sensed $L$ times, with trainable interleaved processing operations, before a single measurement is performed. Due to the specific structure of the time-varying signal, the sensed phases $\theta^{(l)}$ in the $l$th layer are given by:
\begin{align}
    \theta^{(l)} = 
    \begin{cases}
        \theta_1, &l=1,3,5,\ldots, \\
        \theta_2, &l=2,4,6, \ldots.
    \end{cases}
\end{align}
The signal sensed by the quantum computational sensor therefore repeatedly alternates between two values. This structure bears a resemblance to the M-QSP algorithm analyzed recently~\cite{rossi_multivariable_2022}. 

The single qubit measurement probability in the bivariate domain $\theta_1,\theta_2$ for a trained QCS protocol with this architecture is also shown in Fig.~\ref{appfig:mqsp}b. We see that the qubit measurement probability can be engineered to have a nontrivial dependence in this bivariate domain. The classification error for a fixed number of sensing periods $N=2^6$ is shown in Fig.~\ref{appfig:mqsp}c, for both the QS protocol and the single-qubit bivariate sensing protocol. As with other QCS protocols in the main text, we see that the QCS approach can achieve a lower classification error than the QS scheme, and the error is reduced with increasing $L$ for fixed $N$.


\clearpage


\newpage

\section{QCS of spatiotemporal signals using qubit-based quantum computational sensors}
\label{app:meg}

In this Appendix section, we discuss the details of the simulations and results of the binary classification task over time-varying signals, from Sec.~\ref{sec:meg} of the main text. We first discuss the dataset and the task, before moving onto our QCS architecture based on a Quantum Neural Network ansatz, followed by details of the training procedure. Finally, we discuss the improvement in classification accuracy we observe when using QCS protocols with varying degrees of computing resources, over a conventional QS architecture supplemented by classical computation on the measurement outcomes, including supplementary results beyond those included in Fig.~\ref{fig:meg}.

\subsection{Dataset and classification task}

The dataset we use for these simulations is obtained from an open-source experiment~\cite{yeom2023magnetoencephalography}. The dataset is a time-varying, 306-channel magnetoencephalography (MEG) signal, along with a 3-axis accelerometer signals. Each channel consists of data generated by a sensor. There are 3 sensors at 102 spatial locations on a headset. At each location, one of the sensor is a magnetometer, with the other two gradiometers. This dataset is collected for nine patients. Each experiments consists of a visual shown to the patient indicating the (one of four) directions to move their right hand. The movement of their hand is captured by the 3-axis acceloremeter. During the same time, the signals produced by the brain are recorded by the headset. The goal of the dataset is to help train machine learning models to predict the movement of the hand given the MEG signals. Details of the entire experimental setup can be found in~\cite{yeom2023magnetoencephalography}.

Such datasets are crucial in the development of technologies of brain-computer interfaces (BCIs), where the goal is to predict the objective of the patient from the signals produced by their brain. BCIs can help people perform physical mechanisms with the aid of a robot, which otherwise would not be possible. An important aspect in the realization of these applications is to develop sensors to measure the signals produced by the brain with the least amount of noise. For example, SQUID-based sensors are among the most sensitive sensors for magnetic field detection~\cite{fagaly2006superconducting}. However, other quantum sensing platforms have shown tremendous progress in sensitivity, and have the potential to harness quantum resources such as entanglement to improve sensitivity~\cite{aslam2023quantum}. This motivates our idea to pursue a QCSA in the domain of spatio-temporal magnetic field sensing tasks.

\subsection{Dataset preprocessing}

We now discuss the procedure we implement to allow the dataset to be compatible with our simulations. The main goal of this procedure is to reduce the complexity of the simulation of the quantum system, so that they maybe simulated in a reasonable time on a classical computer. We first perform Principal Component Analysis on the data to find the most important component. We then order the channels of the dataset in order of their overlap in this component. We choose the first $N$ channel corresponding to a simulation of an $N$-qubit quantum sensor. Each channel represents the magnetic field experienced by one of the qubit. One can envision this as the qubit sensors being located at the physical locations of the channels on the headset, experiencing the magnetic field signals produced by the brain. We then perform a simple boxcar filtering to reduce the number of time points of the signal. For these simulations, we choose this value to be $T=10$.

We then consider a binary classification task of predicting whether the patient moved their hand to the top-left or bottom-right (by using the data provided by the 3-axis accelerometer), given the MEG signals. We consider the dataset of a single patient, which is of size $60$. We use half the dataset for training the entire quantum sensing processor and the rest of the dataset for the testing phase. In the results below, we provide the performance of the testing phase. We consider the performance of the sensor at different strengths of the magnetic field. This is done by simply scaling the MEG data (which is recorded in the dataset in arbitrary units) to produce different amounts of phases on the qubits. In the following plots, we characterize the strength of the signal by plotting the classification accuracy as a function of the root-mean-squared (RMS) value, averaged across the entire dataset, channels and time bins of the simulation. Physically, this scaling is determined by various factors, such as the strength of the magnetic field at the sensor (which in turn depends on the distance between the sensor and the patient) and the bare sensitivity of the sensing qubits.

In these simulations, we have access to the ``quantum noise-free" dataset of the signal. In other words, the signal is strong enough for the noise added by the (classical) sensor used to collect this data to be small. However, we envision the quantum computational sensors simulated here for settings where the signal is much smaller than the quantum noise (which we simulate by scaling the magnetic field of the dataset to be in the regime where quantum projection noise plays a significant role)

\subsection{Quantum Neural Network architecture for QCS}

\definecolor{meg_r}{RGB}{255, 204, 0}
\definecolor{meg_t}{RGB}{203, 218, 241} 
\definecolor{meg_s}{RGB}{130, 159, 202} 
\definecolor{meg_st}{RGB}{44, 90, 160} 


\begin{table}[h!]
    \centering
    \caption{\textbf{Properties of protocols used for classification of spatiotemporal MEG data.} The protocols are depicted in Fig.~\ref{appfig:meg}. }
    \begin{tabular}{|c||c|c|c|c|} 
     \hline
     \textbf{Quantum System}  & Trainable single-qubit  & Trainable multi-qubit & Layers & Number of \\ 
     \textbf{Architecture} & operations $R$ & operations $V$ & $L$ &  measurements \\ 
     \hline
     \textbf{{\color{meg_r}Conventional QS (R)}} & \CheckmarkBold  & \XSolidBrush & $1$ & $M\cdot T$ \\ 
     \textbf{{\color{meg_t}Temporally-coherent QCS (T)}} & \CheckmarkBold & \XSolidBrush & $T$ & $M$ \\
     \textbf{{\color{meg_s}Spatially-coherent QCS (S)}} & \XSolidBrush & \CheckmarkBold & $1$ & $M\cdot T$\\
     \textbf{{\color{meg_st}Spatiotemporally-coherent QCS (ST)}} & \XSolidBrush & \CheckmarkBold & $T$ & $M$ \\
     \hline
    \end{tabular}
    \label{table:meg_parameters}
\end{table}



\begin{figure}[h]
    \centering
    \includegraphics[width=\linewidth]{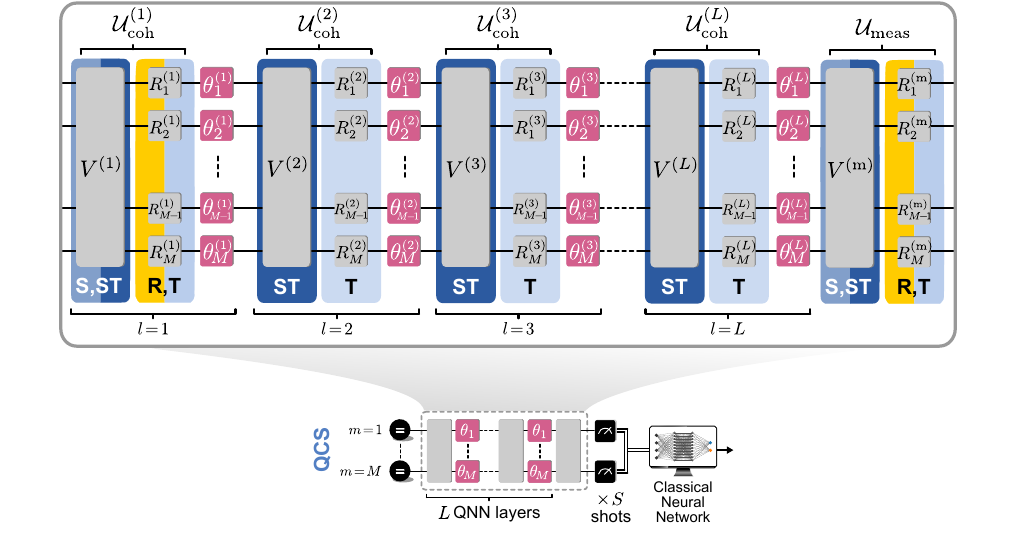}
    \caption{\textbf{Architecture of qubit-based quantum computational sensor for QCS of spatiotemporal signals.} All gray blocks describe trainable operations, whose definitions are included in Eqs.~(\ref{appeq:megunitaries}),~(\ref{appeq:megR}). All operations marked by specific colors are \textit{only} used in the correspondingly-colored protocols in Table~\ref{table:meg_parameters}. }
    \label{appfig:meg}
\end{figure}


In this subsection, we discuss the architecture of the quantum neural network along with how it is trained. We consider four different protocols, depicted compactly in Fig.~\ref{appfig:meg}; the differences between these protocols will be elaborated upon shortly. The various colors denote operations that are only used in specific protocols, as also detailed in Table~\ref{table:meg_parameters}. We simulate these protocols with up to 7 qubits, under different strengths of the magnetic field experienced by the quantum systems. We first discuss in detail the architecture which tends to achieve the highest classification accuracy. For an $M$-qubit sensor, we ae able to sense the $M$ most important channels (as identified using our PCA approach) of the MEG dataset. 

The MEG signal for each qubit is downsampled to $T=10$ timesteps for these simulations, as mentioned earlier; this is to reduce simulation complexity. This downsampling emulates the qubit integrating the magnetic field signal received in that timescale. The action of the magnetic field signal is to impart a phase rotation around the Pauli-Z axis of the qubit. The action of this is represented as a single qubit rotation gate as shown in Fig.~\ref{appfig:meg}. 

The sensor is initialized in its ground state, with all the qubits in the $\ket{0}$ state. Then, the system experiences a series of interleaved gates: the programmable and sensing operations, as defined by Eq.~(\ref{eq:cohprocessing}). The goal of the training procedure is to optimize the parameters of the programmable operations to improve the classification accuracy of the sensor to the machine learning task.

For the spatially and spatiotemporally coherent QCS protocols---which allow entanglement between qubits---we choose the operations $\Ucoh{l}$ to be of the most general form, allowing for arbitrary coupling between the qubits at each stage. This allows the sensor to be sensitive to spatial correlations in the signals among the qubits. Furthermore, by including programmable unitaries in the duration of the signal allows us to be sensitive to correlations of the signals in time.

We parameterize each programmable unitary in the Pauli string basis. For a system of $N$-qubits, the set of $4^N$ Pauli strings generate a complete basis for unitary matrices. Therefore, each programmable unitary $U_l$ is represented as the exponential of all Pauli string with trainable coefficients:
\begin{equation}
    V^{(l)} = \exp\left\{-i\sum_{n=0}^{4^N-1} c^{(l)}_{n}P_n\right\},
    \label{appeq:megunitaries}
\end{equation}
where $l$ indexes the layer of the trainable unitary, $c^{(l)}_{n}$ represents the coefficients of the $n$th Pauli string $P_n$ at the $l$th layer. In total, we have $L+1$ trainable unitaries (since the number of timesteps of the MEG dataset is $T=10=L$, and the final unitary is part of $\Udec$). 

For the QS protocol and the QCS protocol (which allows only temporal coherence), we instead use $\Ucoh{l}$ composed of trainable single-qubit (albeit arbitrary) rotations only. For the $m$th qubit, these have the form,
\begin{align}
    R_m^{(l)} = \exp\left\{ -i \frac{\Omega_m^{(l)}}{2} \left( \sin \theta_m^{(l)} \sin \phi_m^{(l)} \hat{\sigma}_m^x +  \sin \theta_m^{(l)} \cos \phi_m^{(l)} \hat{\sigma}_m^y + \cos\theta_m^{(l)} \hat{\sigma}_m^z \right)  \right\}
    \label{appeq:megR}
\end{align}

At the end of each protocol, we measure the qubits in the computational basis. We assign the qubit measured in the $\ket{0}$ state as a $0$ and in the state $\ket{1}$ as a $1$. The output of the quantum sensor processor is therefore a binary string of length $M$. We then pass this through the classical post processor, which is a multi-layer perceptron (MLP). The input dimension of the classical post processor is the number of measurements from the quantum system, while its output dimension is equal to the number of classes for the discrimination task; as we consider a binary classification task, this is just $2$. The internal structure of the MLP is given by: No. of measurements $ \to 256 \to 256 \to 2$. Therefore, the output of the neural network has dimension $2$ (the number of class labels), with the index of the element with the larger value representing the prediction of the system.

\subsection{Training details}

The simulations are performed with \texttt{PyTorch}~\cite{paszke2019pytorchimperativestylehighperformance}, which allows us to use the conventional method of backpropagation to train the system. All matrices, representing the sensing and programmable unitaries, and the state of the system, are stored as \texttt{PyTorch} tensors (with a dimension set by the Hilbert space size of the system), allowing for them to be differentiable. We initialize the programmable unitaries to be samples from the Haar random distribution of unitaries. To do this, we first generate a sample of a Haar random unitary. Since the Pauli strings represent a complete and orthogonal basis, we compute the overlap of the Pauli strings with the logarithm of the Haar random matrix to obtain the initial values of the coefficients of the Pauli strings, which represent the parameters of the quantum sensing protocol. We sample a new Haar random unitary for each programmable unitary.

Due to the exponentially increasing size of the Hilbert space dimension, our simulations are limited to $M=7$ qubits. However, more efficient representations, such as Matrix Product States~\cite{Schollw_ck_2011}, might allow for efficient training of quantum systems with more number of qubits. We leave this for future work.

At the end of the senssing protocol, we generate the samples of the measurement using \texttt{PyTorch}'s inbuilt Bernoulli sampling function. This sample is then fed as the input vector to the classical post processor. The loss function is the cross entropy loss between the actual class and the predicted class. We train the entire system (both the quantum system and the classical processor) end-to-end. To backpropagate between through the stochastic measurements of the quantum system, we use the ``Straight-Through" estimator. This method simply casts the gradients of the measurement outcomes onto the probabilities of each individual qubit being in the excited state. Standard backpropagation then backpropagates this gradient back till the parameters of the unitaries.

We use \texttt{PyTorch}'s Adam optimizer to train the parameters of the system. We use different learning rates for the parameters of the quantum system (we choose a value of $1\times10^{-4}$) than the parameters of the classical neural network (we choose a value of $3\times10^{-3}$). We find these values as a compromise between speed of training and stability. We notice that the classical neural network is easier to train and can train with a much larger learning rate than for the parameters of the quantum system. The classical post processor we choose is a simple feedforward deep neural network with two hidden dimensions, each of size $256$. We notice negligible change in the performance of the protocol with either increasing the hidden dimension or the depth. We train the entire system for $2000$ epochs, much longer than it takes to effectively train the system. To characterize the performance of the quantum computational sensor, we repeat this simulation several times with different initial configurations of parameters of the quantum system and the classical post processor, and post-select the best results.

\subsection{Additional results}


\begin{figure}[h]
    \centering
    \includegraphics[width=\linewidth]{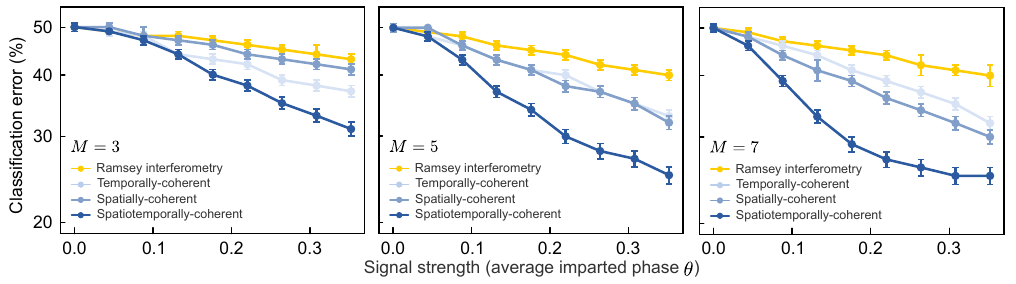}
    \caption{\textbf{Numerical results for the MEG dataset binary classification task for $M=3, 5, 7$ qubits}. We plot the classification accuracy for the $4$ quantum sensing protocols we consider, as a function of the strength of the signal to be sensed. Since this is time dependent distribution which varies across the dataset, we characterize the strength by the average root-mean-squared value of the phase imparted on a qubit within a single sensing unitary. Error bars represent the standard deviation in the estimate of the classification accuracy over different trained protocols.}
    \label{fig_app:meg_results}
\end{figure}


In Fig.~\ref{fig_app:meg_results}, we plot the results of our numerical simulations for the MEG dataset. We consider three different number of qubits $M = 3, 5, 7$. Since we perform simulation in the full Hilbert space of the quantum system, the simulation time increases exponentially in the number of qubits. For each of the $4$ quantum sensing protocols we consider, we vary the amount of phase experienced by the qubits. We repeat the simulation several times with different initial parameters for the trainable unitaries and the classical post processing neural network and choose the performance of the best performing result.

We notice that, for each number of qubits, the spatiotemporal QCS protocol outperforms all other schemes. On the other hand, the conventional Ramsey phase sensing interferometry has the largest classification error. We postulate this happens due to the ability of the spatio-temporal quantum sensing protocol to coherently-process spatial and temporal correlations in the signal, which is not possible in the conventional Ramsey quantum sensing protocol. The difference in the classification accuracy between these two curves represents the quantum computational-sensing advantage. From our simulations, we observe this advantage for all finite signal strengths across the three qubit numbers chosen in the simulations. The performance of the other two QCS protocols, coherently computing either only spatial correlations in the data or only temporal correlations in the data, lie in between the other two protocols. In is interesting to note for the smallest qubit number considered $M=3$, the temporally-coherent quantum sensing protocol outperforms the spatially-coherent quantum sensing protocol. While they have similar performance for $M=5$ qubits, the performance switches for $M=7$, where the spatially-coherent quantum sensing protocol performs better than the temporally-coherent quantum sensing protocol. This suggests the role and importance of the spatial and temporal correlations depends on the dataset (which manifests in this case with the number of qubits in the simulations, since that determines the number of channels of MEG data we consider).

Our results hint at the generality of quantum computational sensing advantage achievable for a wide variety of quantum sensing tasks, including those that could be of practical interest.

\newpage

\section{QCS for function approximation using bosonic quantum computational sensors}
\label{app:bosonicFA}

In this Appendix section we provide details of QCS using sensors comprising bosonic modes, from Sec.~\ref{sec:bosonic} of the main text. In particular we provide details of how the conventional QS protocol using a linear phase-preserving amplifier and our introduced QCS protocol using a quantum nonlinear amplifier can be used for the approximation of arbitrary polynomials of complex-valued displacements. 

\subsection{Architecture: $\Udec$ and quadrature measurements}

The architecture of both the QS and the QCS protocols is defined by our general Eq.~(\ref{eq:cohprocessing}) of the main text, and consists of a pair of bosonic modes in each case. We assume the initial state prior to either protocol to be $\ket{\psi_0} = \ket{0}\otimes\ket{0}$, namely both modes are in the vacuum state.

As with other sensing protocols, we begin by introducing the form of the sensing interaction $\Usense$ for the sensing of displacements, $\bm{u} \to \alpha = \alpha_x + i\alpha_y$, which takes the form:
\begin{align}
    \Usense(\alpha) \equiv D(\alpha) = \exp \{ -i \alpha_x(\hat{a}+\hat{a}^{\dagger})  -i \alpha_y(\hat{a}-\hat{a}^{\dagger}) \},
    \label{appeq:usensebosonic}
\end{align}
which is simply a displacement operator on one of the bosonic modes, here labeled $\hat{a}$, which we consider to be the only mode that directly interfaces with the physical signal. The action of $\Usense$ will simply place mode $\hat{a}$ in the coherent state $\ket{\alpha}$ after sensing.

For protocols using all-bosonic sensors in this work, we choose $L=1$, and also set the only processing operation to be the identity superoperator, $\Ucoh{1} = \Uenc = \mathcal{I}$. As a result, the complexity of both the QS and QCS protocols is provided by the $\Udec$ operation; we will describe these shortly. 

Before doing so, we discuss the measurement procedure following $\Udec$, which is common to both protocols: a heterodyne measurement of the auxiliary mode $\hat{b}$ is performed. We now briefly discuss the statistical properties of the results of heterodyne measurement $X$, which will be essential for determining the fidelity of function approximation using both QS and QCS protocols.

\subsubsection{Initial state and expectation values}

As mentioned above, we assume that prior to the sensing operation, both bosonic modes $\hat{a}$ and $\hat{b}$ are in the vacuum state,
\begin{align}
    \ket{\psi_0} = \ket{0} \otimes \ket{0}
\end{align}
for either protocol. We now define an intermediate state $\ket{\psi_{\rm sense}}$, which defines the state following the sensing operation (and before that, $\Uenc$, although recall that this is just the identity operation), and is again common to both types of sensor. We have:
\begin{align}
    \ket{\psi_{\rm sense}} = \Usense(\alpha)\ket{0} \otimes \ket{0} = \ket{\alpha} \otimes \ket{0},
\end{align}
which simply means that mode $\hat{a}$ is in a coherent state $\ket{\alpha}$ with amplitude $\alpha$ following the sensing operation, as dictated by Eq.~(\ref{appeq:usensebosonic}), while mode $\hat{b}$ remains in the vacuum state.

For the analysis of the evolution of these bosonic sensors following the sensing operation, we will be using the Heisenberg representation: recall that this assumes the quantum state $\ket{\psi_{\rm sense}}$ of the sensor remains unchanged while the operators for modes $\hat{a}$ and $\hat{b}$ evolve. The quantities of interest are always expectation values of general system observables, e.g. $\hat{o}_a$ or $\hat{o}_b$ for modes $\hat{a}$ and $\hat{b}$ respectively. In the Heisenberg representation, the final expectation value for a general observable $\hat{o}_a\hat{o}_b$ at the end of the QS or QCS protocol will be given by:
\begin{align}
    \avg{\hat{o}_a\hat{o}_b} = \bra{\psi_{\rm sense}}\hat{o}_a\hat{o}_b\ket{\psi_{\rm sense}} = \bra{\alpha}\hat{o}_a\ket{\alpha}\!\bra{0}\hat{o}_b\ket{0}
    \label{appeq:bosonicexp}
\end{align}
Provided $\hat{o}_a$, $\hat{o}_b$ are known in the Heisenberg representation, any arbitrary expectation value can be computed using Eq.~(\ref{appeq:bosonicexp}).

\subsubsection{Heterodyne measurements}

At the conclusion of the QS or QCS protocol, information will be extracted via heterodyne measurements of mode $\hat{b}$. For such heterodyne measurements, the corresponding positive-operator-valued-measure (POVM) is given by $\hat{E}_{\beta} = \frac{1}{\pi}\ket{\beta}\bra{\beta}$, noting that $\int d\beta~\hat{E}_{\beta} = \hat{\mathbb{I}}$ as required. We define the final state following the QS or the QCS protocol to be $\rhou = \ket{\psi(\alpha)}\bra{\psi(\alpha)}$. Then, the probability density function that describes the probability of obtaining the outcome $X=\beta$ as a result of the measurement described by $\hat{E}_{\beta}$ is given by:
\begin{align}
    p(\beta) = {\rm Tr}\left( \frac{1}{\pi}\ket{\beta}\bra{\beta} \rhou \right) = \frac{1}{\pi}\bra{\beta} \rhou \ket{\beta} = Q(\beta).
\end{align}
Here we have suppressed the implicit dependence of $\rhou$ on $\alpha$ for simplicity. Note that the probability density function $p(\beta)$ is simply the $Q$ function associated with the quantum state $\rhou$. Moments of the measured stochastic variables $\beta$ can be computed using $p(\beta)$:
\begin{align}
    \mathbb{E}[(\beta^*)^m \beta^n] = \int d\beta ~(\beta^*)^m\beta^n p(\beta) = \int d\beta ~\beta^n(\beta^*)^m Q(\beta) = \avg{\hat{b}^n \hat{b}^{\dagger m}},
    \label{appeq:ehet}
\end{align}
where the final equality follows from the definition of the $Q$ function~\cite{carmichael_statistical_2002}. In particular, we see that the moments of the heterodyne measurement results $\beta$ on mode $\hat{b}$ are related to \textit{anti-normal-ordered} moments of $\hat{b}$ with respect to the quantum state $\rhou$.

\subsection{Function approximation task: target polynomials $\Ft$ and expected mean-squared-error as a metric}

For completeness, we rewrite the target polynomials $\Ft$ considered in the main text,
\begin{align}
    \Ft = \sum_{n,m=0}^D \mathcal{W}_{nm} \alpha^{*n}\alpha^m.
    \label{appeq:ftarget2D}
\end{align}
Recall that $\mathcal{W}_{nm} = \mathcal{W}^*_{mn}$. The above general form can also be used to approximate 1D polynomials, e.g. if $\alpha = \alpha_x$ (where $\alpha_x$ is real). This special case will also be analyzed in this Appendix.

For the target function $\Ft(\bm{u})$, we will be constructing an estimator $\F(\bm{u})$. We therefore define the mean-squared-error (MSE) in the approximation of the target function for inputs $\bm{u}$ over a domain $\mathcal{R}$ as:
\begin{align}
    \MSE = \int_{\mathcal{R}} d\bm{u}~\left( \Ft(\bm{u})-\F(\bm{u})\right)^2     
\end{align}
In the specific case where $\F$ is an unbiased estimator of $\Ft$, namely $\mathbb{E}[\F(\bm{u})] = \Ft(\bm{u})$, we can equivalently write it as $\F(\bm{u}) = \mathbb{E}[\F(\bm{u})] + \left( \F(\bm{u}) - \mathbb{E}[\F(\bm{u})] \right) = \Ft(\bm{u}) + \left( \F(\bm{u}) - \mathbb{E}[\F(\bm{u})] \right)$, following which the $\MSE$ simplifies to:
\begin{align}
    \MSE = \int_{\mathcal{R}} d\bm{u}~\left( \F(\bm{u}) - \mathbb{E}[\F(\bm{u})] \right)^2 
\end{align}
Since $\F$ is a stochastic estimate of the desired target function, the $\MSE$ will also be a random variable. As a result, it is useful to compute the expected value of the MSE,
\begin{align}
    \EMSE &= \int_{\mathcal{R}} d\bm{u}~\mathbb{E}\left[\left( \F(\bm{u}) - \mathbb{E}[\F(\bm{u})] \right)^2 \right] \nonumber \\
    &= \int_{\mathcal{R}} d\bm{u}~\left( \mathbb{E}[ \F(\bm{u})^2]  - (\mathbb{E}[\F(\bm{u})])^2 \right) \nonumber \\
    &= \int_{\mathcal{R}} d\bm{u}~{\rm Var}[\F(\bm{u})],
\end{align}
so that the expected $\MSE$ is simply equal to the integral of the variance of the unbiased estimator $\F$ over the domain of interest.
 
\subsection{Nonlinear function approximation using a linear phase-preserving amplifier}

\subsubsection{Quantum dynamics of a linear phase-preserving amplifier}

We now describe the specific $\Udec$ interaction that defines the conventional QS protocol for an all-bosonic sensor, which takes the form
\begin{align}
    \Udec = \exp \{-i\hat{\mathcal{H}}_{\rm L} \},
\end{align}
where $\hat{\mathcal{H}}_{\rm L}$ describes a linear phase-preserving quantum amplifier, given by the standard Hamiltonian:
\begin{align}
    \hat{\mathcal{H}}_{\rm L} = i\chi \left( \hat{a}^{\dagger}\hat{b}^{\dagger} - \hat{a}\hat{b} \right),
\end{align}
where $\chi$ defines the amplifier interaction strength.

A simple approach to evaluate expectation values with respect to the quantum state of the phase-preserving linear amplifier is in the Heisenberg representation. In this case, expectation values are always computed with respect to the initial amplifier state, while time evolution is encoded in the dynamics of operators. In this case, the final operator for the amplifier is given in terms of the initial operators:
\begin{align}
    \hat{b} = \sqrt{\mathcal{G}-1}~\hat{a}^{\dagger}_0 + \sqrt{\mathcal{G}}~\hat{b}_0 
    \label{appeq:heisppa}
\end{align}
where $\sqrt{\mathcal{G}} = \cosh \chi$.

For completeness, we note that:
\begin{align}
    \hat{a}^{\dagger} = \sqrt{\mathcal{G}-1}~\hat{b}_0 + \sqrt{\mathcal{G}}~\hat{a}_0^{\dagger} 
\end{align}

\subsubsection{Two-variable function estimation using heterodyne measurement}

For estimation of 2D functions, we must perform heterodyne measurements of the amplifier mode $\hat{b}$ yielding a stochastic outcome $\beta$. Using heterodyne measurement results, we now wish to construct an unbiased estimator $\Fqs$ of the 2D target function $\Ft$ defined by Eq.~(\ref{appeq:ftarget2D}), which we define as:
\begin{align}
    \Fqs = \sum_{n,m = 0}^{D} \mathcal{C}_{nm} (\beta^*)^n\beta^m. 
    \label{appeq:Fhetgen}
\end{align}
Requiring $\Fqs$ to be real enforces $\mathcal{C}_{nm} =  \mathcal{C}_{mn}^*$, and we emphasize that $\mathcal{C}_{nm}$ are allowed to be complex-valued for $n \neq m$.

We start by determining the mean of this estimator, which can be related to intracavity expectation values using Eq.~(\ref{appeq:ehet}):
\begin{align}
    \mathbb{E}[\Fqs] = \sum_{n,m = 0}^{D} \mathcal{C}_{nm} \mathbb{E}[(\beta^*)^n\beta^m]  = \sum_{n,m = 0}^{D} \mathcal{C}_{nm} \avg{\hat{b}^m\hat{b}^{\dagger n}}
\end{align}
To also compute the variance of this unbiased estimator, ${\rm Var}[\Fqs] = \mathbb{E}[\Fqs^2] - \mathbb{E}[\Fqs]^2$, we additionally require the quantity:
\begin{align}
    \mathbb{E}[\Fqs^2] &= \sum_{n,m,r,s = 0}^{N} \mathcal{C}_{nm}\mathcal{C}_{rs}\mathbb{E}[(\beta^*)^n\beta^m(\beta^*)^r\beta^s] = \sum_{n,m,r,s = 0}^{N} \mathcal{C}_{nm}\mathcal{C}_{rs} \mathbb{E}[(\beta^*)^{n+r}\beta^{m+s}] \nonumber \\
    &= \sum_{n,m,r,s = 0}^{N} \mathcal{C}_{nm}\mathcal{C}_{rs} \avg{\hat{b}^{m+s}(\hat{b}^{\dagger})^{n+r}}
\end{align}
Both quantities can be computed with knowledge of the general anti-normal-ordered expectation value $\avg{\hat{b}^m\hat{b}^{\dagger n}}$ for the linear amplifier's readout mode. We now proceed to evaluate this quantity. 

Using Eq.~(\ref{appeq:heisppa}), we can immediately write for the general anti-normal-ordered expectation value:
\begin{align}
    \avg{\hat{b}^m\hat{b}^{\dagger n}} = \avg{ (\sqrt{\mathcal{G}-1}~\hat{a}^{\dagger}_0 + \sqrt{\mathcal{G}}~\hat{b}_0)^m(\sqrt{\mathcal{G}-1}~\hat{a}_0 + \sqrt{\mathcal{G}}~\hat{b}^{\dagger}_0)^n  }
\end{align}
Using the binomial theorem and the fact that $[\hat{a}_0,\hat{b}_0] = 0$, we can simplify the above to the form:
\begin{align}
    \avg{\hat{b}^m\hat{b}^{\dagger n}} &=  \left\langle \left( \sum_{k=0}^m {m \choose k} (\sqrt{\mathcal{G}-1})^k\hat{a}^{\dagger k}_0 (\sqrt{\mathcal{G}})^{m-k} \hat{b}^{m-k}_0 \right) \left( \sum_{l=0}^n {n \choose l} (\sqrt{\mathcal{G}-1})^l\hat{a}^{l}_0 (\sqrt{\mathcal{G}})^{n-l} (\hat{b}_0^{\dagger})^{n-l} \right) \right\rangle \nonumber \\
    &= \left\langle \sum_{k=0}^m\sum_{l=0}^n {m \choose k} {n \choose l} (\sqrt{\mathcal{G}-1})^{k+l}(\sqrt{\mathcal{G}})^{m+n-k-l} \hat{a}^{\dagger k}_0 \hat{a}^{l}_0 \hat{b}^{m-k}_0 (\hat{b}_0^{\dagger})^{n-l}  \right\rangle \nonumber \\
    &= \sum_{k=0}^m\sum_{l=0}^n {m \choose k} {n \choose l} (\sqrt{\mathcal{G}-1})^{k+l}(\sqrt{\mathcal{G}})^{m+n-k-l} \avg{\hat{a}^{\dagger k}_0 \hat{a}^{l}_0\hat{b}^{m-k}_0 (\hat{b}_0^{\dagger})^{n-l}}.
\end{align}
To evaluate the expectation value, we can now make use of Eq.~(\ref{appeq:bosonicexp}). In particular, making use of the fact that:
\begin{align}
    (\hat{b}_0^{\dagger})^n \ket{0} = \sqrt{n!} \ket{n}, 
\end{align}
it is clear that the term $\avg{\hat{b}^{m-k}_0 (\hat{b}_0^{\dagger})^{n-l}}$ will only be nonzero if $m-k = n-l$, namely when the number of annihilation operators exactly equals the number of creation operators. It is then straightforward to determine that:
\begin{align}
    \avg{\hat{b}^{m-k}_0 (\hat{b}_0^{\dagger})^{n-l}} = (n-l)!~\delta_{n-l,m-k}
\end{align}
where $\delta_{k,l}$ is the Kronecker $\delta$-function. This collapses one of the two sums, replacing $l=n-m+k$, allowing us to write:
\begin{align}
    \avg{\hat{b}^m\hat{b}^{\dagger n}} = \sum_{k=0}^m {m \choose k} {n \choose n-m+k} (m-k)!(\sqrt{\mathcal{G}-1})^{n-m+2k}(\sqrt{\mathcal{G}})^{2m-2k} \avg{\hat{a}^{\dagger k}_0 \hat{a}^{n-m+k}_0}. 
\end{align}
Now again making use of Eq.~(\ref{appeq:bosonicexp}), which indicates that the expectation value of mode $\hat{a}$ operators is to be computed with respect to the coherent state $\ket{\alpha}$, we have:
\begin{align}
    \avg{\hat{a}^{\dagger k}_0 \hat{a}^{n-m+k}_0} = \bra{\alpha} \hat{a}^{\dagger k}_0 \hat{a}^{n-m+k}_0 \ket{\alpha} = \alpha^{* k}\alpha^{n-m+k},
\end{align}
so that we finally arrive at:
\begin{align}
    \avg{\hat{b}^m\hat{b}^{\dagger n}} = \sum_{k=0}^m \frac{m!}{k!} {n \choose n-m+k} (\sqrt{\mathcal{G}-1})^{n-m+2k}(\sqrt{\mathcal{G}})^{2m-2k} \alpha^{* k}\alpha^{n-m+k}.
    \label{appeq:momentsQ}
\end{align}
We note that for the above expression to be valid for arbitrary $n,m$, we define the second binomial coefficient term as:
\begin{align}
    {n \choose n-m+k} = 
    \begin{cases}
        \frac{n!}{(m-k)!(n-m+k)!}, &k \geq m-n. \\
        0, &{\rm otherwise}.
    \end{cases}
\end{align}
If $n\geq m$, $k$ is always larger than $m-n$, and so the standard definition of the binomial coefficient holds. If $n<m$, only specific (large enough) $k$ values will survive.

The coefficients $\mathcal{C}_{nm}$ are determined by the requirement that $\Fqs$ must be an unbiased estimator of $\Ft$, $\mathbb{E}[\Fqs] = \Ft$. Enforcing this condition yields:
\begin{align}
    \mathbb{E}[\Fqs] = \sum_{n,m=0}^D \mathcal{C}_{nm} \avg{\hat{b}^m\hat{b}^{\dagger n}} = \sum_{n,m=0}^D \mathcal{W}_{nm} \alpha^{*n}\alpha^m.
\end{align}
Having computed the general expression for the anti-normal-ordered expectation values, we can simplify the above equation:
\begin{align}
    \sum_{n,m=0}^D \mathcal{C}_{nm} \left( \sum_{k=0}^m \frac{m!}{k!} {n \choose n-m+k} (\sqrt{\mathcal{G}-1})^{n-m+2k}(\sqrt{\mathcal{G}})^{2m-2k} \alpha^{* k}\alpha^{n-m+k} \right)   = \sum_{n,m=0}^D \mathcal{W}_{nm} \alpha^{*n}\alpha^m.
    \label{appeq:ppacoeffs}
\end{align}
Both sides of the above equation are polynomials in $\alpha^{*n}\alpha^m$; the coefficients $\mathcal{C}_{nm}$ are therefore determined by the set of equations obtained by comparing coefficients of each such term on both sides. An example of this for the \XOR{} task is analyzed explicitly in due course.

\subsection{Nonlinear function approximation using a nonlinear amplifier}

\subsubsection{Quantum dynamics of a nonlinear amplifier}

The $\Udec$ interaction that defines the QCS protocol for a bosonic nonlinear amplifier is given by
\begin{align}
    \Udec = \exp \{-i\hat{\mathcal{H}}_{\rm NL} \},
\end{align}
where $\hat{\mathcal{H}}_{\rm NL}$ takes the form
\begin{align}
    \hat{\mathcal{H}}_{\rm NL} = -i g \hat{f} ( \hat{b}-\hat{b}^{\dagger})
    \label{appeq:hnl}
\end{align}
where $\hat{f}=\hat{f}^{\dagger}$ is a Hermitian operator defined in the Hilbert space of the sensing mode $\hat{a}$, and $\chi$ is the nonlinear amplifier strength. 

To determine expectation values of the monitored mode of a nonlinear amplifier described by the Hamiltonian given by Eq.~(\ref{appeq:hnl}), it once again proves useful to use the Heisenberg representation, which yields for the final state of the readout mode~\cite{epstein_quantum_2021}:
\begin{align}
    \hat{b} = g \hat{f} + \hat{b}_0
    \label{appeq:nlh}
\end{align}

\subsubsection{Two-variable function estimation using heterodyne measurement}

We again consider heterodyne measurements of the nonlinear amplifier mode $\hat{b}$, yielding a stochastic outcome $X_{\rm QCS} = \beta$, which we use to construct an unbiased estimator $\Fqcs$ of the 2D target function $\Ft$ defined by Eq.~(\ref{appeq:ftarget2D}). The form of the unbiased estimator is given by:
\begin{align}
    \Fqcs = \frac{w}{2} \left( \beta + \beta^* \right)
    \label{appeq:Fhetnl}
\end{align}
The above can be compared against Eq.~(\ref{appeq:Fhetgen}), the unbiased estimator for the linear phase-preserving amplifier. We only allow linear processing of the measurement result for the QCS protocol using the nonlinear amplifier.

\begin{align}
    \mathbb{E}[\Fqcs] = \frac{w}{2}\left( \mathbb{E}[\beta] + \mathbb{E}[\beta^*] \right) = \frac{w}{2} \left( \avg{\hat{b}} + \avg{\hat{b}^{\dagger}} \right) = w g \avg{\hat{f}}  
\end{align}
where we have used $\avg{\hat{b}_0} = 0$. For convenience we will now set $w = \frac{1}{g}$. 

To compute polynomials in powers of $\alpha^{* n}\alpha^m$, we propose the following form for the operator $\hat{f}$:
\begin{align}
    \hat{f} = \sum_{n,m=0} \mathcal{Q}_{nm} \hat{a}^{\dagger n} \hat{a}^m,
\end{align}
where we again require $\mathcal{Q}_{nm} = \mathcal{Q}^*_{mn}$ to enforce that $\hat{f}$ is Hermitian. 

For this specifically-chosen normal-ordered form of $\hat{f}$, its expectation value follows immediately, using Eq.~(\ref{appeq:bosonicexp}):
\begin{align}
    \avg{\hat{f}} = \sum_{n,m=0} \mathcal{Q}_{nm} \avg{\hat{a}^{\dagger n} \hat{a}^m} = \sum_{n,m=0} \mathcal{Q}_{nm} \alpha^{* n}\alpha^m.
    \label{appeq:fnl}
\end{align}
Finally, obtaining the unbiased estimator requires determining the coefficients $\mathcal{Q}_{nm}$, by setting $\mathbb{E}[\Fqcs] = \Ft$,
\begin{align}
    \mathbb{E}[\Fqcs] = \sum_{n,m=0} \mathcal{Q}_{nm} \alpha^{* n}\alpha^m = \sum_{n,m=0} \mathcal{W}_{nm} \alpha^{* n}\alpha^m \implies \mathcal{Q}_{nm} = \mathcal{W}_{nm}.
    \label{appeq:nlacoeffs}
\end{align}
which recovers the simple relation introduced in the main text.

To compute the MSE, we will also require the variance of the unbiased estimator, ${\rm Var}[\Fqcs] = \mathbb{E}[\Fqcs^2]-\mathbb{E}[\Fqcs]^2$. We therefore also require:
\begin{align}
    \mathbb{E}[\Fqcs^2] &= \frac{1}{4g^2} \mathbb{E}[(\beta+\beta^*)^2] = \frac{1}{4g^2} \left( \mathbb{E}[\beta^2] + \mathbb{E}[\beta^{*2}] + 2\mathbb{E}[\beta^*\beta] \right) = \frac{1}{4g^2} \left( \avg{\hat{b}^2} + \avg{\hat{b}^{\dagger 2}} + 2\avg{\hat{b}\hat{b}^{\dagger}}  \right) \nonumber \\
    &= \frac{1}{4g^2}\left( g^2\avg{\hat{f}^2} + g^2\avg{\hat{f}^2} + 2g^2\avg{\hat{f}^2} + 2\avg{\hat{b}_0\hat{b}_0^{\dagger}}  \right) =  \avg{\hat{f}^2} + \frac{1}{2g^2} 
\end{align}
where we have again used $\avg{\hat{f}\hat{b}_0} = \avg{\hat{f}}\avg{\hat{b}_0} = 0$, $\avg{\hat{b}_0^2}=0$, and $\avg{\hat{b}_0\hat{b}_0^{\dagger}} = 1$. Therefore, we must also compute $\avg{\hat{f}^2}$:
\begin{align}
    \avg{\hat{f}^2} = \sum_{n,m=0}^D\sum_{r,s=0}^D \mathcal{Q}_{nm}\mathcal{Q}_{rs} \avg{\hat{a}^{\dagger n} \hat{a}^m\hat{a}^{\dagger r} \hat{a}^s}
\end{align}
The above can once again be easily evaluated provided we are able to normal-order the required expectation value. To do so, we use the relationship~\cite{deepak_general_2023}:
\begin{align}
     \hat{a}^m\hat{a}^{\dagger r} = \sum_{l=0}^{{\rm min}(m,r)} \frac{m!r!}{l!(m-l)!(r-l)!} \hat{a}^{\dagger(r-l)}\hat{a}^{m-l}
     \label{appeq:ordera}
\end{align}
Then, using Eq.~(\ref{appeq:bosonicexp}) one more time,
\begin{align}
    \avg{\hat{f}^2} = \sum_{n,m,r,s=0}^D \mathcal{Q}_{nm}\mathcal{Q}_{rs}  \sum_{l=0}^{{\rm min}(m,r)} \avg{\hat{a}^{\dagger n} \hat{a}^{\dagger(r-l)}\hat{a}^{m-l} \hat{a}^s} = \sum_{n,m,r,s=0}^D \mathcal{Q}_{nm}\mathcal{Q}_{rs}  \sum_{l=0}^{{\rm min}(m,r)} \alpha^{*(n+r-l)}\alpha^{m+s-l}.
    \label{appeq:f2nl}
\end{align}
Eqs.~(\ref{appeq:fnl}) and (\ref{appeq:f2nl}) together allow us to compute the variance of the unbiased estimator $\Fqcs$ for an arbitrary $\Ft$.

\subsection{Worked example: \XOR{} task}
\label{app:xor}

For the \XOR{} task we wish to construct the target function $\Ft$:
\begin{align}
    \Ft = -i\alpha^2 + i\alpha^{*2} 
    \label{appeq:xortarget}
\end{align}
In this subsection, we will work through an example of constructing an unbiased estimator of the above target function for both the conventional QS approach using a phase-preserving linear quantum amplifier, and QCS approach using the quantum nonlinear amplifier. Writing the target polynomial in the form defined by Eq.~(\ref{appeq:ftarget2D}), the degree $D$ required is $D=2$, and the coefficients $\mathcal{W}_{nm}$ are summarized in Table~\ref{tab:xortarget}.

\begin{table}[h]
    \centering
    \begin{tabular}{c||c|c|c}
         $\mathcal{W}_{nm}$ & $m\!=\!0$ & $m\!=\!1$ & $m\!=\!2$  \\
         \hline 
         \hline
         $n\!=\!0$ & 0 & 0 & $-i$  \\ 
         \hline
         $n\!=\!1$ & 0 & 0 & 0  \\ 
        \hline
         $n\!=\!2$ & $i$ & 0 & 0  \\ 
    \end{tabular}
    \caption{Coefficients $\mathcal{W}_{nm}$ of the target polynomial for the \XOR{} classification task, given by Eq.~(\ref{appeq:xortarget}), and considered in Fig.~\ref{fig:bosonic}c of the main text. }
    \label{tab:xortarget}
\end{table}

\subsubsection{Conventional QS approach: phase-preserving linear amplifier}

For the \XOR{} task using a phase-preserving linear amplifier, the classical postprocessing coefficients $\mathcal{C}_{nm}$ to be applied to heterodyne measurement results are determined using Eq.~(\ref{appeq:ppacoeffs}). Writing this equation out explicitly,
\begin{align}
    &\left( \mathcal{C}_{00} + \mathcal{G}\mathcal{C}_{11} + 2\mathcal{G}^2\mathcal{C}_{22}\right) + \left[ (\mathcal{G}-1)\mathcal{C}_{11} + 4\mathcal{G}(\mathcal{G}-1)\mathcal{C}_{22} \right]|\alpha|^2 + (\mathcal{G}-1)^2\mathcal{C}_{22}|\alpha|^4 + \nonumber \\
    &\left[ \left(\sqrt{\mathcal{G}-1}~\mathcal{C}_{01} + 2\mathcal{G}\sqrt{\mathcal{G}-1}~\mathcal{C}_{12} \right)\alpha^* + (\mathcal{G}-1)\mathcal{C}_{02}\alpha^{*2} +  (\mathcal{G}-1)\sqrt{\mathcal{G}-1}~\mathcal{C}_{12} \alpha^{*2}\alpha + c.c. \right] = -i \alpha^2 + i \alpha^{* 2},
\end{align}
where $c.c.$ indicates complex conjugation of the first set of terms in square brackets. Comparing coefficients of terms on each side yields $(D+1)^2$ equations in total, which define a matrix system for $\mathcal{C}_{nm}$. Note that the requirement $\mathcal{C}_{nm} = \mathcal{C}_{mn}^*$ ensures that there are only 6 unique coefficients: $\mathcal{C}_{00},\mathcal{C}_{11},\mathcal{C}_{22},\mathcal{C}_{01},\mathcal{C}_{12},\mathcal{C}_{02}$. For the simple polynomial under consideration here, the comparison can be performed directly. Comparing coefficients of $\alpha^{* 2}$ immediately yields $\mathcal{C}_{02} = +\frac{i}{\mathcal{G}-1}$. It is easy to see that all other coefficients vanish, $\mathcal{C}_{00} = \mathcal{C}_{11} = \mathcal{C}_{22} = \mathcal{C}_{01} =\mathcal{C}_{12} = 0$. The required coefficients are summarized in Table~\ref{tab:xorqs}.


\begin{table}[h]
    \centering
    \begin{tabular}{c||c|c|c}
         $\mathcal{C}_{nm}\beta^{* n}\beta^m$ & $m\!=\!0$ & $m\!=\!1$ & $m\!=\!2$  \\
         \hline 
         \hline
         $n\!=\!0$ & 0 & 0 & $+\frac{i}{\mathcal{G}-1}$  \\ 
         \hline
         $n\!=\!1$ & 0 & 0 & 0  \\ 
        \hline
         $n\!=\!2$ & $-\frac{i}{\mathcal{G}-1}$ & 0 & 0  \\ 
    \end{tabular}
    \caption{Coefficients $\mathcal{C}_{nm}$ for classical nonlinear postprocessing of phase-preserving amplifier heterodyne output for the \XOR{} classification task considered in Fig.~\ref{fig:bosonic}d of the main text. Note that $\mathcal{C}_{nm} = \mathcal{C}_{mn}^*$. }
    \label{tab:xorqs}
\end{table}


Using Eq.~(\ref{appeq:Fhetgen}), the unbiased estimator is therefore given by 
\begin{align}
    \Fqs = \frac{-i}{\mathcal{G}-1} \beta^{* 2} + \frac{i}{\mathcal{G}-1} \beta^{2}.
\end{align}
To verify this construction, we can compute $\mathbb{E}[\Fqs]$ explicitly:
\begin{align}
    \mathbb{E}[\Fqs] = \frac{-i}{\mathcal{G}-1} \mathbb{E}[\beta^{* 2}] + \frac{i}{\mathcal{G}-1} \mathbb{E}[\beta^{2}]= \frac{-i}{\mathcal{G}-1} \avg{\hat{b}^{\dagger 2}} + \frac{i}{\mathcal{G}-1} \avg{\hat{b}^{2}} 
\end{align}
To simplify the above further, we use Eq.~(\ref{appeq:momentsQ}) to write:
\begin{align}
    \avg{\hat{b}^{\dagger 2}} = (\sqrt{\mathcal{G}-1})^2 \alpha^2 ,
\end{align}
which yields
\begin{align}
    \mathbb{E}[\Fqs] &= \frac{-i}{\mathcal{G}-1} \left( (\sqrt{\mathcal{G}-1})^2 \alpha^2  \right) + \frac{i}{\mathcal{G}-1} \left( (\sqrt{\mathcal{G}-1})^2 \alpha^{* 2}  \right) = -i \alpha^2 +  i\alpha^{*2} = \Ft
\end{align}
as required.

This analysis is incomplete without also computing the variance of this unbiased estimator, which determines the error in the function approximation. To compute the variance, we require:
\begin{align}
    \mathbb{E}[\Fqs^2] &= \mathbb{E}[ \mathcal{C}_{02}^2\beta^{* 4} + \mathcal{C}_{20}^2  \beta^{4} + 2 \mathcal{C}_{02}\mathcal{C}_{20} \beta^{* 2}\beta^2] \nonumber \\
    &= \mathcal{C}_{02}^2\mathbb{E}[\beta^{* 4}] + \mathcal{C}_{20}^2 \mathbb{E}[\beta^{4}] + 2 \mathcal{C}_{02}\mathcal{C}_{20} \mathbb{E}[\beta^{* 2}\beta^2] \nonumber \\
    &= \mathcal{C}_{20}^2 \avg{\hat{b}^{\dagger 4}} + \mathcal{C}_{02}^2 \avg{\hat{b}^{4}} + 2 \mathcal{C}_{02}\mathcal{C}_{20} \avg{\hat{b}^{\dagger 2}\hat{b}^{\dagger 2}}
\end{align}
The required anti-normal-ordered expectation values can be computed using Eq.~(\ref{appeq:momentsQ}) again:
\begin{align}
    \avg{\hat{b}^{\dagger 4}} &= (\mathcal{G}-1)^2\alpha^4 \nonumber \\
    \avg{\hat{b}^{\dagger 2}\hat{b}^{\dagger 2}} &= 4 \sum_{k=0}^2 \frac{1}{(k!)^2(2-k)!} (\sqrt{\mathcal{G}-1})^{2k}(\sqrt{\mathcal{G}})^{4-2k}|\alpha|^{2k} \nonumber \\
    &= 2\mathcal{G}^2 + 4\mathcal{G}(\mathcal{G}-1)|\alpha|^2 + (\mathcal{G}-1)^2|\alpha|^4
\end{align}
We therefore find:
\begin{align}
    \mathbb{E}[\Fqs^2] = -\alpha^4 - \alpha^{* 4} + \frac{2}{(\mathcal{G}-1)^2} \left( 2\mathcal{G}^2 + 4\mathcal{G}(\mathcal{G}-1)|\alpha|^2 + (\mathcal{G}-1)^2|\alpha|^4 \right)
\end{align}
Furthermore, using $(\mathbb{E}[\Fqs])^2 = -\alpha^4 - \alpha^{* 4} + 2|\alpha|^4$, we can finally calculate the variance of $\Fqs$,
\begin{align}
    {\rm Var}[\Fqs] = \mathbb{E}[\Fqs^2] - \mathbb{E}[\Fqs]^2 = \frac{2}{(\mathcal{G}-1)^2} \left( 2\mathcal{G}^2 + 4\mathcal{G}(\mathcal{G}-1)|\alpha|^2 \right).
\end{align}
In the large gain limit $\mathcal{G} \gg 1$, the variance reduces to:
\begin{align}
    {\rm Var}[\Fqs] \stackrel{\mathcal{G} \gg 1}{\simeq} 4 + 8 |\alpha|^2
    \label{appeq:qsxorvar}
\end{align}

\subsubsection{QCS approach: nonlinear amplifier}

For the QCS approach using a nonlinear amplifier, we must determine the nonlinear interaction coefficients $\mathcal{Q}_{nm}$; as derived in Eq.~(\ref{appeq:nlacoeffs}), these are simply related to $\mathcal{C}_{nm}$, as summarized in Table~\ref{tab:xorqcs}. 

\begin{table}[h]
    \centering
    \begin{tabular}{c||c|c|c}
         $\mathcal{Q}_{nm}\hat{a}^{\dagger n}\hat{a}^m$ & $m\!=\!0$ & $m\!=\!1$ & $m\!=\!2$  \\
         \hline 
         \hline
         $n\!=\!0$ & 0 & 0 & $-i$  \\ 
         \hline
         $n\!=\!1$ & 0 & 0 & 0  \\ 
        \hline
         $n\!=\!2$ & $i$ & 0 & 0  \\ 
    \end{tabular}
    \caption{Coefficients $\mathcal{Q}_{nm}$ for quantum nonlinear processing using a nonlinear amplifier for the \XOR{} classification task considered in Fig.~\ref{fig:bosonic}(c) of the main text. Note that $\mathcal{Q}_{nm} = \mathcal{Q}_{mn}^*$. Also, the target matrix coefficients are exactly equal to the quantum nonlinear processing coefficients, $\mathcal{W}_{nm} = \mathcal{Q}_{nm}$. }
    \label{tab:xorqcs}
\end{table}

These coefficients mean the operator $\hat{f}$ takes the form:
\begin{align}
    \hat{f} = -i \hat{a}^2 + i\hat{a}^{\dagger 2}.
\end{align}
We can verify the construction of the resulting unbiased estimator $\Fqcs$ explicitly:
\begin{align}
    \mathbb{E}[\Fqcs] = \avg{\hat{f}} = -i \avg{\hat{a}^2} + i\avg{\hat{a}^{\dagger 2}} = -i\alpha^2 + i\alpha^{*2} = \Ft,
\end{align}
as required.

To compute the variance of $\Fqcs$, 
\begin{align}
    \mathbb{E}[\Fqcs^2] &= \avg{\hat{f}^2} + \frac{1}{2g^2} = \frac{1}{2g^2} + \langle \left( -\hat{a}^4 - \hat{a}^{\dagger 4} + \hat{a}^2\hat{a}^{\dagger 2} + \hat{a}^{\dagger 2}\hat{a}^{2} \right) \rangle \nonumber \\
    &= \frac{1}{2g^2} - \avg{\hat{a}^4} - \avg{\hat{a}^{\dagger 4}} + \avg{\hat{a}^{\dagger 2}\hat{a}^{2}} + \avg{\hat{a}^2\hat{a}^{\dagger 2} } \nonumber \\
    &= \frac{1}{2g^2} - \alpha^4 - \alpha^{* 4} + |\alpha|^4 +\avg{\hat{a}^2\hat{a}^{\dagger 2} }
\end{align}
To simplify the above, we can use Eq.~(\ref{appeq:ordera}), 
\begin{align}
    \hat{a}^2\hat{a}^{\dagger 2} = \sum_{l=0}^{2} \frac{2!~2!}{l!(2-l)!(2-l)!} \hat{a}^{\dagger (2-l)}\hat{a}^{(2-l)} = \hat{a}^{\dagger 2}\hat{a}^2 + 4\hat{a}^{\dagger}\hat{a} + 2 
\end{align}
Using this we arrive at:
\begin{align}
    \mathbb{E}[\Fqcs^2] = \frac{1}{2g^2} - \alpha^4 - \alpha^{* 4} + 2|\alpha|^4 + 4|\alpha|^2 + 2
\end{align}
It can be verified that the above gives the same result as Eq.~(\ref{appeq:f2nl}). Using $\mathbb{E}[\Fqcs]^2 = -\alpha^{4} -\alpha^{* 4} + 2|\alpha|^4$, the variance of $\Fqcs$ is finally given by:
\begin{align}
    {\rm Var}[\Fqcs] = \mathbb{E}[\Fqcs^2] - \mathbb{E}[\Fqcs]^2 = \frac{1}{2g^2} + 4|\alpha|^2 + 2
\end{align}
Again in the large gain limit, $g \gg 1$, we arrive at:
\begin{align}
    {\rm Var}[\Fqcs] \stackrel{g \gg 1}{\simeq} 2 + 4|\alpha|^2
    \label{appeq:qcsxorvar}
\end{align}
For the target polynomial required for this \XOR{} task, comparing Eqs.~(\ref{appeq:qsxorvar}),~(\ref{appeq:qcsxorvar}) immediately implies that ${\rm Var}[\Fqs] = 2{\rm Var}[\Fqcs]$: the QCS protocol reduces the error in function approximation in comparison to the QS protocol by a factor of two. The extent of the advantage can depend on $\Ft$, as shown in Fig.~\ref{fig:bosonic} of the main text.

\subsection{QS and QCS protocols: 1D polynomial approximation}

As we have considered a large number of 1D polynomial approximation tasks, listing all the required coefficients is slightly cumbersome; these have instead been provided in the form of \texttt{Python} code, which can also be used to reproduce the expected mean-squared-error in their approximation using both protocols (see Data and Code Availability).

\subsection{QS and QCS protocols: Higher-order \spirals{} task}

In Table~\ref{tab:spiralsqcs} we provide the quantum nonlinear processing coefficients $\mathcal{Q}_{nm}$ required for the Higher-order \spirals{} classification task in the main text. The classical postprocessing coefficients for a linear phase-preserving amplifier for the same task are included in Table~\ref{tab:spiralsqs}.

\begin{turnpage}

\begin{table}[h]
    \centering
    \begin{tabular}{c||c|c|c|c|c}
         $\mathcal{Q}_{nm}\hat{a}^{\dagger n}\hat{a}^m$ & $m\!=\!0$ & $m\!=\!1$ & $m\!=\!2$ & $m\!=\!3$ & $m\!=\!4$    \\
         \hline 
         \hline
         $n\!=\!0$ & $-4.3$ & $1.5-170i$ & $0.85-0.20i$ & $2.0-13i$ & $-0.19+0.23i$ \\ 
         \hline
         $n\!=\!1$ & $1.5+170i$ & $-0.069$ & $-37-27i$ & $0.36+0.080i$ & $-0.70+0.80i$ \\ 
        \hline
         $n\!=\!2$ & $0.85+0.20i$ & $-37+27i$ & $-0.039$ & $5.0+11i$ & $-0.0060-0.065i$ \\ 
        \hline
         $n\!=\!3$ & $2.0+13i$ & $0.36-0.080i$ & $5.0-11i$ & $-0.076$ & $-0.10-0.43i$ \\ 
         \hline
         $n\!=\!4$ & $-0.19-0.23i$ & $-0.70-0.80i$ & $-0.006+0.065i$ & $-0.10+0.43i$ & $0.0059$ \\ 
    \end{tabular}
    \caption{Coefficients $\mathcal{Q}_{nm}$ for quantum nonlinear processing using a nonlinear amplifier for the \spirals{} classification task considered in Fig.~\ref{fig:bosonic}(d) of the main text. Note that $\mathcal{Q}_{nm} = \mathcal{Q}_{mn}^*$. Also, the target matrix coefficients are exactly equal to the quantum nonlinear processing coefficients, $\mathcal{W}_{nm} = \mathcal{Q}_{nm}$. }
    \label{tab:spiralsqcs}
    \vspace{25mm}
    \centering
    \begin{tabular}{c||c|c|c|c|c}
         $\mathcal{C}_{nm}\beta^{* n}\beta^m$ & $m\!=\!0$ & $m\!=\!1$ & $m\!=\!2$ & $m\!=\!3$ & $m\!=\!4$    \\
         \hline 
         \hline
         $n\!=\!0$ & $\frac{-3.8 \mathcal{G}^4+17\mathcal{G}^3-26\mathcal{G}^2+17\mathcal{G}-4.3}{(\mathcal{G}-1)^4}$ & $\frac{\mathcal{G}^3 (39i+110)+\mathcal{G}^2 (-340i-180)+\mathcal{G} (460i+78)-170i-1.5}{(\mathcal{G}-1)^{7/2}}$ & $\frac{\mathcal{G}^2 (1.2i-0.31)+\mathcal{G} (-0.63i-0.62)+0.20i+0.85}{(\mathcal{G}-1)^3}$ & $\frac{\mathcal{G} (17i+4.6)-13i-1.8}{(\mathcal{G}-1)^{5/2}}$ & $\frac{-0.23i-0.19}{(\mathcal{G}-1)^2}$ \\
         \hline
         $n\!=\!1$ & - & $\frac{-1.9 \mathcal{G}^3+1.3 \mathcal{G}^2-0.051 \mathcal{G}+0.069}{(\mathcal{G}-1)^4}$ & $\frac{\mathcal{G}^2 (110i-71)+\mathcal{G} (100-120i)+27i-37}{(\mathcal{G}-1)^{7/2}}$ & $\frac{\mathcal{G} (0.41-0.60i)+0.079i-0.36}{(\mathcal{G}-1)^3}$ & $\frac{-0.83i-0.72}{(\mathcal{G}-1)^{5/2}}$ \\ 
        \hline
         $n\!=\!2$ & - & - & $\frac{1.1 \mathcal{G}^2-0.61 \mathcal{G}-0.039}{(\mathcal{G}-1)^4}$ & $\frac{\mathcal{G} (6.4-17i)+11i-5.1}{(\mathcal{G}-1)^{7/2}}$ & $\frac{0.065i-0.0063}{(\mathcal{G}-1)^3}$ \\ 
        \hline
         $n\!=\!3$ & - & - & - & $\frac{0.076-0.17 \mathcal{G}}{(\mathcal{G}-1)^4}$ & $\frac{0.43i-0.10}{(\mathcal{G}-1)^{7/2}}$ \\ 
         \hline
         $n\!=\!4$ & - & - & - & - & $\frac{0.0059}{(\mathcal{G}-1)^4}$ \\ 
    \end{tabular}
    \caption{Coefficients $\mathcal{C}_{nm}$ for classical nonlinear postprocessing of phase-preserving amplifier heterodyne output for the \spirals{} classification task considered in Fig.~\ref{fig:bosonic}(d) of the main text. For clarity, only coefficients for $m \geq n$ are shown; the remaining coefficients are simply related to these via conjugation, as $\mathcal{C}_{nm} = \mathcal{C}_{mn}^*$.   }
    \label{tab:spiralsqs}
\end{table}
    
\end{turnpage}

\clearpage

\newpage

\section{QCS using hybrid quantum computational sensors}
\label{app:bqsp}

In this Appendix section we provide additional details of the hybrid quantum computational sensors considered in Sec.~\ref{sec:hybrid} of the main text, including explicit forms of the operations used and information about Hilbert space truncation.

\subsection{Architecture for hybrid qubit-cavity quantum computational sensors}

In Fig.~\ref{appfig:bqsp}, we present the structure of the hybrid qubit-bosonic mode quantum computational sensor analyzed in Sec.~\ref{sec:hybrid} of the main text. 


\begin{figure}[h!]
    \centering
    \includegraphics{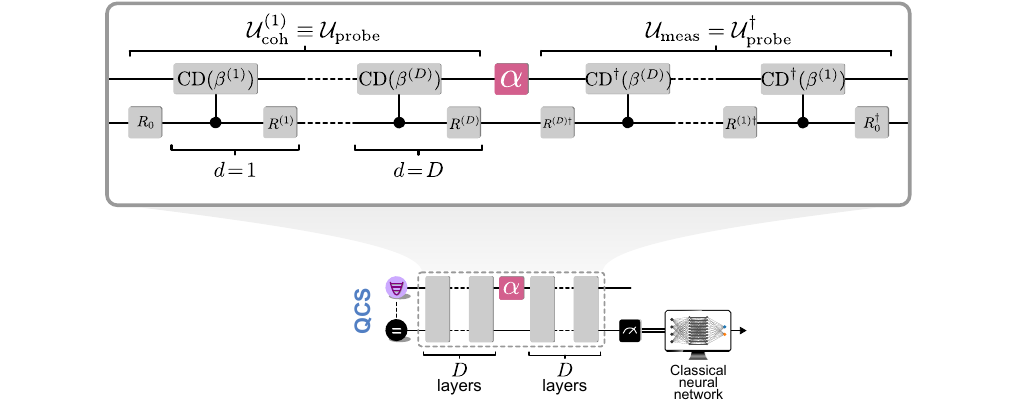}
    \caption{\textbf{Hybrid qubit-cavity quantum computational sensor architecture.} All gray blocks describe trainable operations, whose precise definitions are included in Eqs.~(\ref{appeq:bqspCD}),~(\ref{appeq:bqspR}). Note that only a single sensing operation is performed prior to measurement; this QCS protocol is therefore defined by $L=1$ in Eq.~(\ref{eq:cohprocessing}) of the main text. }
    \label{appfig:bqsp}
\end{figure}


Before describing this architecture in detail, we introduce the sensing interaction $\Usense$ for the sensing of displacements, $\bm{u} \to \alpha = \alpha_x + i\alpha_y$, which takes the form:
\begin{align}
    \Usense(\alpha) = \exp \{ -i \alpha_x(\hat{a}+\hat{a}^{\dagger})  -i \alpha_y(\hat{a}-\hat{a}^{\dagger}) \}.
    \label{appeq:usensebqsp}
\end{align}
Note that this interaction is the same as the sensing interaction used for sensors comprising bosonic modes only, Eq.~(\ref{appeq:usensebosonic}).

The QCS protocol takes the form given by Eq.~(\ref{eq:cohprocessing}), as always. The quantum computational sensor is initialized in the state $\ket{0,g}$, namely with the cavity in vacuum and the qubit in the ground state. We define:
\begin{align}
    \Uenc = \left( \prod^D_{d=1} R^{(d)}{\rm CD}(\beta^{(d)}) \right) R^{(0)},
\end{align}
where $d$ indexes the layer of the encoding unitary operation.

The operator $R^{(d)}$ describes general single qubit rotations around a trainable axis defined by the unit vector $\bm{n}^{(d)} = \begin{psmallmatrix} 
\sin \theta^{(d)} \sin \phi^{(d)} \\ 
\sin \theta^{(d)} \cos \phi^{(d)} \\
\cos\theta^{(d)} 
\end{psmallmatrix}$; defining the vector of Pauli operators as $\hat{\bm{\sigma}} = \begin{psmallmatrix}
    \hat{\sigma}^x \\
    \hat{\sigma}^y \\
    \hat{\sigma}^z 
\end{psmallmatrix}$, the general rotation operator takes the standard form:
\begin{align}
    R^{(d)} \equiv R(\Omega^{(d)},\theta^{(d)},\phi^{(d)}) &= \exp\left\{ -i \frac{\Omega^{(d)}}{2}\bm{n}^{(d)} \cdot \hat{\bm{\sigma}} \right\} \nonumber \\
    &= \exp\left\{ -i \frac{\Omega^{(d)}}{2} \left( \sin \theta^{(d)} \sin \phi^{(d)} \hat{\sigma}^x +  \sin \theta^{(d)} \cos \phi^{(d)} \hat{\sigma}^y + \cos\theta^{(d)} \hat{\sigma}^z \right)  \right\}
    \label{appeq:bqspR}
\end{align}

The operator ${\rm CD}^{(d)}$ defines a displacement of the bosonic mode conditioned on the state of the qubit:
\begin{align}
     {\rm CD}^{(d)} \equiv {\rm CD}(x^{(d)},y^{(d)}) = \exp\left\{ -i \chi_{\rm d} \hat{\sigma}^z \left( x^{(d)} \hat{q} + y^{(d)} \hat{p} \right) \right\}
    \label{appeq:bqspCD}
\end{align}
where we have defined the standard bosonic quadrature operators $\hat{q} = \frac{1}{\sqrt{2}}\left(\hat{a} + \hat{a}^{\dagger}\right)$, $\hat{p} = -\frac{i}{\sqrt{2}}\left( \hat{a} -\hat{a}^{\dagger} \right)$, and $\chi_{\rm d}$ is the strength of the qubit-bosonic mode coupling. In the absence of losses (as is the case here), $\chi_{\rm d}$ is simply a scaling factor for the conditional displacement values.

Following the action of $\Uenc$, the physical signal is sensed. We note again that for this protocol we choose $L=1$, so that the signal is sensed only once prior to measurement, and the single coherent processing operation is equivalent to a probe state preparation operation, $\Ucoh{1} \equiv \Uenc$. Post-sensing, we apply the measurement basis preparation unitary $\Udec$, which here takes the particularly simple form $\Udec = \Uenc^{\dagger}$, and perform a measurement of the qubit in the computational basis to conclude a single shot of the protocol. Recall that in the main text we use this protocol for single-shot sensing, so that repeated access to the displacement $\alpha$ is not assumed.

The set of trainable parameters for the qubit rotations is therefore $\{\Omega^{(d)},\theta^{(d)},\phi^{(d)}\}$, for $d=0,1,\ldots,D$, while the set of trainable parameters for the conditional displacement gates is $\{x^{(d)},y^{(d)}\}$ for $d=1,\ldots,D$, leading to a total of $3\times (D+1) + 2\times D = 5D + 3$ trainable parameters for a protocol of depth $D$.

We note that the architecture here is in part inspired by the bosonic QSP protocol introduced in Ref.~\cite{sinanan-singh_single-shot_2024}, making use of interleaved single qubit rotations and conditional displacements in $\Uenc$. However, there are a couple of key differences. First, we consider completely arbitrary single-qubit rotations. Secondly, and much more importantly, we allow for conditional displacements along \textit{both} $\hat{q}$ and $\hat{p}$ quadratures. This ensures the measured qubit probability can be a nontrivial function of both $\alpha_x$ and $\alpha_y$, which is necessary for the bivariate classification tasks we consider in the main text. Finally, as we will discuss next, the trainable parameters are all optimized via supervised learning on a training dataset, unlike Ref.~\cite{sinanan-singh_single-shot_2024} where the optimization is for specific, one-dimensional target functions.

\subsection{Hilbert space truncation}

An important consideration specific to the training of quantum computational sensors comprising modes with infinite dimensional Hilbert spaces (e.g. bosonic modes) is to guard against numerical artefacts that may arise due to Hilbert space truncation effects. We ensure that such truncation effects do not affect our results using several safeguards.

First, we use relatively large sensing displacements $\alpha$, which equivalently mean that conditional displacements of a relatively smaller amplitude are required (inspired by the BQSP protocol~\cite{sinanan-singh_single-shot_2024}). We still use a relatively large fixed Hilbert space truncation of $N_{\rm Fock} = 70$ for our simulations, but find that the average occupation number of the quantum computational sensor (prior to the sensing operation) remains low (below 5). Secondly, we include a penalty term in the loss function that becomes large if statevector elements at large Fock state occupation numbers start to become nontrivially occupied (see Appendix~\ref{app:train}); we have observed that this guides the trained QCS protocols away from parameter regimes that yield large bosonic mode occupations. Finally, and most importantly, we verify all our trained models by simulating their results with a larger Hilbert space cutoff than that used during training. The invariance of the results with increasing cutoff number ensures that our results are not limited by any truncation errors.

\subsection{Comparing hybrid quantum computational sensors against phase-preserving amplifiers}

In this subsection, we compare the hybrid QCS protocols analyzed in Sec.~\ref{sec:hybrid} of the main text to a simpler conventional QS protocol for displacement sensing, again starting with the \circles{} classification task depicted in Fig.~\ref{appfig:bqsp}a. This protocol, which is depicted in Fig.~\ref{appfig:bqsp}b, differs from the TMS protocol in that no probe state preparation step precedes the single sensing operation, and the sensing operation is followed by a linear phase-preserving quantum amplifier (this protocol is the same as that used in Sec.~\ref{sec:bosonic} of the main text). Heterodyne measurement of the bosonic mode $\hat{b}$ allows estimation of the complex-valued displacement: in particular the average measurement results $x_{\rm QS}$ are linearly-related to the real and imaginary parts of the sensed displacement, as depicted in Fig.~\ref{appfig:bqsp}b; in particular, ${\rm Re}\{x_{\rm QS} \} \propto \sqrt{\mathcal{G}-1}~\alpha_x$ and ${\rm Im}\{x_{\rm QS} \} \propto -\sqrt{\mathcal{G}-1}~\alpha_y$. Therefore, heterodyne measurement of the mode $\hat{b}$ can be used to obtain an estimate of the displacement $\alpha$, just like the TMS-based conventional QS protocol. However, relaxing the requirement of preparing an entangled probe state makes such a protocol easier to implement in experiment; it is regularly used for field quadrature readout. 

The stochastic measurement results $X_{\rm QS}$ are passed through a classical neural network that is trained to predict the class label using supervised learning. For the circles task, the output of the classical neural network is depicted in the right panel of Fig.~\ref{appfig:bqsp}b. The classification error is now about 28\%, more than twice the error of the TMS-based QS protocol. The QCS protocol we use is the same as in Sec.~\ref{sec:hybrid} of the main text, and has a much lower error as shown in Fig.~\ref{appfig:bqsp}c.

As with qubit-based schemes, the extent of QCSA depends not only on the complexity of the quantum computational sensor but also on the task itself. We consider instances of the \spirals{} dataset and compare the performance of both protocols in Fig.~\ref{appfig:bqsp}d. The first \spirals{} instance is a simpler task and the conventional QS protocol is able to perform it using a lower classification error than the other tasks. The quantum computational sensor can be trained to perform this task and eventually outperform the QS protocol. However, due to the good performance of the latter, a larger value of $d$ is required. The second task is a slightly more complex instance of the \spirals{} task, which leads to worse QS performance. The \circles{} task turns out to yield the largest classification error for the QS protocol out of the tasks considered here. On the other hand, the quantum computational sensing protocol reaches a much lower error with a small value of $D$.


\begin{figure}
    \centering
    \includegraphics{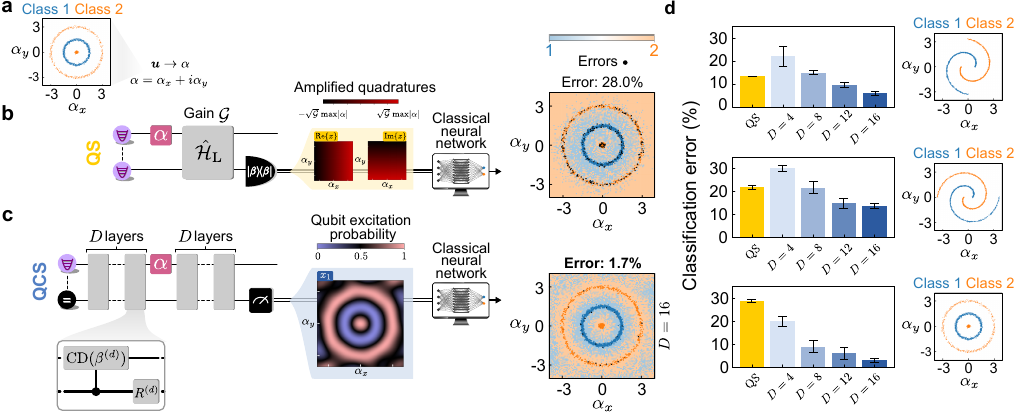}
    \caption{\textbf{Quantum computational sensing using a hybrid qubit-bosonic mode sensor.} \textbf{a}, Example of the binary classification tasks analyzed, here an instance of the \circles{} dataset. \textbf{b}, Conventional bosonic quantum sensor for sensing a complex-valued displacement $\alpha$, same as Fig.~\ref{fig:bosonic}b. \textbf{c}, Quantum computational sensor realized by coupling the sensing bosonic mode to a single qubit, whose state is read out and passed through a classical postprocessor, here a neural network. \textbf{d}, Comparison of QS and QCS performance, the latter for varying $D$, on the various binary classification tasks shown. The extent of QCSA can vary with the task, owing to both the complexity of the task for the conventional QS approach, as well as its ease for the QCS approach. }
    \label{appfig:bqsp}
\end{figure}


\newpage

\section{General training details}
\label{app:train}

In this Appendix section we include some training details that are common to the various QCS protocols trained in this work. We also discuss some general findings about training using finitely-sampled measurement results.

\subsection{Trainable models, classical postprocessing, and loss functions}

\subsubsection{Differentiable models and measurement results}

Differentiable models for all the trainable quantum computational sensors used in this work are written in \texttt{PyTorch}~\cite{paszke2019pytorchimperativestylehighperformance}, and have been made publicly available (see Data and Code availability). Starting from an initial state, the models perform evolution of the quantum state for gate-based dynamics; individual models for the various quantum computational sensors are provided in the corresponding Appendices. We have tested the evolution of our models against exact \texttt{QuTiP}~\cite{johansson_qutip_2013} simulations to verify the accuracy of our code. 

The output of the quantum computational sensor is taken to be obtained from finite measurements, instead of training directly on elements of the final quantum state (or equivalently, density matrix). The result of single measurement for a given input $\bm{u}$ is given by $\bm{X}_{\rm QCS}(\bm{u})$, and the average over $S$ measurements for the same input is defined as $\bm{\bar{X}}_{\rm QCS}(\bm{u}) = \sum_s^S \bm{X}_{\rm QCS}^{s}(\bm{u})$ where the plain superscript $s$ indexes independent shots (to avoid overloading of notation, we do not specify this label $s$ for measurements in what follows). To incorporate sampling and hence the role of quantum sampling noise in simulations, our models incorporate a sampling layer to describe measurement. This layer simulates the action of appropriate POVMs on the final quantum state of the quantum computational sensor, to simulate the execution of a finite number of measurements $S$. 

To train these models, gradients with respect to trainable parameters are calculated using backpropagation. An important factor to note here is that the aforementioned sampling layer is not in-principle a differentiable operation. To allow gradient calculation in the presence of the sampling layer, we therefore use a ``Straight-through'' estimator, which sets the gradient of the sampling layer to be the identity map. We have found that the use of this straight-through estimator has allowed us to successfully train the various QCS protocols analyzed in this paper.

\subsubsection{Training datasets and loss functions}

We consider multi-class discrimination tasks, so each input $\bm{u}$ to the quantum computational sensor is indexed b
Training data point is given by a sensed signal $\bm{u}_{(j)}^{(n)}$, with two indices. The subscript in parentheses denotes the class label $j$ of the input, where $j \in 1,\ldots, C$ for $C$-class discrimination tasks. The superscript in parentheses denotes the sample within this class, where $n \in 1,\ldots,N_{\rm samp}$. For each input $\bm{u}_{(j)}^{(n)}$, corresponding measurement results $\bm{\bar{X}}_{\rm QCS}(\bm{u}_{(j)}^{(n)})$ are obtained from the quantum computational sensor. This allows us to compile the full training dataset which consists of pairs of inputs and quantum system outputs,
\begin{align}
    \mathcal{D}_{\bm{u}} = \Big\{ \left[ \bm{u}_{(j)}^{(n)}, \bm{\bar{X}}_{\rm QCS}(\bm{u}_{(j)}^{(n)})\right] \Big\}_{n \in 1,\ldots,N_{\rm samp},~j\in 1,\ldots,C}
\end{align}
The total number of training data pairs in the training set is therefore $C\times N_{\rm samp}$.

We now define the loss function $\mathcal{L}$ that is to be minimized during the training procedure. We have considered two types of loss functions:
\begin{enumerate}[(1),noitemsep,leftmargin=3\parindent] 
\item For the majority of QCS protocols, we use end-to-end training, including a classical postprocessor as part of the trainable parameters. Here a classical postprocessing step $\mathcal{C}(\cdot)$ is applied to finitely-sampled measurement results (details in the next subsection), following which the cross-entropy loss is computed over the entire training dataset: 
\begin{align}
    \mathcal{L} = \mathcal{L}_{\rm ce}\left[  \left\{ \mathcal{C} \left( \bm{\bar{X}}_{\rm QCS}(\bm{u}_{(j)}^{(n)}) \right) \right\}_{n \in 1,\ldots,N_{\rm samp},~j\in 1,\ldots,C} \right].
    \label{appeq:lossce}
\end{align}
This cross-entropy loss $\mathcal{L}_{\rm ce}[\cdot]$, which is implemented using the \texttt{CrossEntropyLoss()} function in \texttt{PyTorch}, is minimized during end-to-end training. 
\item For the special case of training hybrid quantum computational sensors for binary classification tasks, we have used a slightly more tailored loss function. The loss function directly takes in finitely-sampled measurement results $X_{\rm QCS}$ obtained in the single-shot regime from single qubit readout of the quantum computational sensor (namely, without any classical postprocessing), and computes the quantity
\begin{align}
    \mathcal{L} = \frac{1}{2N_{\rm samp}}\left( \sum_{n}^{N_{\rm samp}} (X_{\rm QCS}(\bm{u}^{(n)}_{(1)})-0)^2 + \sum_{n}^{N_{\rm samp}} (X_{\rm QCS}(\bm{u}^{(n)}_{(2)})-1)^2 \right) + \mathcal{L}_{\rm tr}.
    \label{appeq:losserr}
\end{align}
Here the first term is simply the classification error over the training dataset. To see this, note that the summand in the first (second) term vanishes when the quantum computational sensor output $X_{\rm QCS}$ approaches $0$ ($1$) if the input belongs to class $j=1$ ($j=2$). Every time these conditions are not met, each term in the summand will contribute $1$; in other words, counting the occurrence of an error in the classification. The second term is a penalty term added to mitigate against Fock truncation effects. For a Fock cutoff of $N_{\rm tr}$, this takes the form $\mathcal{L}_{\rm tr} = 10\times{\rm ReLU}( \lambda - 10^{-3})$, where $\lambda = |\langle N_{\rm tr}-1|\psi(\alpha)\rangle|^2 + |\langle N_{\rm tr}|\psi(\alpha)\rangle|^2 $ is the total occupation of the two highest Fock number states for the final sensor state, $\ket{\psi(\alpha)}$, and ${\rm ReLU}(y)$ returns $y$ if $y>0$, and $0$ otherwise. Note that $\mathcal{L}_{\rm tr}$ is positive-definite, and becomes large if $\lambda > 10^{-3}$, namely if the occupation of the highest Fock states starts to become large. Minimizing this loss function therefore reduces the classification error (over the training dataset), while also avoiding very high occupation of the bosonic mode.
\end{enumerate}
For hybrid quantum computational sensors, we have also tried using end-to-end training using Eq.~(\ref{appeq:lossce}) (with $\mathcal{L}_{\rm tr}$ also included), and have not found a substantial difference in training performance. More broadly, we note that Eq.~(\ref{appeq:lossce}) can be used for general forms of outputs $X_{\rm QCS}$ from a quantum computational sensor, and allows for use of postprocessing layers, making it more versatile.

\subsubsection{Classical postprocessing}

For end-to-end training as described above, the trainable classical postprocessing layer $\mathcal{C}(\cdot)$ in Eq.~(\ref{appeq:lossce}) is implemented as a classical neural network based on a multilayer perceptron (MLP). The standard MLP we use has the structure $2^M \to 10 \to 10 \to C$, namely an input dimension of $2^M$ which is the measurement space of the quantum computational sensor comprising $M$ qubits, two hidden layers of 10 neurons each, and an output dimension of $C$ for $C$-class discrimination. The activation functions applied after each layer (apart from the output layer) are \texttt{ReLU} functions. For time-varying tasks using qubit-based quantum computational sensors, a slightly different MLP structure is used (see Appendix~\ref{app:meg} for details).

\subsection{Performance of training using finitely-sampled measurements}


\begin{figure}[t]
    \centering
    \includegraphics{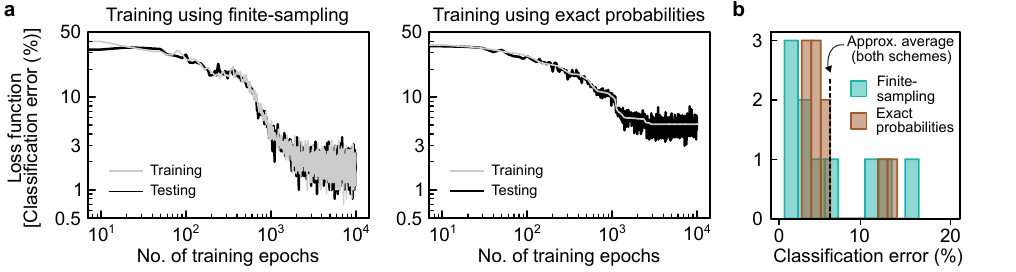}
    \caption{\textbf{Comparison of training using finite samples and exact probabilities.} We consider the training of a hybrid qubit-bosonic mode quantum computational sensor for the \circles{} task considered in Sec.~\ref{sec:hybrid} of the main text. \textbf{a}, Descent of loss function---here the classification error over the training dataset---as a function of number of training epochs, for training using finitely-sampled measurement results (left panel) and using exact probabilities (right panel). Also shown is the classification error over the testing set during the training procedure. \textbf{b}, Histogram of testing set classification error of ten trained QCS protocols using both training procedures. Dashed line shows average classification error, which here is approximately the same for both schemes. Training using finitely-sampled measurements exhibits larger variance in performance, but also provides trained instances with the lowest classification errors. }
    \label{appfig:fstrain}
\end{figure}


In this section we provide a comparison of training using finitely-sampled measurements against training using exact expectation values, e.g. probabilities for qubit state measurement. Of course, the performance of training in general depends on many factors, including the specific quantum system being trained, the classical postprocessor and whether it is included in training, the loss function, learning rates, initial parameter values, and so on. Nevertheless, we discuss some general findings in the context of one of our QCS examples: the training of a hybrid quantum computational sensor (Sec.~\ref{sec:hybrid} of the main text).

In Fig.~\ref{appfig:fstrain}a we plot the typical evolution of the loss function (given by Eq.~(\ref{appeq:losserr}) here) as a function of the number of training epochs for the training of hybrid quantum computational sensors. In the left panel, we show the result for training using finitely-sampled measurements, while the right panel shows the loss function when using exact probabilities. Immediately, we note that loss function for finitely-sampled measurement results is much more stochastic, owing to quantum sampling noise. In contrast, when using exact probabilities the loss function exhibits a much smoother evolution. Crucially, however, the loss function using finitely-sampled measurement results also exhibits a decrease over multiple training epochs. 

We also show the classification error over the testing set, i.e. during inference, which is stochastic for both cases since we always use a finite number of shots $S$ for inference. For the specific example shown, the training and testing accuracies achieved are in fact lower than that achieved for the example shown using exact probabilities. However, this is just one example: in Fig.~\ref{appfig:fstrain}b we plot a histogram of classification errors over the testing set for ten (10) QCS protocols trained using our finite-sampling procedure, and also using exact probabilities. We note that the spread of performance is larger for training using finite-sampling, an observation we have noted generally: the stochasticity due to quantum sampling noise appears to lead to a greater variability in training performance. However, this variability is not necessarily a detriment: we often find that while the finite-sampling procedure can produce QCS protocols that perform worse than those trained using exact probabilities, it is also likelier to produce QCS protocols that are outliers in terms of better performance. The introduction of stochasticity to training procedures is often used in machine learning, for example to escape local minima: here we find that noise that is intrinsic to quantum measurements appears to play a similar role.

\subsection{Number of measurement samples for training versus inference}

Our use of finitely-sampled measurement results for training raises a question that does not present itself when training using exact probabilities: how many samples should be used to calculate measurement results $\bm{\bar{X}}_{\rm QCS}$ for a given input $\bm{u}$? Should one choose the same $S$ as are intended to be used during inference? Or is there an advantage to training using as many measurement samples for a given input as possible (eventually approaching the `exact probabilities' limit), even if this far exceeds the number of shots that would be available during the inference phase?

\begin{figure}[t]
    \centering
    \includegraphics{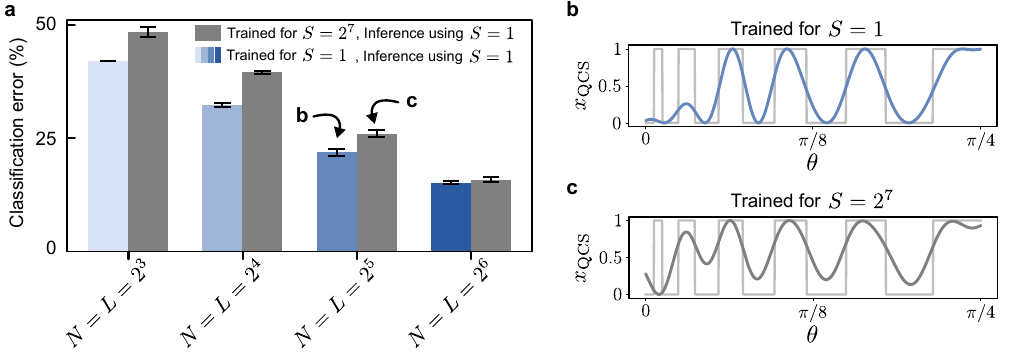}
    \caption{\textbf{Training using more shots than used during inference.} \textbf{a}, Classification error of single-shot ($S=1$) inference for QCS protocols trained using $S=1$ (color) and for QCS protocols trained using $S=2^7$, for varying $N=L$. \textbf{b}, Expected measurement outcomes $x_{\rm QCS}(\theta)$ for a QCS protocol trained using $S=1$. \textbf{c}, Expected measurement outcomes $x_{\rm QCS}(\theta)$ for a QCS protocol trained using $S=2^7$. Note the reduced contrast in comparison to \textbf{b}, allowed due to the uniform reduction in sampling noise at large $S$.   }
    \label{appfig:1Dfs}
\end{figure}

To answer this question, we consider single-qubit quantum computational sensors for binary classification tasks from Sec.~\ref{subsec:1q} of the main text, which have been trained using end-to-end training and cross-entropy-loss minimization, Eq.~(\ref{appeq:lossce}). For the results in the main text, we have used the same number of samples $S$ during training and inference. We reproduce the results of these QCS protocols with $L=N$ and for $S=1$ (single-shot training and inference), for a varying number of sensing periods $N$, in Fig.~\ref{appfig:1Dfs}a. Additionally, we now also calculate the classification error for QCS protocols trained using a much larger number of shots per input---here, $S=2^7$---but using only a single shot for inference; these results are shown in Fig.~\ref{appfig:1Dfs}a in gray. Note that QCS protocols trained using larger number of shots but used for inference at at $S=1$ always perform worse than the QCS protocols trained using $S=1$, even exhibiting a difference that can be as large as $8$-$10$ percentange points. We have found qualitatively similar results for two-qubit quantum computational sensors, for both binary and multi-class discrimination tasks.

This at-first surprising result implies that at least for end-to-end training and cross-entropy-loss minimization, one should train using the same number of shots as one intends to use during the inference phase. Some additional insight can be gleaned from the actual qubit response for the trained QCS protocols, $x_{\rm QCS}(\theta)$, learned during training. In Fig.~\ref{appfig:1Dfs}b, we plot $x_{\rm QCS}(\theta)$ for the single-variable binary classification task for a QCS protocol trained using $S=1$, while in Fig.~\ref{appfig:1Dfs}c, we plot $x_{\rm QCS}(\theta)$ for a QCS protocol trained using a much larger number of shots, $S=2^7$. Note that for lower shots during training, the learned response of the quantum computational sensor is directed towards the values $x_{\rm QCS}\to 0,1$, points at which sampling noise is reduced. The QCS protocol trained using more shots, on the other hand, exhibits a much lower contrast, and $x_{\rm QCS}$ values are much close to $0.5$ (where single-shot sampling noise would be largest). 

The reason for this is simple: at larger shots, the effect of quantum sampling noise is reduced. As a result, even for $x_{\rm QCS}$ values near $0.5$ where sampling noise would be large for a single shot, useful information can be obtained when $S=2^7$, allowing the QCS protocol to perform classification. For training using just a single shot, on the other hand, quantum sampling noise dominates. Therefore it appears that minimizing sampling noise is essential to reduce the loss function and to subsequently perform successful classification in this low-shot regime. We have observed a similar reduction in contrast for QCS protocols trained at larger number of shots for two-qubit quantum computational sensors.


It is quite possible that the outcomes of the training procedures observed here depend on the type of loss function used. For example, the loss function based on classification error, Eq.~(\ref{appeq:losserr}), directly attempts to force measurement results $X_{\rm QCS} \to 0,1$, which would therefore lead to trained protocols more like Fig.~\ref{appfig:1Dfs}b. However, for more general loss functions, or if we wish to incorporate training of a classical postprocessor with the training of the quantum system in an end-to-end fashion for more general tasks, matching the number of shots used during training and inference may be important to obtain the best possible QCS protocols.

\clearpage

\newpage

\section{Limitations of classical postprocessing in the presence of sampling noise}
\label{app:postp}

For all tasks considered in the main text, we allow a classical postprocessing layer to be applied to measurements from both the quantum sensor and the quantum computational sensor. In this Appendix section, we show that the performance difference between QS and QCS protocols observed across tasks in the main text (namely, the achieved QCSA) is \textit{not} due to a limitation of the complexity of these classical postprocessing resources, but is a fundamental limitation of the amount of information that can be extracted from noisy quantum measurement data.

To demonstrate this, we analyze the performance of both QS and QCS protocols for a selection of tasks analyzed in this work (tasks in Sec.~\ref{subsec:2q}, Sec.~\ref{sec:hybrid}, and Appendix~\ref{app:bqsp}), but now with increasing complexity of the classical postprocessing applied to measurement results from either protocol. To quantify the complexity of postprocessing, we consider the measurement outputs for both protocols to be passed through a classical neural network with the architecture of a multilayer perceptron. The complexity of the neural network is then given by the number of hidden layers with each layer including one application of a ReLU activation function; the structure is characterized by the total number of ReLU applications, $\texttt{nReLU}$. Note that within this structure, $\texttt{nReLU}=0$ corresponds to a postprocessing layer that is completely linear. This MLP structure with $\texttt{nReLU}=2$ (which is therefore nonlinear) is used for the classical neural networks considered in the main text.

\begin{figure}[h!]
    \centering
    \includegraphics{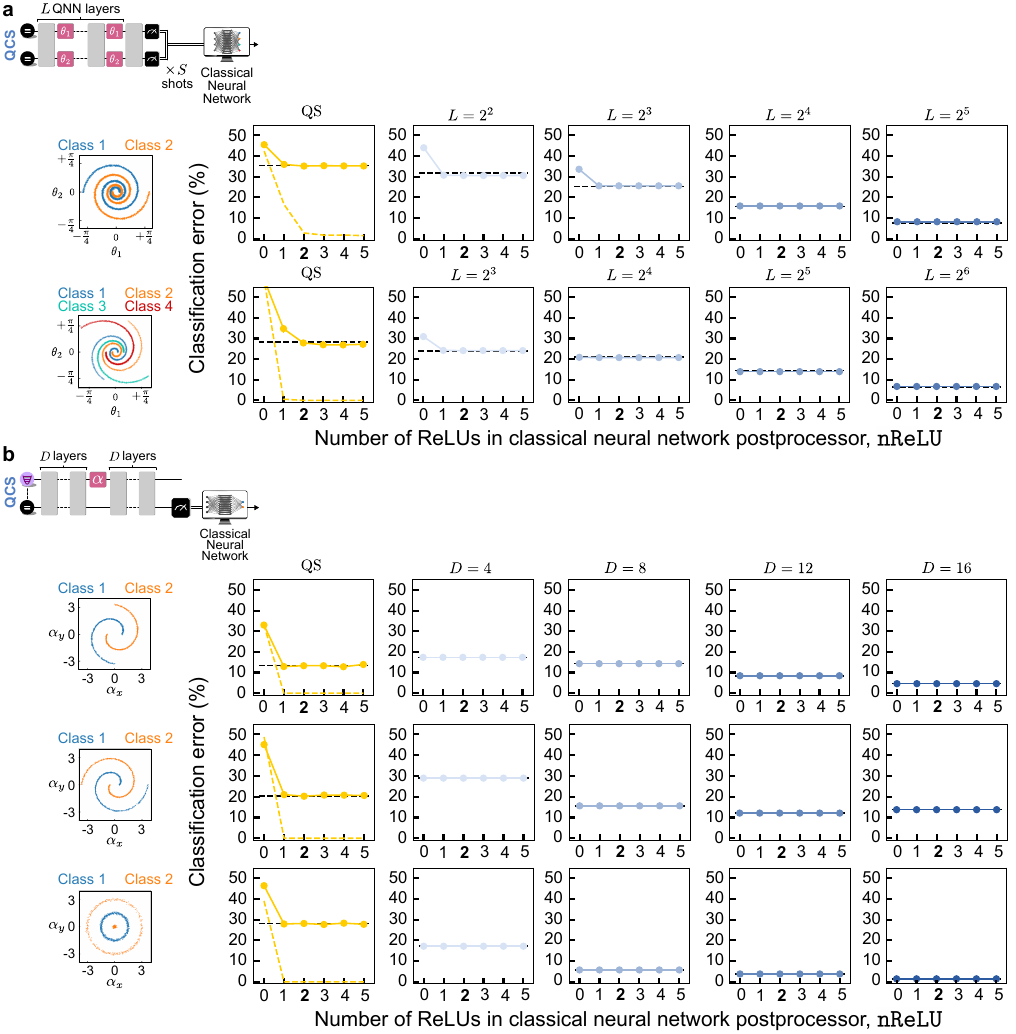}
    \caption{\textbf{Convergence of QS and QCS performance versus classical postprocessing resources.} We present classification errors of conventional QS schemes and QCS protocols for several classification tasks considered in the main text, as a function of the expressivity of the classical postprocessor applied to measurement results, to ensure that performance limitations are not due to the limitations of the postprocessing step. \textbf{a}, Classification error for a qubit-based conventional QS scheme, and QCS protocols for specific $L$, for the two classification tasks considered in Sec.~\ref{subsec:2q} of the main text.  \textbf{b}, Same as \textbf{a} but for hybrid quantum computational sensors considered in Sec.~\ref{sec:hybrid} of the main text, for specific $D$. Each row of plots corresponds to the classification task depicted on the left. Black dashed line is the best performance from the main text. Yellow dashed line for conventional QS (first column) is the classification error if the conventional QS protocol was allowed $N=S\to \infty$, hence neglecting noise in measurement; this quickly approaches 0\% error for the same $\texttt{nReLU}$, further emphasizing the role of sampling noise in limiting classification performance.  }
    \label{appfig:postp}
\end{figure}

We plot the performance of the two-qubit QS and QCS protocols from Sec.~\ref{subsec:2q}) as a function of $\texttt{nReLU}$ in Fig.~\ref{appfig:postp}a. The performance of the hybrid qubit-cavity quantum computational sensor and the QS protocol based on the phase-preserving amplifier, analyzed in Sec.~\ref{sec:hybrid}), are similarly plotted in Fig.~\ref{appfig:postp}b. Note that performance is unchanged beyond $\texttt{nReLU} = 2$; this is the neural network structure we have used for classical postprocessing in the static tasks considered in the main text. For QCS schemes where the nonlinear processing is done via the dynamics of the quantum computational sensor itself, even linear postprocessing ($\texttt{nReLU} = 0$) suffices to extract maximal information from the measurement data. For the QS scheme, the dashed yellow line shows the performance in the limit of infinite sampling; the classification error approaches zero, indicating that in the absence of noise, the postprocessing is sufficient to perform the considered tasks. This reiterates that the restriction on classification performance is due to noise in sampling.


\end{document}